\documentclass[
 reprint,
 amsmath,amssymb,
 aps,
]{revtex4-2}

\usepackage[dvipdfmx]{graphicx}
\usepackage{dcolumn}
\usepackage{bm}
\usepackage{color}
\usepackage[colorlinks=true, linkcolor=blue, citecolor=red, urlcolor=blue]{hyperref}
\usepackage{physics}
\usepackage{orcidlink}
\usepackage{tikz}
\usetikzlibrary{positioning, shapes.geometric}
\usetikzlibrary{arrows.meta}
\usetikzlibrary{calc}

\newcommand{\dbeta}{\Delta\!\beta}

\newcommand{\xprojection}{$\times$-projection}

\begin{document}

\preprint{APS/123-QED}

\title{Fast modeling of the shear three-point correlation function}

\author{Sunao Sugiyama\orcidlink{0000-0003-1153-6735}}\email{ssunao@sas.upenn.edu}
\author{Rafael C. H. Gomes \orcidlink{0000-0002-3800-5662}}
\author{Mike Jarvis \orcidlink{0000-0002-4179-5175}}
\affiliation{Department of Physics and Astronomy, University of Pennsylvania, Philadelphia, PA 19104, USA}

\date{\today}

\begin{abstract}
    The three-point correlation function (3PCF) of a weak lensing shear field contains information that is complementary to that in the two-point correlation function (2PCF), which can help improve the cosmological parameters and calibrate astrophysical and observational systematics parameters. However, the application of the 3PCF to observed data has been limited due to the computational challenges of calculating theoretical predictions of the 3PCF from a bispectrum model. In this paper, we present a new method to compute the shear 3PCF efficiently and accurately. We employ the multipole expansion of the bispectrum to compute the shear 3PCF, and show that the method is substantially more efficient than direct numerical integration. We found that the multipole-based method can compute the shear 3PCF with 5\% accuracy in 10 (40) seconds for the single (four) source redshift bin setup. The multipole-based method can be also used to compute the third-order aperture mass statistics quickly and accurately, accounting for the bin-averaging effect on the shear 3PCF. Our method provides a fast and robust tool for probing the underlying cosmological model with third-order statistics of weak lensing shear.
\end{abstract}

\maketitle

\section{Introduction}\label{sec:introduction}
In the analysis of wide field galaxy survey data, most of the works have relied on the two-point correlation function (2PCF) to extract cosmological information. The 2PCF is a powerful statistic that characterizes the clustering of galaxies or matter through weak lensing, and provides a wealth of information about the underlying cosmological model. In particular, the recent 2PCF analysis of weak lensing survey data put a tight constraint on the matter clustering parameter $S_8\equiv \sigma_8\sqrt{\Omega_m/0.3}$ at the level of several percent precision \cite{Heymans.Velander.2013, Hildebrandt.Waerbeke.2016, Troxel.Zhang.2018, Hikage.Yamada.2019, Hamana.Tanaka.2019, Asgari.Valentijn.2020, Amon.Weller.2021, Secco.To.2021, Dalal.Wang.2023, Li.Wang.2023}. This tight constraint is now getting comparable to the cosmological constraint from CMB data \citep{Collaboration.Zonca.2020}, and starts to indicate the so called $S_8$ tension, referring to the discrepancy between the late-time universe probed by galaxy surveys and the early universe probed by CMB data. 

However, the 2PCF is limited in its ability to capture the full information content of the weak-lens shear field. If the shear field is Gaussian distributed, the 2PCF contains the full information of the field, while in reality the field is highly non-Gaussian due to non-linear structure formation in the late-time Universe. Because of this non-Gaussianity of the weak lensing field, the 2PCF is not enough and one needs to use higher-order statistics in addition to the 2PCF to capture the full information of the field, e.g. higher-order moments \cite{Chang.Collaboration.2018,Gatti.collaboration.2020, Gatti.Collaboration.2022,Peel.Baldi.2018, Petri.Kratochvil.2015,Porth.Smith.2021, Waerbeke.Velander.2013,Vicinanza.Er.2016, Vicinanza.Er.2018}, peak counts \cite{Ajani.Liu.2020, Dietrich.Hartlap.2010,Harnois-Deraps.Reischke.2021, Kacprzak.Collaboration.2016,Kratochvil.May.2010,Liu.May.2015,Martinet.Nakajima.2017,Peel.Baldi.2018,Shan.Wang.2017,Zurcher.Kacprzak.2023,Zurcher.Refregier.2020}, one-point probability distributions \cite{Barthelemy.Gavazzi.2020,Boyle.Baldi.2021,Thiele.Smith.2020}, Minkowski functionals \cite{Grewal.Amon.2022,Kratochvil.Huffenberger.2012,Parroni.Scaramella.2019,Petri.Kratochvil.2015,Vicinanza.Tereno.2019}, Betti numbers \cite{Feldbrugge.Vegter.2019,Parroni.Scaramella.2020}, persistent homology \cite{Heydenreich.Martinet.2022,Heydenreich.Schneider.2022}, scattering transform coefficients \cite{Cheng.Bruna.2020,Valogiannis.Dvorkin.2022,Valogiannis.Dvorkin.2022m5}, wavelet phase harmonics \cite{Allys.Mallat.2020}, and kNN and CDF \cite{Anbajagane.Wiseman.2023,Banerjee.Abel.2022}, map-level inference \cite{Boruah.Hudson.2022,Porqueres.Lavaux.2021}, and machine-learning methods \cite{Fluri.Schneider.2019,Fluri.Hofmann.2018,Jeffrey.Lanusse.2020,Lu.Li.2023,Ribli.Csabai.2019}.
These higher-order statistics offers additional statistical power on the cosmological parameters that the 2PCF constrains and also offers different sensitivities to cosmological parameters compared to the 2PCF and can be used to test aspects of cosmological models that the 2PCF cannot adequately probe.

One of the most natural statistics beyond the 2PCF is the three-point correlation function (3PCF). The 3PCF has been extensively studied in the context of galaxy surveys\citep{Schneider.Lombardi.2002,Schneider.Lombardi.2003,Zaldarriaga.Scoccimarro.2003} and it has been shown that the information content of the 3PCF is complementary to the 2PCF to give a tighter constraint on the cosmological parameters\citep{Takada.Jain.2004, Heydenreich.Schneider.2022}. With stage-III weak lensing survey data, the shear 3PCF and third-order aperture mass statistics are measured \citep{Secco.Weller.2022,Porth.Schneider.2023} and combined with the 2PCF to more tightly constrain the cosmological parameters\citep{Burger.Martinet.2023}. However, the application of the shear 3PCF analysis for cosmology inference has been limited until recently due to the computational challenges associated with measuring the 3PCF from galaxy shape catalogs and calculating its theoretical predictions. The naive computational cost of measuring the 3PCF from the shapes of $N$ galaxies scales as $\mathcal{O}(N^3)$, making it impractical for large datasets. \citet{Porth.Schneider.2023} developed a new algorithm to compute the 3PCF based on the multipole expansion, reducing the computational cost to $\mathcal{O}(N \log N)$\footnote{Formally, the implementation given in \citet{Porth.Schneider.2023} is $O(N^2)$, however, in practice using hierarchical grids or a tree structure, the algorithm can be converted to $O(N \log N)$}. This method builds on previous work by \cite{Chen.Szapudi.2005, Slepian.Eisenstein.2015, Philcox.Eisenstein.2021} on the 3PCF of galaxy clustering, and extends their method to the spin-2 shear 3PCF.

The theoretical prediction of the shear 3PCF faces similar computational challenges. The 3PCF is the Fourier transform of the bispectrum, involving a three-dimensional integral. Additionally, the shear field is a spin-2 field, adding complexity due to oscillating phase factors in the integrand, which complicates numerical integration. When we estimate the cosmological parameters by fitting the theoretical model to the measured 3PCF, we evaluate the 3PCF at $\sim \!10^{3\text{-}4}$ Monte-Carlo sampling points in the parameter space, so it is crucial to have a fast and accurate method to compute the theoretical prediction of the 3PCF. 

In this paper, we employ the same idea of a multipole expansion as \citet{Porth.Schneider.2023} used for the efficient measurement algorithm. This approach has been previously developed and investigated in the context of the galaxy clustering 3PCF \cite{Umeh.Umeh.2020, Guidi.Carbone.2022, Aviles.Niz.2023}. We extend the multipole-based method for theoretical predictions to the 3PCF of spin-2 shear fields. This multipole-based approach allows for the separation of the opening angle dependence of the triangle in both real and Fourier space, enabling analytical integration over the angle. This method is substantially more efficient than direct numerical integration, with evaluations taking only approximately 10 seconds. 

This paper is organized as follows. In section~\ref{sec:shear-3pcf}, we review the basic properties of weak lensing shear statistics. In section~\ref{sec:formalism}, we present the formalism of the multipole-based method to compute the shear 3PCF efficiently. We also discuss the binning effect of the 3PCF signal. In section~\ref{sec:result}, we present the results of the multipole-based method: we discuss the stability of the computation, the accuracy of the method, and the computational efficiency compared to existing methods. Finally we conclude the paper in section~\ref{sec:conclusion}.

\section{Third-order statistics of cosmic shear}\label{sec:shear-3pcf}
\subsection{Connection of shear and matter fields}
\begin{figure}
    \centering
    \includegraphics[width=\linewidth]{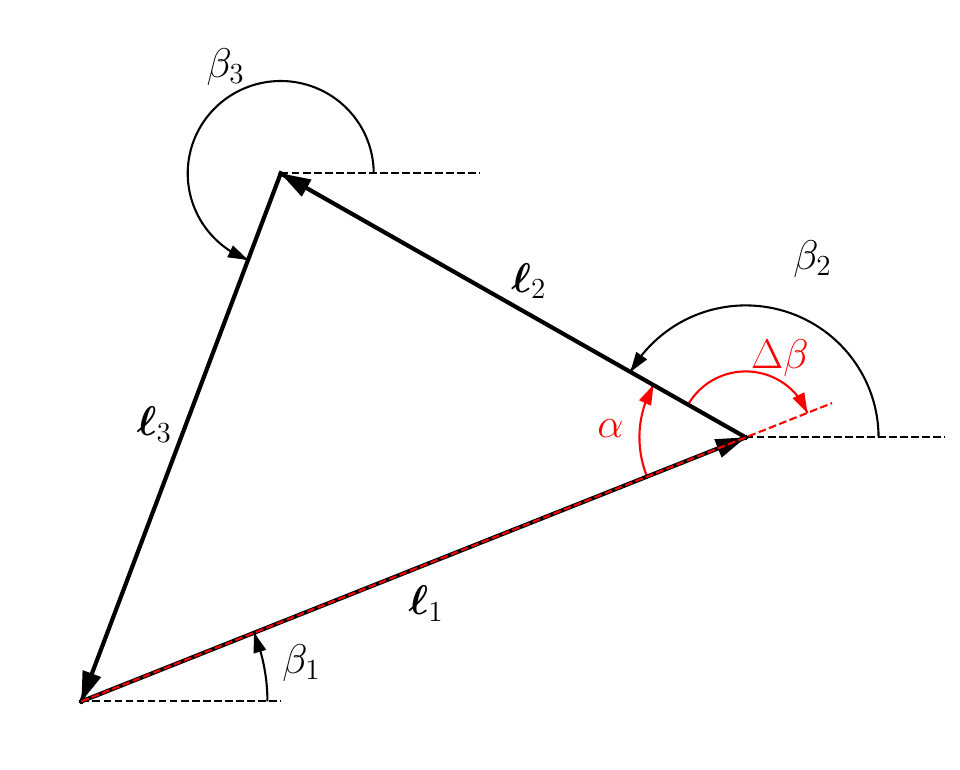}
    \caption{The parametrization of triangle in Fourier space. Because of the statistical isotropy, three vectors satisfy $\sum_{i=1}^3\bm{\ell}_i=0$, forming a triangle. The polar angle of each Fourier vector, $\bm{\ell}_i$, measured from the x-axis is denoted by $\beta_i$. The multipole expansion is applied to the inner angle $\alpha$, measured from $\ell_1$ side to $\ell_2$ side. The outer angle is defined by $\Delta\beta\equiv\beta_1-\beta_2=\pi-\alpha$.}
    \label{fig:triangle-config}
\end{figure}

Weak lensing is the effect of the shape distortion of the background image due to the intervening matter density field of large-scale structure. The two most fundamental quantities of weak lensing effect is the convergence $\kappa$ and the shear $\gamma$, which quantify the isotropic stretch and shape distortion of the background image respectively. The convergence field at a position $\bm{X}$ on the sky plane is given by the line-of-sight projection of the matter density contrast $\delta_{\rm m}$ weighted by the lensing efficiency function,
\begin{align}
    \kappa(\bm{X}) &= \frac{3\Omega_{\rm m}H_0^2}{2c^2}\int_0^{\infty}\dd\chi ~q(\chi)\frac{\delta_{\rm m}\left(\chi\bm{X}; z(\chi)\right)}{a(\chi)}\\
    q(\chi) &= \int_\chi^\infty \dd\chi' p(\chi')\frac{\chi'-\chi}{\chi'},\label{eq:lensing-efficiency}
\end{align}
where the lensing efficiency $q(\chi)$ is averaged with the source probability distribution $p(\chi)$, satisfying the normalization $\int_0^\infty\dd\chi~p(\chi)=1$.
Weak lensing shear is a spin-2 quantity, which in a Cartesian frame can be described by the complex form: $\gamma_{\rm c}=\gamma_1+i\gamma_2$, with $\gamma_{1}$ and $\gamma_2$ denoting the shear components along the x-axis and an axis rotated 90 degrees from the x-axis, respectively. When the shear is projected onto another reference frame which is rotated by $\zeta$ from the Cartesian frame, the shear field follows the spin-2 transformation,
\begin{align}
    \gamma(\bm{X};\zeta)
    \equiv \gamma_{\rm t}(\bm{X};\zeta)+i\gamma_\times(\bm{X};\zeta)
    = -\gamma_{\rm c}(\bm{X}) e^{-2i\zeta}.
\end{align}
We defined the Fourier transformation of shear as 
\begin{align}
    \gamma_{\rm c}(\bm{X}) = \int\frac{\dd^2\bm{\ell}}{(2\pi)^2}\hat{\gamma}_{\rm c}(\bm{\ell})e^{-i\bm{\ell}\cdot\bm{X}},
\end{align}
and same for convergence field $\kappa$. The convergence and shear fields are related to each other in Fourier space:
\begin{align}
    \hat{\gamma}_{\rm c}(\bm{\ell}) = \hat{\kappa}(\bm{\ell})e^{2i\beta},
    \label{eq:shear-convergence-relation}
\end{align}
where $\beta$ is the polar angle of Fourier mode $\bm{\ell}$.

The power spectrum and bispectrum of the convergence field are defined by
\begin{align}
    \langle\hat{\kappa}(\bm{\ell}_1)\hat{\kappa}(\bm{\ell}_2)\rangle 
    =& (2\pi)^2\delta^{\rm D}(\bm{\ell}_1+\bm{\ell}_2)P_\kappa(\ell_1)\\
    \langle\hat{\kappa}(\bm{\ell}_1)\hat{\kappa}(\bm{\ell}_2)\hat{\kappa}(\bm{\ell}_3)\rangle 
    =& (2\pi)^2\delta^{\rm D}\left(\sum\nolimits_{i=1}^3\bm{\ell}_i\right)\nonumber\\
    &\times B_\kappa(\bm{\ell}_1, \bm{\ell}_2)
\end{align}
Here the Dirac delta arises because of the statistical uniformity of the convergence field. Because of this Dirac delta function, the three Fourier vectors forms a triangle for the bispectrum, which is depicted in Fig.~\ref{fig:triangle-config}. The dependence of the bispectrum on its Fourier modes can be reduced further by using statistical isotropy and by noting that the triangle can be uniquely identified by the two side lengths and the \textit{inner} angle between them; to clearly express this dependency, we introduce side-angle-side (SAS) notation as follows.
\begin{align}
    B_\kappa(\bm{\ell}_1, \bm{\ell}_2) = b_\kappa(\ell_1, \ell_2, \alpha).
\end{align}
Note that the inner angle $\alpha$ is related to the polar angles of Fourier vectors $\bm{\ell}_{1,2}$ as
\begin{align}
    \alpha = \beta_2 - (\beta_1 - \pi) \equiv \pi - \dbeta,
\end{align}
where $\dbeta$ is the \textit{outer} angle. Weak lensing produces only E-mode and consequently, the bispectrum due to weak lensing is mirror symmetric,
\begin{align}
    b_\kappa(\ell_1, \ell_2, \alpha)=b_\kappa(\ell_1, \ell_2, -\alpha),
    \label{eq:bispectrum-mirror-even}
\end{align}
meaning that the bispectrum depends on its inner angle $\alpha$ only through its cosine.

The convergence power spectrum and bispectrum are related to those of the matter density field under the flat sky and Limber approximations as 
\begin{align}
    P_\kappa(\ell) 
    &= \left(\frac{3\Omega_{\rm m}H_0^2}{2c^2}\right)^2\int_0^\infty \dd\chi\frac{q^2(\chi)}{a^2(\chi)}\nonumber\\
    &\hspace{4em}\times P_{\rm m}\left(\frac{\ell}{\chi};z(\chi)\right)\\
    B_\kappa(\ell_1, \ell_2, \ell_3)
    &= \left(\frac{3\Omega_{\rm m}H_0^2}{2c^2}\right)^3\int_0^\infty \dd\chi\frac{q^3(\chi)}{a^3(\chi)\chi}\nonumber\\
    &\hspace{4em}\times B_{\rm m}\left(\frac{\ell_1}{\chi}, \frac{\ell_2}{\chi}, \frac{\ell_3}{\chi};z(\chi)\right),\label{eq:kbispectrum-mbispectrum}
\end{align}
where $P_{\rm m}$ and $B_{\rm m}$ are the matter power- and bispectrum in three dimensional space defined as
\begin{align}
    \langle\hat{\delta}_{\rm m}(\bm{k}_1;z)\hat{\delta}_{\rm m}(\bm{k}_2;z)\rangle=
    &(2\pi)^3\delta^{\rm D}(\bm{k}_1+\bm{k}_2)\nonumber\\
    &\times P_{\rm m}(k_1;z)\\
    \langle\hat{\delta}_{\rm m}(\bm{k}_1;z)\hat{\delta}_{\rm m}(\bm{k}_2;z)\hat{\delta}_{\rm m}(\bm{k}_3;z)\rangle=
    &(2\pi)^3\delta^{\rm D}\left(\sum\nolimits_{i=1}^3\bm{k}_i\right)\nonumber\\
    &\times B_{\rm m}(\bm{k}_1,\bm{k}_2;z).
\end{align}
In this paper, we consider Bihalofit for the model of the matter bispectrum \citep{Takahashi.Shirasaki.2019}. Bihalofit predicts the matter bispectrum at a given redshift and a set of three comoving Fourier modes in a given cosmology based on a fitting formula calibrated with $N$-body simulations. We perform the line-of-sight integration in Eq.~(\ref{eq:kbispectrum-mbispectrum}) to obtain the convergence bispectrum. Throughout this paper, we use the cosmology from Planck 2018, where $\sigma_8=0.8102, \Omega_m=0.3111, \Omega_{\Lambda}=0.6889, h=0.6766$.

\subsection{Cosmic shear 3PCF}\label{subsec:shear-3pcf}
\begin{figure}
    \centering
    \includegraphics[width=\linewidth]{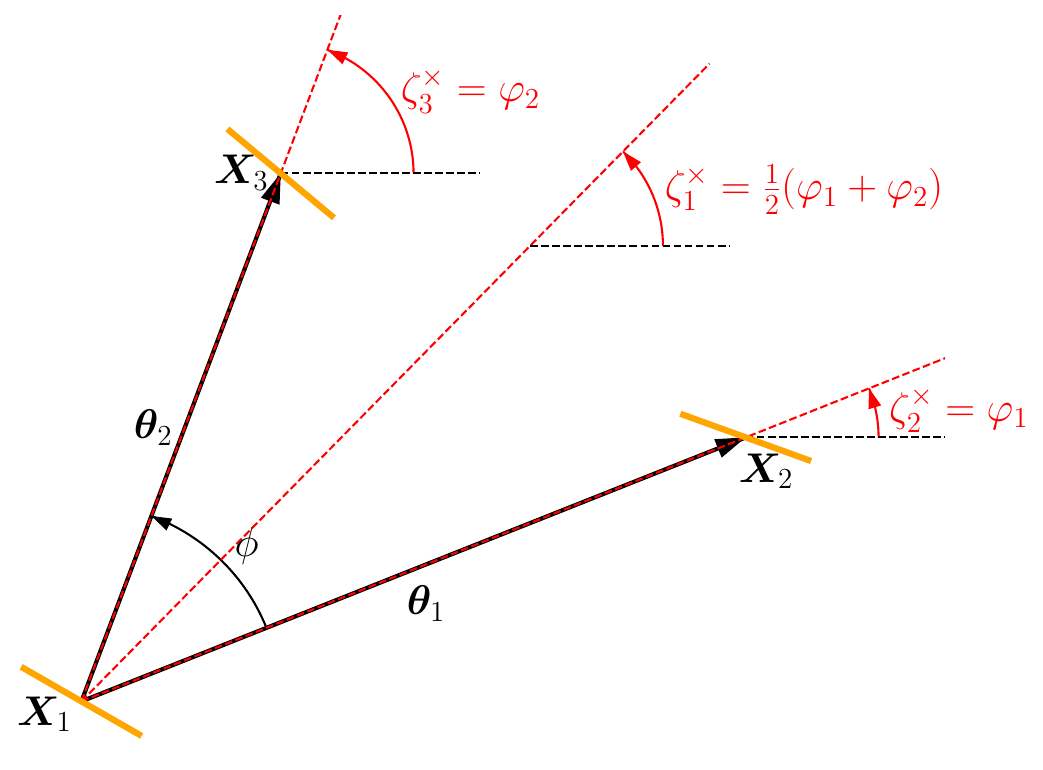}
    \caption{The definition of \xprojection{}. The shear is located at the three different positions $\bm{X}_1, \bm{X}_2$ and $\bm{X}_3$ on the sky, forming an triangle. The triangle is uniquely specified by the two side lengths, $\theta_1$ and $\theta_2$, and the inner angle between them $\phi$, (side-angle-side notation) where we define the angle measured from side $\theta_1$ to $\theta_2$. The \xprojection{} is defined by the angles, $\zeta_1^\times, \zeta_2^\times$ and $\zeta_3^\times$, denoted in red, and the shear on each vertex of the triangle is projected the corresponding \xprojection{} angle.}
    \label{fig:x-projection}
\end{figure}

Because the shear has two components, there is some freedom in choice of shear components with which we construct the three-point correlation function. \citet{Schneider.Lombardi.2003} developed the \textit{natural components} of the shear 3PCF. There are four natural components of the shear 3PCF, which do not mix together under any rotations of the triangle. For a fixed shape of triangle, we still have a degree of freedom for the choice of reference point on which we define the shear components, called shear projection. They showed that each of the natural components is invariant under a change of shear projections up to a phase factor. With arbitrary choice of shear projection $\mathcal{P}$, the natural components of the shear 3PCF are defined as
\begin{align}
    \label{eq:Gamma0-P}
    \Gamma_0^\mathcal{P}(\theta_1, \theta_2, \phi) = 
    \langle 
    \gamma\hspace{.1cm}(\bm{X}_1; \zeta_1^\mathcal{P})
    \gamma\hspace{.1cm}(\bm{X}_2; \zeta_2^\mathcal{P})
    \gamma\hspace{.1cm}(\bm{X}_3; \zeta_3^\mathcal{P})
    \rangle,\hspace{-.5em} \\
    \label{eq:Gamma1-P}
    \Gamma_1^\mathcal{P}(\theta_1, \theta_2, \phi) = 
    \langle 
    \gamma^{*}\hspace{-.05cm}(\bm{X}_1; \zeta_1^\mathcal{P})
    \gamma\hspace{.1cm}(\bm{X}_2; \zeta_2^\mathcal{P})
    \gamma\hspace{.1cm}(\bm{X}_3; \zeta_3^\mathcal{P})
    \rangle,\hspace{-.5em} \\
    \label{eq:Gamma2-P}
    \Gamma_2^\mathcal{P}(\theta_1, \theta_2, \phi) = 
    \langle
    \gamma\hspace{.1cm}(\bm{X}_1; \zeta_1^\mathcal{P})
    \gamma^{*}\hspace{-.05cm}(\bm{X}_2; \zeta_2^\mathcal{P})
    \gamma\hspace{.1cm}(\bm{X}_3; \zeta_3^\mathcal{P})
    \rangle,\hspace{-.5em} \\
    \label{eq:Gamma3-P}
    \Gamma_3^\mathcal{P}(\theta_1, \theta_2, \phi) = 
    \langle 
    \gamma\hspace{.1cm}(\bm{X}_1; \zeta_1^\mathcal{P})
    \gamma\hspace{.1cm}(\bm{X}_2; \zeta_2^\mathcal{P})
    \gamma^{*}\hspace{-.05cm}(\bm{X}_3; \zeta_3^\mathcal{P})
    \rangle.\hspace{-.5em}
\end{align}
In Fig.~\ref{fig:x-projection}, we show the parametrization of the triangle. In this paper, we specify the triangle with the side-angle-side (SAS) notation, $\theta_1, \theta_2$ and $\phi$ because this parametrization is suitable for the method of multipole expansion developed in this paper. Note that the inner angle $\phi$ is measured from side $\theta_1$ to $\theta_2$. 

Under the exchange of the first two arguments of the natural components, they transforms to each other:
\begin{align}
    \label{eq:Gamma0-reflect}
    &\Gamma_0^\mathcal{P}(\theta_1, \theta_2, \phi) = \Gamma_0^\mathcal{P}(\theta_2, \theta_1, -\phi)\\
    \label{eq:Gamma1-reflect}
    &\Gamma_1^\mathcal{P}(\theta_1, \theta_2, \phi) = \Gamma_1^\mathcal{P}(\theta_2, \theta_1, -\phi)\\
    \label{eq:Gamma2-reflect}
    &\Gamma_2^\mathcal{P}(\theta_1, \theta_2, \phi) = \Gamma_3^\mathcal{P}(\theta_2, \theta_1, -\phi)\\
    \label{eq:Gamma3-reflect}
    &\Gamma_3^\mathcal{P}(\theta_1, \theta_2, \phi) = \Gamma_2^\mathcal{P}(\theta_2, \theta_1, -\phi)
\end{align}
Note that these will be useful later to reduce the number of independent quantities that need to be calculated.

We follow \citet{Porth.Schneider.2023} for the 3PCF triangle notation. The public measurement code TreeCorr\footnote{\url{https://github.com/rmjarvis/TreeCorr}} \cite{Jarvis.Jain.2003} parametrizes the triangle in a different way and care is needed for the comparison of the predictions by this paper and the measurements by TreeCorr. TreeCorr defines $d_2$ ($d_3$) as the side connecting vertices $\bm{X}_1$ and $\bm{X}_3$ ($\bm{X}_1$ and $\bm{X}_2$), and the opening angle $\phi^{\rm TC}$ to sweep from $d_2$ to $d_3$. Comparing with our parametrization, we have $\theta_1=d_3$ $\theta_2=d_2$ and $\phi=-\phi^{\rm TC}$. Especially, using the relations in Eqs.~(\ref{eq:Gamma0-reflect})--(\ref{eq:Gamma3-reflect}), we have
\begin{align}
    \Gamma_0^{\mathcal{P}}(d_2, d_3, \phi^{\rm TC}) = \Gamma_0^{\mathcal{P},{\rm TC}}(d_2, d_3, \phi^{\rm TC}) \nonumber\\
    \Gamma_1^{\mathcal{P}}(d_2, d_3, \phi^{\rm TC}) = \Gamma_1^{\mathcal{P},{\rm TC}}(d_2, d_3, \phi^{\rm TC}) \nonumber\\
    \Gamma_2^{\mathcal{P}}(d_2, d_3, \phi^{\rm TC}) = \Gamma_3^{\mathcal{P},{\rm TC}}(d_2, d_3, \phi^{\rm TC}) \nonumber\\
    \Gamma_3^{\mathcal{P}}(d_2, d_3, \phi^{\rm TC}) = \Gamma_2^{\mathcal{P},{\rm TC}}(d_2, d_3, \phi^{\rm TC}) \nonumber
\end{align}
where we denote the 3PCF defined in TreeCorr as $\Gamma_\mu^{\mathcal{P},{\rm TC}}$.

\subsection{Choice of shear projection}\label{subsec:shear-projection}
In this paper, we mainly focus on the \xprojection{} of the shears\cite{Porth.Schneider.2023}:
\begin{align}
    \zeta_1^\times = \frac{1}{2}(\varphi_1+\varphi_2), \hspace{1em}
    \zeta_2^\times = \varphi_1, \hspace{1em}
    \zeta_3^\times = \varphi_2
\end{align}
as shown in Fig.~\ref{fig:x-projection}. The natural components with \xprojection{} are written in terms of the shears in the Cartesian frame as
\begin{align}
    \label{eq:Gamma0-x}
    \Gamma_0^{\times}(\theta_1, \theta_2, \phi) = &
    -\langle
    \gamma_{\rm c}(\bm{X}_1)
    \gamma_{\rm c}(\bm{X}_2)
    \gamma_{\rm c}(\bm{X}_3)
    \rangle \nonumber\\
    &\hspace{2em}\times e^{-3i(\varphi_1+\varphi_2)},\hspace{-.5em}\\
    \label{eq:Gamma1-x}
    \Gamma_1^{\times}(\theta_1, \theta_2, \phi) = &
    -\langle
    \gamma_{\rm c}^{*}(\bm{X}_1)
    \gamma_{\rm c}(\bm{X}_2)
    \gamma_{\rm c}(\bm{X}_3)
    \rangle \nonumber\\
    &\hspace{2em}\times e^{-i(\varphi_1+\varphi_2)},\hspace{-.5em}\\
    \label{eq:Gamma2-x}
    \Gamma_2^{\times}(\theta_1, \theta_2, \phi) = &
    -\langle
    \gamma_{\rm c}(\bm{X}_1)
    \gamma_{\rm c}^{*}(\bm{X}_2)
    \gamma_{\rm c}(\bm{X}_3)
    \rangle\nonumber\\
    &\hspace{2em}\times e^{i(\varphi_1-3\varphi_2)},\hspace{-.5em}\\
    \label{eq:Gamma3-x}
    \Gamma_3^{\times}(\theta_1, \theta_2, \phi) = &
    -\langle
    \gamma_{\rm c}(\bm{X}_1)
    \gamma_{\rm c}(\bm{X}_2)
    \gamma_{\rm c}^{*}(\bm{X}_3)
    \rangle \nonumber\\
    &\hspace{2em}\times e^{i(-3\varphi_1+\varphi_2)}.\hspace{-.5em}
\end{align}
Although we extensively use the \xprojection{} in this paper, one can always convert the natural components in Eqs.~(\ref{eq:Gamma0-x})~-~(\ref{eq:Gamma3-x}) to those with arbitrary projection by multiplying each component by the appropriate phase factor. In particular, one can use Eqs.~(12)~-~(15) in \cite{Porth.Schneider.2023} to obtain natural components with the more typical centroid projection from the \xprojection{}.

\section{Formalism}\label{sec:formalism}
\subsection{Derivation of the formalism}\label{subsec:derivation}
In this work we develop the multipole expansion of third-order cosmic-shear statistics. We first focus on the $\Gamma_0^\times$, and show the results for other projections later. We start with the bispectrum multipole expansion. As we have seen in the last section, the bispectrum due to weak lensing is the mirror symmetric function which depends on the inner angle $\alpha$ only through its cosine, we choose Legendre Polynomial as the basis of the multipole expansion:
\begin{align}
    b_\kappa(\ell_1, \ell_2, \alpha) = \sum_{L=0}^{\infty} b_\kappa^{(L)}(\ell_1, \ell_2)P_L(\cos\alpha)
    \label{eq:bispectrum-multipole-expansion}.
\end{align}
The multipoles of the function can be obtained from the convolution with the Legendre polynomial basis,
\begin{align}
\begin{split}
    b_\kappa^{(L)}(\ell_1, \ell_2) &= \frac{2L+1}{2}\int_{-1}^{1}{\dd}(\cos\alpha)\\
    &\hspace{3em}\times b_\kappa(\ell_1, \ell_2, \alpha)P_L(\cos\alpha)
    \label{eq:bispectrum-multipole}.
\end{split}
\end{align}
We note that we can use the multipole expansion with Legendre polynomials only for the mirror symmetric bispectrum. If the bispectrum model also has a mirror asymmetric part, e.g. a tidal alignment and tidal torquing (TATT) model of intrinsic alignment\citep{Schmitz.Krause.2018}, we need to use another basis including an odd function of $\alpha$ for multipole expansion, e.g. Fourier basis $e^{iL\alpha}$. This paper focuses on the weak lensing bispectrum and uses the multipole expansion with Legendre polynomials, but the extension to the case including mirror asymmetric term using Fourier basis is straightforward and discussed in Appendix~\ref{sec:other-coupling-functions}.

We then express $\Gamma_0^\times$ in terms of the convergence bispectrum to derive its multipole-expansion form. Using the relationship between the shear field and the convergence field in Fourier space [Eq.~(\ref{eq:shear-convergence-relation})], we obtain
\begin{align}
\begin{split}
    \Gamma_0^{\times}(\theta_1, \theta_2, \phi) = 
    &-\int\frac{\dd^2\bm{\ell_1}}{(2\pi)^2}\frac{\dd^2\bm{\ell_2}}{(2\pi)^2}
    e^{-i\bm{\ell_1}\cdot\bm{\theta}_1-i\bm{\ell_2}\bm{\theta}_2} \\
    &\times
    b_\kappa(\ell_1, \ell_2, \alpha) 
    e^{2i\sum_i\beta_i} 
    e^{-3i(\varphi_1+\varphi_2)}. 
    \label{eq:Gamma0-by-bispectrum}
\end{split}
\end{align}
Now we will go through a series of substitutions to reduce the above four dimensional integral to a simpler two dimensional integral and the multipole re-summation.
By substituting the expression of the bispectrum multipole expansion in Eq.~(\ref{eq:bispectrum-multipole-expansion}) and the plane wave expansion using several identities of special functions summarized in Appendix~\ref{sec:identities} into the above equation, we obtain
\begin{widetext}
\begin{align}
\begin{split}
    \Gamma_0^{\times}(\theta_1, \theta_2, \phi) 
    &= -\sum_{L,m,n}i^{-(m+n)}
    \frac{1}{(2\pi)^4}\int\dd\ln\ell_1\dd\ln\ell_2\ell_1^2\ell_2^2~
    b_\kappa^{(L)}(\ell_1,\ell_2)J_m(\ell_1\theta_1)J_n(\ell_2\theta_2)\\
    &\hspace{2em}\times
    \int\dd\beta_1\dd\beta_2P_L(\cos\alpha)
    e^{2i\sum_i\beta_i}
    e^{im(\beta_1-\varphi_1)}
    e^{in(\beta_2-\varphi_2)}
    e^{-3i(\varphi_1+\varphi_2)}.
    \label{eq:Gamma0-temp}
\end{split}
\end{align}
\end{widetext}
Here the second line of the above equation is the integral over two polar angles of Fourier modes, and the integrand includes only mathematical functions, i.e. it is independent of the cosmological model, and hence enables us further calculation. Following ~\citet{Schneider.Lombardi.2003}, we first change the integral variables from two polar angles to the rotational angle and the outer angle so that we can carry out the integral over the rotation angle,
\begin{align}
    \begin{cases}
    \beta_1 &= \beta + \dbeta/2 \\
    \beta_2 &= \beta - \dbeta/2 \\
    \beta_3 &= \beta + \bar{\beta}.
    \end{cases}
\end{align}
Because of the triangle condition of three Fourier modes ($\sum_{i=1}^3\bm{\ell}_i=0$), $\bar\beta$ is not an independent variable but a function of $\dbeta$ and the ratio of two Fourier modes,
\begin{align}
    \begin{cases}
    \cos[2\bar\beta(\psi, \dbeta)]
    &= {\cal N}[(\ell_1^2+\ell_2^2)\cos\dbeta + 2\ell_1\ell_2] \\
    &= {\cal N}\ell^2[\cos\dbeta + \sin(2\psi)] \\
    \sin[2\bar\beta(\psi, \dbeta)]
    &= {\cal N}(\ell_1^2-\ell_2^2)\sin\dbeta\\
    &= {\cal N}\ell^2\cos(2\psi)\sin\dbeta
    \end{cases}
    \label{eq:angbar-sincos}
\end{align}
where we defined $\ell_1=\ell\cos\psi$ and $\ell_2=\ell\sin\psi$, and $\mathcal{N}$ is a normalization. Note that the range of $\psi$ is $[0,\pi/2]$. In a signed (mirror-distinguished) triangle, the exchange of two sides $\ell_1\leftrightarrow\ell_2$  (or equivalently $\psi\rightarrow\pi/2-\psi$) is equivalent to flipping the sign of the outer angle between them $\dbeta\rightarrow-\dbeta$, which leads to the following symmetry of $\bar\beta$ function:
\begin{align}
    e^{2i\bar\beta(\pi/2-\psi,~\dbeta)} = e^{2i\bar\beta(\psi,~-\dbeta)} = e^{-2i\bar\beta(\psi,~\dbeta)}.
    \label{eq:angbar-symmetry}
\end{align}
By changing the integration variables in the second line of Eq.~(\ref{eq:Gamma0-temp}), we obtain
\begin{align}
    &\int\dd(\dbeta)\dd\beta P_L(\cos\alpha)e^{i(6+m+n)[\beta-(\varphi_1+\varphi_2)/2]}\nonumber\\
    &\hspace{3em}\times
    e^{i[2\bar\beta+(m-n)(\dbeta+\phi)/2]}\nonumber\\
    =&(-1)^Le^{iM\phi}\delta^{K}_{6+m+n,0}\nonumber\\
    &\hspace{2em}\times4\pi\int_0^{\pi}\dd(\dbeta) P_L(\cos\dbeta) \cos(2\bar\beta+M\dbeta)\nonumber\\
    \equiv& (-1)^Le^{iM\phi}\delta^{K}_{6+m+n,0}G_{LM}(\psi),
    \label{eq:GLM}
\end{align}
where we performed the $\beta$ integral to obtain a constraint condition $6+m+n=0$ from the first to second line, defined $M\equiv3+m$ and the multipole coupling function $G_{LM}(\psi)$. Using the symmetry of the $\bar\beta$ function in Eq.~(\ref{eq:angbar-symmetry}), one can find a corresponding symmetry for the multipole coupling function $G_{LM}(\psi)$
\begin{align}
    G_{LM}(\pi/2-\psi) = G_{L(-M)}(\psi). 
    \label{eq:GLM-symmetry}
\end{align}
As we will see below, $M$ represents the index of multipoles of the 3PCF components, and hence the function $G_{LM}$ describe the coupling of multipoles between the bispectrum and the 3PCF. 
By substituting these expressions into Eq.~(\ref{eq:Gamma0-temp}), we now reach to the multipole expansion of the zero-th component:
\begin{align}
    \Gamma_0^{\times}(\theta_1, \theta_2, \phi) 
    &= \frac{1}{2\pi}\sum_{M=-\infty}^{\infty} 
    e^{iM\phi}\Gamma_0^{\times, (M)}(\theta_1, \theta_2)
    \label{eq:Gamma0-multipole-expansion}\\
    \Gamma_0^{\times, (M)}(\theta_1, \theta_2) 
    &= \frac{1}{(2\pi)^3}
    \int\dd\ln\ell_1\dd\ln\ell_2\ell_1^2\ell_2^2 H_M(\ell_1, \ell_2)\nonumber\\
    &\hspace{2em}\times
    J_{M-3}(\ell_1\theta_1)J_{-M-3}(\ell_2\theta_2),
    \label{eq:Gamma0-multipole}
\end{align}
where we define the kernel function $H_M(\ell_1,\ell_2)$
\begin{align}
    H_M(\ell_1, \ell_2) = \sum_{L=0}^\infty(-1)^L b_\kappa^{(L)}(\ell_1, \ell_2)G_{LM}(\psi).
    \label{eq:HM}
\end{align}
We note that the 3PCF is expanded as multipoles indexed by $M$ in the Fourier basis with respect to the inner angle $\phi$ of the real space triangle. From Eq.~(\ref{eq:GLM-symmetry}), we find
\begin{align}
    H_M(\ell_2,\ell_1) = H_{(-M)}(\ell_1, \ell_2).
    \label{eq:HM-symmetry}
\end{align}

The multipole expansion for other components can be derived in the same way. We can obtain the same form of multipole expansion for $\Gamma_{1,2,3}^\times$ as in Eq.~(\ref{eq:Gamma0-multipole-expansion}) but the multipoles are given by
\begin{align}
    \Gamma_1^{\times,(M)}(\theta_1, \theta_2) 
    &= \frac{1}{(2\pi)^3}
    \int\dd\ln\ell_1\dd\ln\ell_2\ell_1^2\ell_2^2 H_M({\color{red}\ell_2, \ell_1})\nonumber\\
    &\hspace{2em}\times
    J_{\color{red}M-1}(\ell_1\theta_1)
    J_{\color{red}-M-1}(\ell_2\theta_2),
    \label{eq:Gamma1-multipole}\\
    \Gamma_2^{\times,(M)}(\theta_1, \theta_2) 
    &= \frac{1}{(2\pi)^3}
    \int\dd\ln\ell_1\dd\ln\ell_2\ell_1^2\ell_2^2 H_M(\ell_1, \ell_2)\nonumber\\
    &\hspace{2em}\times
    J_{\color{red}M+1}(\ell_1\theta_1)
    J_{\color{red}-M-3}(\ell_2\theta_2),
    \label{eq:Gamma2-multipole}\\
    \Gamma_3^{\times,(M)}(\theta_1, \theta_2) 
    &= \frac{1}{(2\pi)^3}
    \int\dd\ln\ell_1\dd\ln\ell_2\ell_1^2\ell_2^2 H_M(\ell_1, \ell_2)\nonumber\\
    &\hspace{2em}\times
    J_{\color{red}M-3}(\ell_1\theta_1)
    J_{\color{red}-M+1}(\ell_2\theta_2),
    \label{eq:Gamma3-multipole}
\end{align}
We highlighted the difference in red font compared to the multipoles of $\Gamma_0^{\times,(M)}$ in Eq.~(\ref{eq:Gamma0-multipole}) for the ease of comparison. There are several properties of the multipoles of the 3PCF. The multipoles of the zero-th and first components have the following symmetry in itself
\begin{align}
    \Gamma_\mu^{\times,(M)}(\theta_1, \theta_2) = \Gamma_\mu^{\times,(-M)}(\theta_2, \theta_1),
\end{align}
for $\mu=0$ and $1$. Similarly, the multipoles of the second and the third components are related as
\begin{align}
    \Gamma_2^{\times,(M)}(\theta_1, \theta_2) = \Gamma_3^{\times,(-M)}(\theta_2, \theta_1),
\end{align}
which can be checked directly comparing Eqs.~(\ref{eq:Gamma1-multipole}) and (\ref{eq:Gamma2-multipole}). The above properties of the component multipoles stem from the properties in Eqs.~(\ref{eq:Gamma0-reflect})~-~(\ref{eq:Gamma3-reflect}). These are especially useful because we can reduce the computational time by a factor of two just by exchanging the arguments $\theta_1$ and $\theta_2$ in one multipole to get the corresponding multipole for a different component.

It is interesting to compare the formalism obtained in this section for weak lensing to the similar formalism developed in the context of galaxy clustering \cite{Umeh.Umeh.2020, Guidi.Carbone.2022, Aviles.Niz.2023}. In these papers, authors introduced multipole expansion using Legendre polynomials both in real and Fourier space, with a focus on the 3PCF of galaxy number density contrast, i.e. the 3PCF of spin-0 scalar fields. The 3PCF multipoles defined in these papers for the galaxy-clustering 3PCF has one-to-one correspondence to the bispectrum multipoles and there is no coupling between the 3PCF and bispectrum multipoles.  In contrast, the multipole formalism developed in this paper for the 3PCF of spin-2 shear fields has coupling between the 3PCF and bispectrum multipoles, which essentially originates from the presence of spin-2 phase factors, the second-to-last factor in Eq.~(\ref{eq:Gamma0-by-bispectrum}).

The formalism developed here has several advantages for the efficiency of numerical evaluation from a given model of the bispectrum. First, the bispectrum in the context of standard cosmology is usually a smooth function of the inner angle $\alpha$, and thus the multipole expansion is efficient even after truncating the summation at a reasonable choice of maximum multipole $L_{\rm max}$. As we will see later in section~\ref{sec:result}, we can safely set $L_{\rm max}\sim 30$ to get a robust estimate of the 3PCF, which enable us quick resummation of bispectrum multipoles. Second, because the function $G_{LM}(\psi)$ is independent of any cosmological/astrophysical model, we can evaluate the functional form for all desired $L$ and $M$ as a function of $\psi$ prior to the actual time-consuming parameter inference.
Third, the integral over two Fourier-mode amplitudes $\ell_{1,2}$ including two oscillating Bessel functions, i.e. double Hankel transformation, can be robustly and efficiently performed using a two-dimensional version of Fast Fourier Transformation on logarithmic scale (2D-FFTLog), which was originally developed for the efficient estimate of covariances of two-point correlation functions \cite{Fang.Krause.2020}.

It is noteworthy that the multipole expansion method developed in this paper has synergy with the recent paper by \citet{Porth.Schneider.2023}. In that paper, they invented an efficient algorithm to estimate the 3PCF based on a multipole expansion with \xprojection{}, where the estimator they directly measure is multipoles of the 3PCF as in our paper. Because of this agreement in the quantities we measure/predict from data/theory, the effect of the finite truncation of multipoles in the multipole expansion can be alleviated by using the same maximum multipoles in the measurement and theory.

\subsection{Multipole coupling function}\label{sec:multipole-coupling}
\begin{figure}
    \centering
    \includegraphics[width=1.0\linewidth]{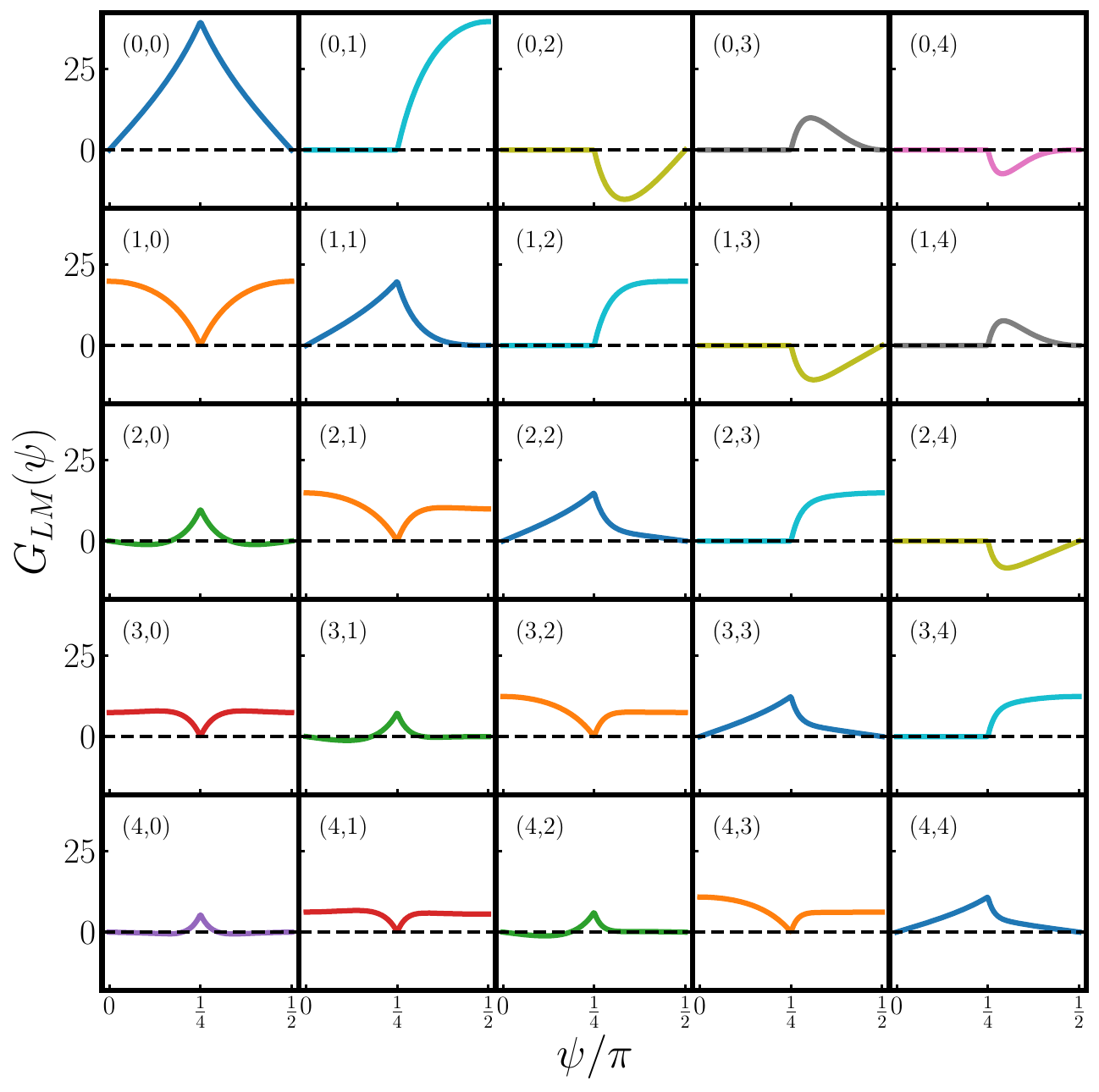}
    \caption{The behaviour of the function $G_{LM}(\psi)$ for $L=0,\dots, 4$ and $M=0,\dots, 4$ in each panel, where $(L,M)$ is indicated at the top left within each panel. The functions with the same $L-M$ value have similar $\psi$-dependence, which are visualized by the same color on diagonal panels. Because of the symmetry $G_{LM}(\pi/2-\psi)=G_{L(-M)}(\psi)$ we only show the function for $M\geq0$.}
    \label{fig:GLM}
\end{figure}
In this section, we present the features of the multipole-coupling function $G_{LM}(\psi)$ defined in Eq.~(\ref{eq:GLM}). Fig.~\ref{fig:GLM} shows the behaviour of the function for $L=0,\dots, 4$ and $M=0,\dots,4$. The functions with the same $L-M$ value have a similar shape for their $\psi$ dependence. The overall amplitude of the function decreases as $L$ or $M$ gets larger. Especially by comparing the functions in a column (or row), we can see that the amplitude of the function decreases as the difference between $L$ and $M$ increases, and the amplitude is largest when $L=M$. This means that the bispectrum and 3PCF multipoles couple most tightly between similar multipole indices ($L\approx M$). Lastly, we list other minor but nontrivial features of the multipole-coupling function: 
\begin{align}
    \label{eq:GLM-property-half-zero}
    &G_{LM}\left(\psi\in[0,\pi/4]\right) = 0 \text{ for $L < M$}\\
    \label{eq:GLM-property-even-zero}
    &G_{LM}(0) = 0 \text{ for $L-M=2n$}\\
    \label{eq:GLM-property-odd-zero}
    &G_{LM}(\pi/4) = 0 \text{ for $L-M=2n+1$},
\end{align}
where $n\geq0$ is an integer. In Appendix~\ref{sec:other-coupling-functions}, we discuss more details of the multipole coupling function through the expansion of the Legendre polynomial in a Fourier basis.

\subsection{Bin-averaging effect}\label{sec:bin-averaging}
In real observations, we measure the signal of the 3PCF in binned triangles, which means the underlying signal is averaged within each bin of the measurement. This effect alters the scale dependence and triangle shape dependence of the signal. For a robust comparison of theory to the measured signal to extract unbiased cosmological/astrophysical parameters, we need to take into account the bin-averaging effect on the theory side as well.

The bin averaging effect on theoretical prediction is formulated as
\begin{align}
\begin{split}
    &\bar\Gamma_0^\times(\theta_{1i}, \theta_{2j}, \phi_k)\\
    &= \int_{\theta_{1i}} \frac{\dd\ln \theta_1~\theta_1^2}{A_{1i}}\int_{\theta_{2j}}\frac{\dd\ln \theta_2~\theta_2^2}{A_{2j}}\int_{\phi_k}\frac{\dd\phi}{\Delta\phi_k}\Gamma_0^\times(\theta_1,\theta_2,\phi)
\end{split}
\end{align}
where $i,j$ and $k$ are the labels of the $\theta_1, \theta_2$ and $\phi$ bins respectively. The integration runs over $\theta_1\in[\theta_{1i,\rm min}, \theta_{1i,\rm max}]$, $\theta_2\in[\theta_{2j,\rm min}, \theta_{2j,\rm max}]$, and $\phi\in[\phi_{k,\rm min}, \phi_{k,\rm max}]$, and $A_{1i}=(\theta_{1i,\rm max}^2-\theta_{1i,\rm min}^2)/2$, $A_{2j}=(\theta_{2j,\rm max}^2-\theta_{2j,\rm min}^2)/2$, and $\Delta\phi_k=\phi_{k, \rm max} - \phi_{k,\rm min}$.

The bin-averaging operator on the angle $\phi$ acting on the basis $e^{iM\phi}$ in Eq.~(\ref{eq:Gamma0-multipole-expansion}) can be performed analytically, and the result is equivalent to replacing
\begin{align}
    e^{iM\phi}
    ~\rightarrow~ 
    e^{iM\phi_{k,\rm min}}\frac{e^{iM\Delta\phi_k}-1}{iM\Delta\phi_k}.
\end{align}

Similar the averaging effect on $\theta_1$ and $\theta_2$ can be easily taken into account by replacing the basis of 2D-FFTLog. While we do not go into detail about the 2D-FFTLog algorithm, we briefly summarize it to show how the averaging effect can be taken into account. The algorithm basically expands the target function as the sum of a double-power law with coefficients as
\begin{align}
    \Gamma_0^{\times,(M)}(\theta_1,\theta_2) = \sum_{p,q} c_{pq}\theta_1^{-i\eta_p-\nu_1} \theta_2^{-i\eta_q-\nu_2}
\end{align}
where $p$ and $q$ are the multipole indices, and $\eta_p, \eta_q, \nu_1, \nu_2$ are the real values. Then the action of bin-averaging operators is equivalent to replacing
\begin{align}
    \theta_1^{-i\eta_p-\nu_1}
    ~\rightarrow~
    &\theta_{1i,\rm min}^{-i\eta_p-\nu_1} 
    \frac{s(2-i\eta_p-\nu_1, \Delta\!\ln x_{1i})}{s(2,\Delta\!\ln x_{1i})}
\end{align}
where $\Delta\ln \theta_1\equiv \ln(\theta_{1i,\rm max}/\theta_{1i,\rm min})$, $s(z,\lambda)=[\exp(z\lambda)-1]/z$, and the exact analogy holds for $\theta_2$. This is already implemented in the public code, and we refer the reader to \cite{Fang.Krause.2020} for the complete formalism of 2D-FFTLog. In this way, the bin-averaging effect can be taken into account without any additional computational cost.

Note that the way to incorporate the bin-averaging effect by replacing the bases with the bin-averaged ones described above is only valid for the multipole expansion with \xprojection{}. In a different projection, we need a phase factor to convert from \xprojection{} to the projection under consideration, and it breaks the form of the component to be expressed with analytic bases because the multiplied phase factor depends on the triangle. For this reason, \xprojection{} has the advantage that it can include the bin-average effect without any additional computational cost compared to other choice of projections.

\section{Results}\label{sec:result}
\subsection{Numerical implementation}\label{subsec:numerical-implementation}
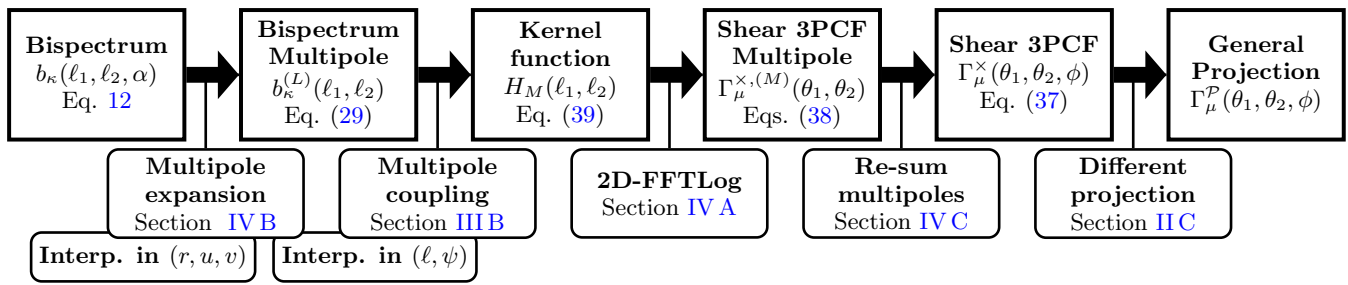
\begin{figure*}
    \centering
    \begin{tikzpicture}[
            node distance=0.7cm and 0.7cm,
            every node/.style={draw, rectangle, minimum width=3em, align=center, font=\sffamily},
            product/.style={fill=white, text width=6.5em, line width=1.5pt, minimum height=5.3em, font=\bfseries},
            arrow/.style={black, fill=black, -{Triangle[width = 18pt, length = 8pt]}, line width = 6pt]},
            process/.style={rectangle, rounded corners, fill=white, minimum width=8em, minimum height=3.6em, font=\bfseries, line width=1pt},
            line/.style={black, line width=1pt},
            sub/.style={rectangle, rounded corners, fill=white, minimum height=2em, font=\bfseries, line width=1pt}
        ]
        \node[product              ] (bl)  {Bispectrum $b_\kappa(\ell_1, \ell_2, \alpha)$\\ \normalfont{Eq.~\ref{eq:kbispectrum-mbispectrum}}};
        \node[product, right=of bl ] (blm) {Bispectrum Multipole $b_\kappa^{(L)}(\ell_1, \ell_2)$\\ \normalfont{Eq.~(\ref{eq:bispectrum-multipole})}};
        \node[product, right=of blm] (hm)  {Kernel function $H_M(\ell_1, \ell_2)$\\ \normalfont{Eq.~(\ref{eq:HM})}};
        \node[product, right=of hm ] (gmm) {Shear 3PCF Multipole $\Gamma_\mu^{\times,(M)}(\theta_1, \theta_2)$\\ \normalfont{Eqs.~(\ref{eq:Gamma0-multipole})}};
        \node[product, right=of gmm] (gm3) {Shear 3PCF $\Gamma_\mu^{\times}(\theta_1, \theta_2,\phi)$\\ \normalfont{Eq.~(\ref{eq:Gamma0-multipole-expansion})}};
        \node[product, right=of gm3] (proj){General Projection $\Gamma_\mu^{\mathcal{P}}(\theta_1, \theta_2, \phi)$};
        \node[sub, below=6.4em of $(bl)!0.2!(blm)$  ] (process1sub) {Interp. in $(r,u,v)$};
        \node[sub, below=6.4em of $(blm)!0.2!(hm)$  ] (process2sub) {Interp. in $(\ell, \psi)$};
        \node[process, below=3em of $(bl)!0.47!(blm)$  ] (process1) {Multipole\\ expansion\\ \normalfont{Section~ \ref{sec:bispectrum-multipole-expansion}}};
        \node[process, below=3em of $(blm)!0.47!(hm)$  ] (process2) {Multipole\\ coupling\\ \normalfont{Section~\ref{sec:multipole-coupling}}};
        \node[process, below=3em of $(hm)!0.47!(gmm)$  ] (process3) {2D-FFTLog\\ \normalfont{Section~\ref{subsec:numerical-implementation}}};
        \node[process, below=3em of $(gmm)!0.47!(gm3)$ ] (process4) {Re-sum\\ multipoles\\ \normalfont{Section~\ref{sec:validation}}};
        \node[process, below=3em of $(gm3)!0.47!(proj)$] (process5) {Different\\ projection\\ \normalfont{Section~\ref{subsec:shear-projection}}};
        \draw[arrow] (bl)  -- (blm);
        \draw[arrow] (blm) -- (hm);
        \draw[arrow] (hm)  -- (gmm);
        \draw[arrow] (gmm) -- (gm3);
        \draw[arrow] (gm3) -- (proj);
        \draw[line] (process1.north) |- (blm.west);
        \draw[line] (process2.north) |- (hm.west);
        \draw[line] (process3.north) |- (gmm.west);
        \draw[line] (process4.north) |- (gm3.west);
        \draw[line] (process5.north) |- (proj.west);
    \end{tikzpicture}
    \caption{An overview of the numerical implementation of the formalism developed in this paper. Square boxes indicate the quantities we evaluate in this method, and the round boxes indicate the operations we perform to obtain the quantity on the right from the quantity on the left. Starting from the bispectrum model $b_\kappa(\ell_1, \ell_2, \alpha)$ on the left, we compute the bispectrum multipole $b_\kappa^{(L)}(\ell_1,\ell_2)$, the kernel function $H_M(\ell_1, \ell_2)$, the multipole components $\Gamma_\mu^{\times,(M)}(\theta_1, \theta_2)$, and finally $\Gamma_\mu^{\times}(\theta_1, \theta_2, \phi)$ on the right. The calculation is done on FFT grids $(\ell_1, \ell_2)$ and $(\theta_1, \theta_2)$. For the sake of computational efficiency, we use interpolations of bispectrum and bispectrum multipoles in the multipole expansion and multipole coupling operations respectively.}
    \label{fig:procedure}
\end{figure*}

\begin{table}
    \centering
    \caption{The fiducial choice of hyper parameters. From top to bottom blocks, the hyper parameters of the FFT grid, bispectrum multipole expansion, and 2D-FFTLog are tabulated. The range of the FFT grid are optimized for the study of cosmology, especially for BiHalofit, and one should set an appropriate scale for other choices of bispectrum.}
    \begin{tabular}{p{0.4\linewidth} p{0.4\linewidth}}
        \hline\hline
        \multicolumn{2}{l}{\bf FFT grid}\\
        $\ell_{12, \rm min}$ & $10^{-1}$\\
        $\ell_{12, \rm max}$ & $10^{5}$\\
        $n_{\ell_{12}}$ & 200 \\
        \hline
        \multicolumn{2}{l}{\bf 2D-FFTLog}\\
        $\nu_1=\nu_2$ & 1.01\\
        \texttt{c\_window\_width} & 0.25\\
        \hline
        \multicolumn{2}{l}{\bf Bispectrum multipole}\\
        $\epsilon_\mu$ & $10^{-7}$ \\
        $n_{\mu, \rm log}$ & 30\\
        $n_{\mu, \rm lin}$ & 50\\
        \hline
        \multicolumn{2}{l}{\bf Maximum multipole}\\
        $L_{\rm max}$ & 30\\
        $M_{\rm max}$ & 30\\
        \hline\hline
    \end{tabular}
    \label{tab:setup}
\end{table}

\begin{figure*}
    \centering
    \includegraphics[width=1.0\linewidth]{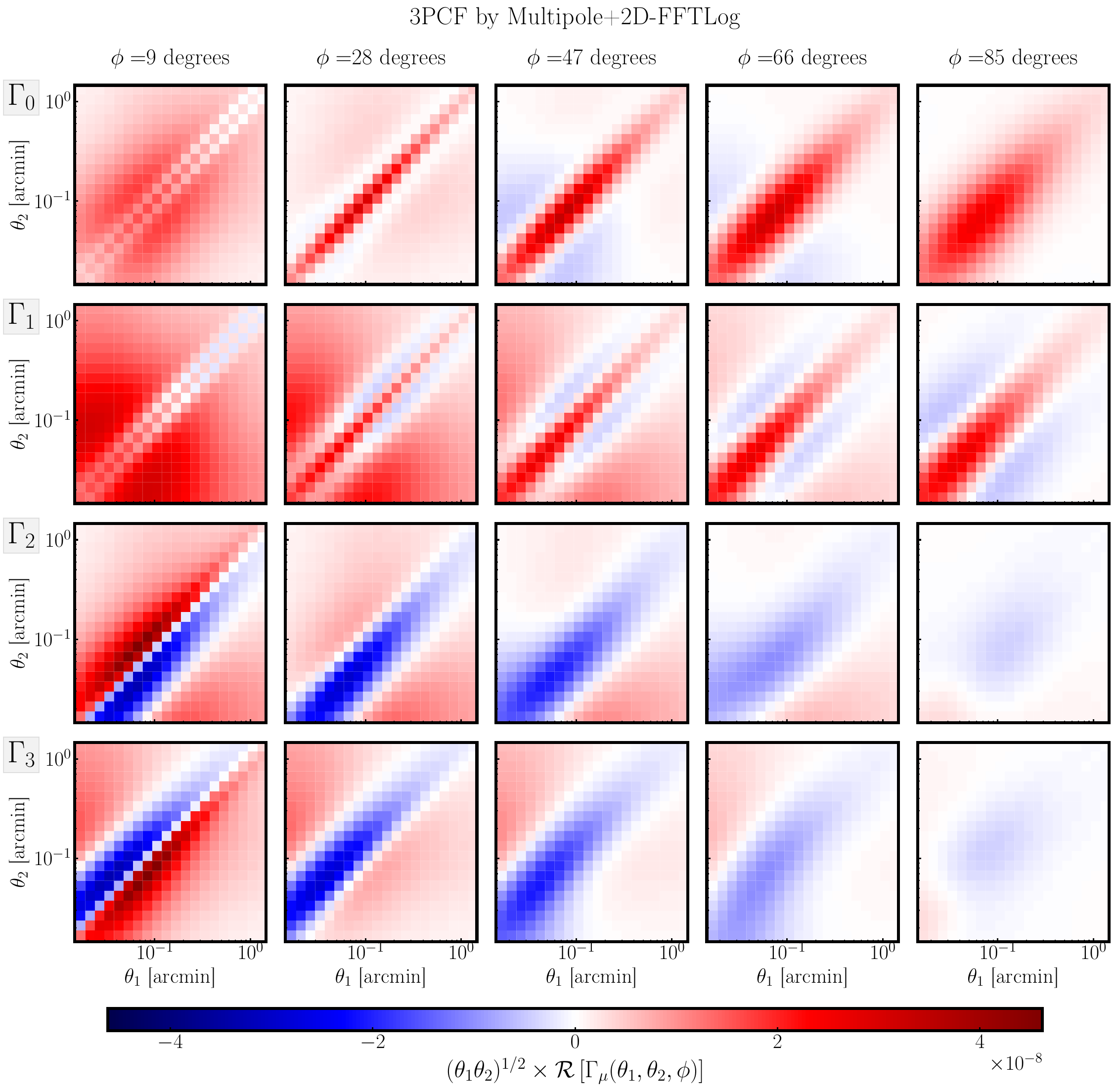}
    \caption{A demonstration of the method developed in this paper based on multipole expansion and 2D-FFTLog. From top to bottom, the $\mu$-th components ($\mu=0,1,2,3$) are plotted as a function of $\theta_1$ and $\theta_2$. The opening angle $\phi$ between sides $\theta_1$ and $\theta_2$ is increasing from let to right. Here the real parts of the 3PCF components are plotted. }
    \label{fig:fastnc-demo}
\end{figure*}

In this section, we discuss some details of the numerical implementation of the formalism described in the last section. Fig.~\ref{fig:procedure} shows a flow chart of the calculations. We start from a model of bispectrum $b_\kappa(\ell_1, \ell_2, \alpha)$, and compute the bispectrum multipole $b_\kappa^{(L)}(\ell_1,\ell_2)$ on an FFT grid $(\ell_1, \ell_2)$. We then calculate the kernel function $H_M(\ell_1, \ell_2)$ on the FFT grid using Eq.~(\ref{eq:HM}). We then perform the double Hankel transformation to obtain the 3PCF multipole $\Gamma_\mu^{\times,(M)}(\theta_1,\theta_2)$ on the real-space FFT grid. Re-summing the 3PCF multipoles with the Fourier basis using Eq.~(\ref{eq:Gamma0-multipole-expansion}), we calculate $\Gamma_\mu^\times(\theta_1, \theta_2, \phi)$ on the FFT grid. If we want to obtain the 3PCF with different shear projection $\mathcal{P}$, we multiply the corresponding phase factor to convert the \xprojection{} to the desired projection $\mathcal{P}$. 

Table~\ref{tab:setup} summarize the fiducial choice of FFT grid parameters. For 2D-FFTLog, we use $200^2$ FFT grid values in $(\ell_1, \ell_2)$ and $(\theta_1, \theta_2)$. 
The range of the FFT grid depends on the model of the bispectrum. For the case of BiHalofit that we use in this paper, we found the scales $\ell_{12}\in[10^{-1}, 10^5]$ capture the most significant bispectrum signals, and hence we use $\ell_{12, {\rm min}}=10^{-1}$ and $\ell_{12, {\rm max}}=10^5$ as the fiducial choice. 

The algorithm of 2D-FFTLog has its own hyper parameters, the bias parameters $\nu_{i}$ (i=1,2) and the width of the window \texttt{c\_window\_width}. The former is the real part of the power law of the Fourier expansion and its choice is arbitrary as long as the following condition is satisfied: the bias parameter must be $-|m|<\nu_i<2$ for the Hankel transformation with $J_{m}(\theta_i\ell_i)$, and $\nu_i$ must be non-integer. The latter hyper parameter controls the width of window multiplied on the Fourier coefficients in the unit of Nyquist frequency, which is introduced to reduce the ringing effect in \cite{McEwen.Blazek.2016}. In this paper, we follow the default choice of 2D-FFTLog for these parameters as summarized in Table~\ref{tab:setup}, which we found gives stable results.

In Section~\ref{sec:formalism}, we introduced the multipole expansion of triangles in real and Fourier spaces. For the practical implementation of Eqs.~(\ref{eq:HM}) and (\ref{eq:Gamma0-multipole-expansion}), we need to truncate the multipole summation at a finite maximum multipole $L\leq L_{\rm max}$ and $|M|\leq M_{\rm max}$. The choice of $L_{\rm max}$ and $M_{\rm max}$ is a trade-off between the computational cost and the accuracy of the prediction. We use $L_{\rm max}=30$ and $M_{\rm max}=30$ as the fiducial choice, which we found gives a robust estimate of the 3PCF as we will see in Section~\ref{sec:validation}. We note that this choice of $L_{\rm max}$ and $M_{\rm max}$ is tested for BiHalofit, and they can be optimized differently for the other specific model of bispectrum and the desired accuracy of the prediction.

We implemented the formalism with Python using the implementation of 2D-FFTLog available at \footnote{\url{https://github.com/xfangcosmo/2DFFTLog}}. The code developed in this paper is publicly available on GitHub\footnote{\url{https://github.com/git-sunao/fastnc}} and installable with {\tt pip}\footnote{\url{https://pypi.org/project/fastnc/}} or {\tt conda}\footnote{\url{https://anaconda.org/ssunao/fastnc}}.

In Fig.~\ref{fig:fastnc-demo}, we demonstrate the method developed in this paper based on multipole expansion and 2D-FFTLog. Using this code, we can evaluate a 3PCF within $\sim 10$ seconds. While we show only the real part of the 3PCFs and for specific choice of $\phi$ values in this plot for simplicity, the imaginary part and those for other $\phi$ values are also obtained at the same time without any extra runtime. We can also clearly see the symmetry properties of Eqs.~(\ref{eq:Gamma2-reflect}) and (\ref{eq:Gamma3-reflect}) by comparing the lowest and second lowest panels. The ease of the visualization of the shear 3PCF as a function of two side lengths $\theta_1$ and $\theta_2$ plus the opening angle $\phi$ is an important advantage of this method for understand the dependence on the triangle configuration, as well as the speed of the evaluation of the shear 3PCF.

\subsection{Bispectrum multipole expansion}\label{sec:bispectrum-multipole-expansion}
\begin{figure*}[t]
    \centering
    \includegraphics[width=0.49\linewidth]{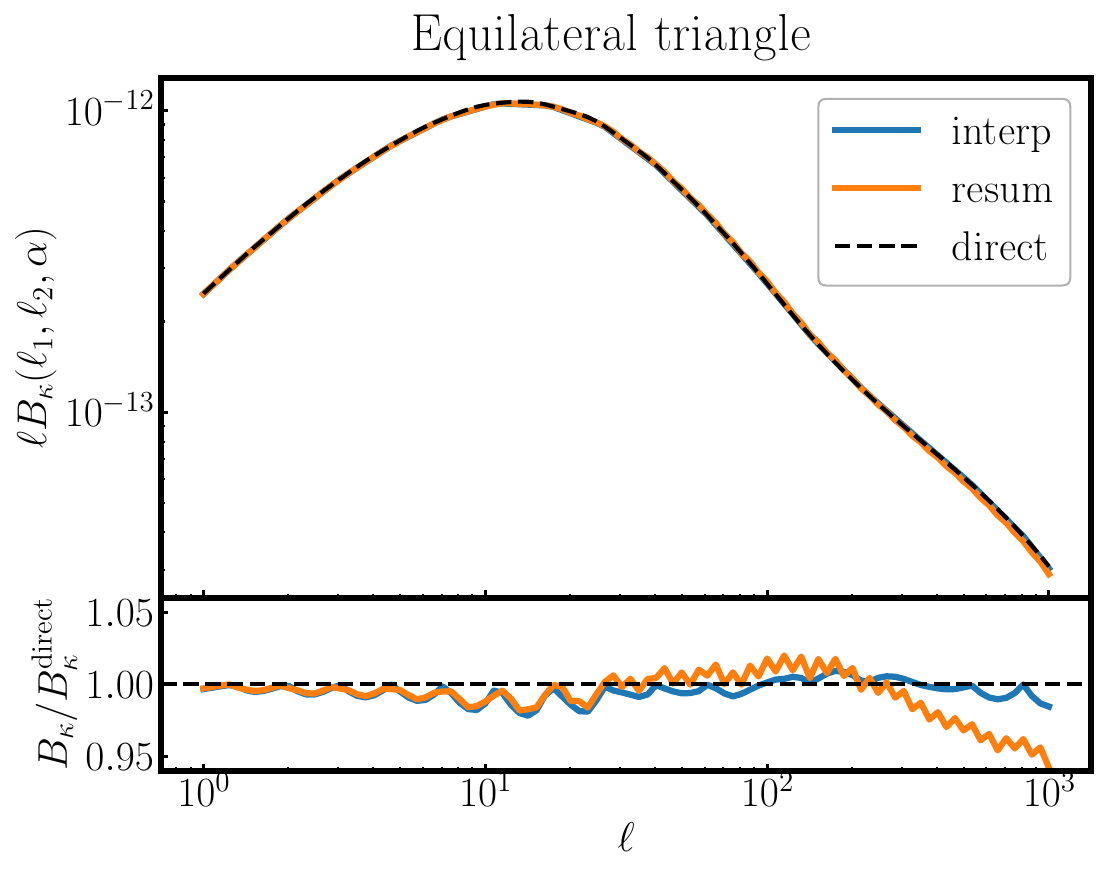}
    \includegraphics[width=0.48\linewidth]{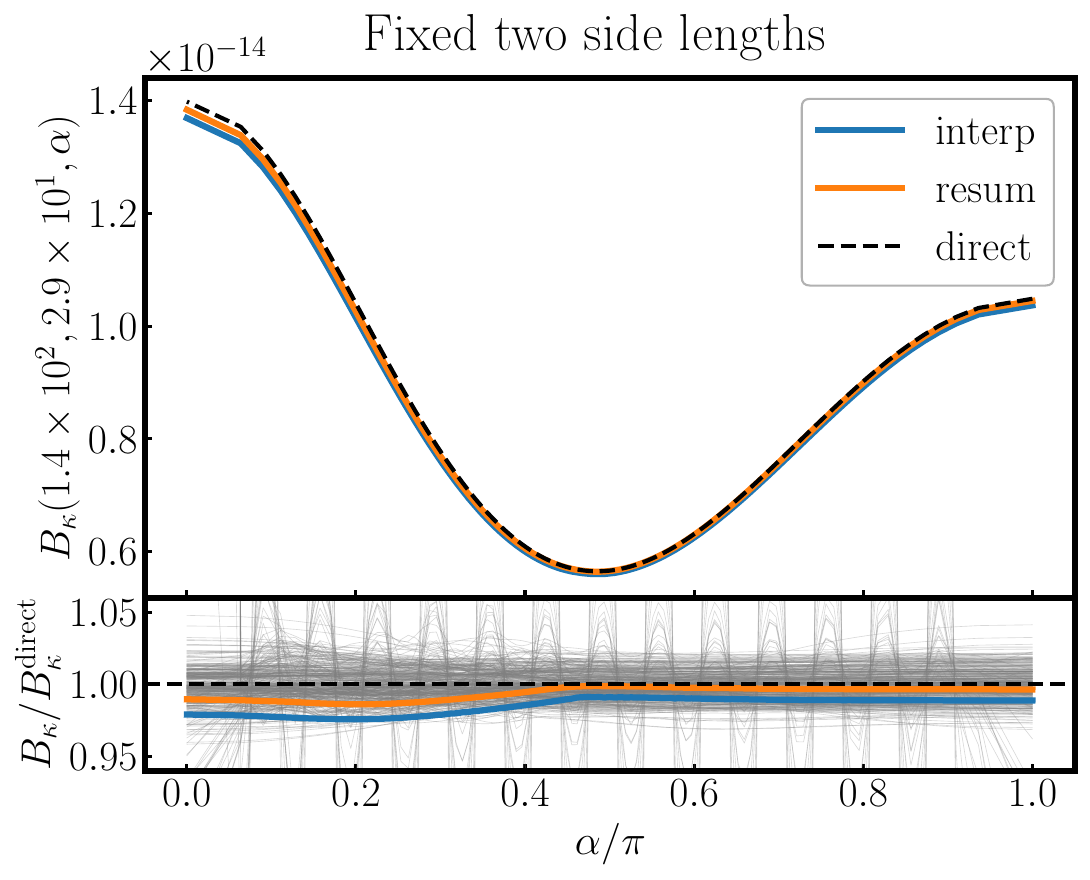}
    \caption{Performance of bispectrum interpolation and multipole expansion. Here the bispectra are functions of two sides and an opening angle. The interpolated bispectrum is obtained by the interpolation of the bispectrum on the sampling points discussed in Section~\ref{sec:bispectrum-multipole-expansion} and shown as a blue solid line. For comparison with the multipole-based method, we reconstructed the bispectrum from the bispectrum multipoles obtained by re-summing the multipole with the Legendre basis (Eq.~\ref{eq:bispectrum-multipole-expansion}), shown as an orange solid line. Here we used $L_{\rm max}=30$ for re-summed bispectrum. {\it Left-hand side}: The upper panel shows the bispectrum for the equilateral triangle with varying scale $\ell$, and the lower panel shows the ratios of the interpolated and re-summed bispectrum to the direct bispectrum. {\it Right-hand side}: The upper panel shows the bispectrum for the fixed-side triangle with varying opening angle $\alpha$, and the lower panel shows the ratios of the interpolated and re-summed bispectrum to the direct bispectrum. The gray lines in the lower panel are the ratios for different triangle configurations. The residuals larger than 5\% are for isosceles triangles, $\ell_1\sim\ell_2$.}
    \label{fig:bispectrum-resum}
\end{figure*}
Accurate estimation of bispectrum multipoles is an important part of this formalism. We find that the naive discretization of Eq.~(\ref{eq:bispectrum-multipole}), i.e. approximation with Riemann sum, fails, and gives artificially higher multipoles depending on the choice of number of bins on the opening angle, $\cos\alpha$. To overcome this difficulty, we instead use the interpolation-coefficient based approach. Based on the fact that the bispectrum of interest is usually a smooth function, we can sufficiently capture the functional form of the bispectrum with a small number of sampling points on the opening angle $\alpha$ and their interpolation. Once the functional form of the bispectrum is given with the interpolation coefficients, which are calculated from the small number of sampling points, we can analytically perform the multipole expansion in Eq.~(\ref{eq:bispectrum-multipole}) and then the result is stable to the choice of bin number as long as the number of bins is large enough to capture the opening-angle dependence of the bispectrum. Appendix~\ref{sec:numerical-multipole-expansion} gives more details about the practical implementation and accuracy of this approach.

In addition to the discretization of the multipole expansion described above, we need to be careful about the choice of sampling points on the cosine of the opening angle, $\cos\alpha$. The bispectrum has very sharp peak at $\cos\alpha\sim1$ when $\ell_1\sim\ell_2$, because this domain corresponds to the squeezed-limit triangle which is known to give high amplitude of the bispectrum. Therefore, a linearly equally spaced bins on $\cos\alpha$ fails to capture the sharp peak at $\cos\alpha\sim1$ and consequently gives smaller amplitude of multipoles. To obtain unbiased bispectrum multipoles, taking care of such squeezed-limit triangles, we combine two different binning choices: linearly equally-spaced bins over $\mu\equiv\cos\alpha\in[-1, 1-0.05]$ and logarithmically equally-spaced bins over $\mu\in[1-0.05, 1-\epsilon_{\mu}]$, where $\epsilon_\mu$ determines the most squeezed-limit triangle taken into account. The fiducial choice of $\epsilon_\mu$ and the numbers of linear and logarithmic bins are summarized in Table~\ref{tab:setup}, which we validated to give stable results.

When we evaluate the bispectrum multipoles through Eq.~(\ref{eq:bispectrum-multipole}), we need to evaluate the bispectrum for various triangle configurations $(\ell_1, \ell_2,\alpha)$. To speed up the computation, we use an interpolation function for evaluating the bispectrum multipoles. We also note that our interpolating function parametrizes the triangles using the $ruv$ notation introduced in \cite{Jarvis.Jain.2003} rather than SAS notation. In the SAS parametrization, a given triangle shows up multiple times with different parameters, depending on which vertex is taken to be the ``A'' vertex. The $ruv$ parametrization only includes each triangle once, so it avoids this redundancy. 

We also use interpolation for bispectrum multipoles $b_\kappa^{(L)}(\ell_1, \ell_2)$ for the sake of numerical efficiency. Again to avoid redundancy of the bispectrum multipole evaluation, we use the $(\ell, \psi)$ notation with $0<\psi<\pi/4$ of the bispectrum multipole rather than $(\ell_1, \ell_2)$ notation. The interpolation of bispectrum multipole is done on the $(\ell, \psi)$ grid.

Fig.~\ref{fig:bispectrum-resum} shows the performance of the bispectrum multipole expansion. To evaluate the performance, we reconstructed the bispectrum as a function of $\cos\alpha$ from the bispectrum multipoles using Eq.~(\ref{eq:bispectrum-multipole-expansion}) up to $L_{\rm max}=30$ and compared it to the original bispectrum. The left panel shows the bispectrum for equilateral triangles with varying scale $\ell$, and the right panel shows the bispectrum for fixed-side triangles with varying opening angle $\alpha$.
We find that the bispectrum reconstructed from the bispectrum multipoles obtained by the multipole expansion is in good agreement with the original bispectrum, at the level of 3\% accuracy. For the isosceles triangles, $\ell_1\sim\ell_2$, the residuals of the reconstructed bispectrum sometimes exceed 3\%. We found that this is due to leakage of the higher multipoles which tries to fit the steep increase of the bispectrum toward $\alpha\rightarrow0$ for squeezed limit triangles. The figure also shows the error of the bispectrum interpolation, where we can see that the error is mostly introduced in the interpolation step. Here we use $35\times35\times25$ bins on $ruv$ for bispectrum interpolation. We can achieve higher accuracy by using a finer grid on $ruv$ at the expense of computational efficiency.

\subsection{Validation}\label{sec:validation}
\begin{figure*}
    \centering
    \includegraphics[width=0.48\linewidth]{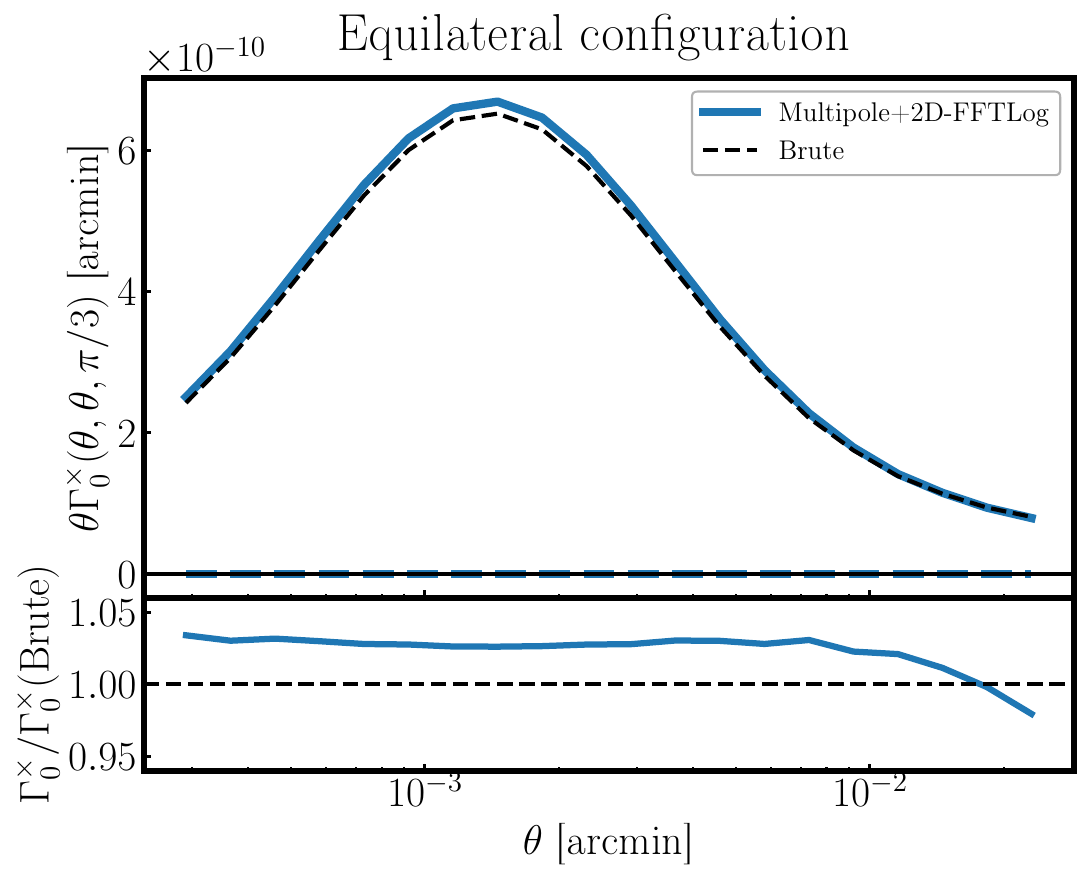}
    \includegraphics[width=0.48\linewidth]{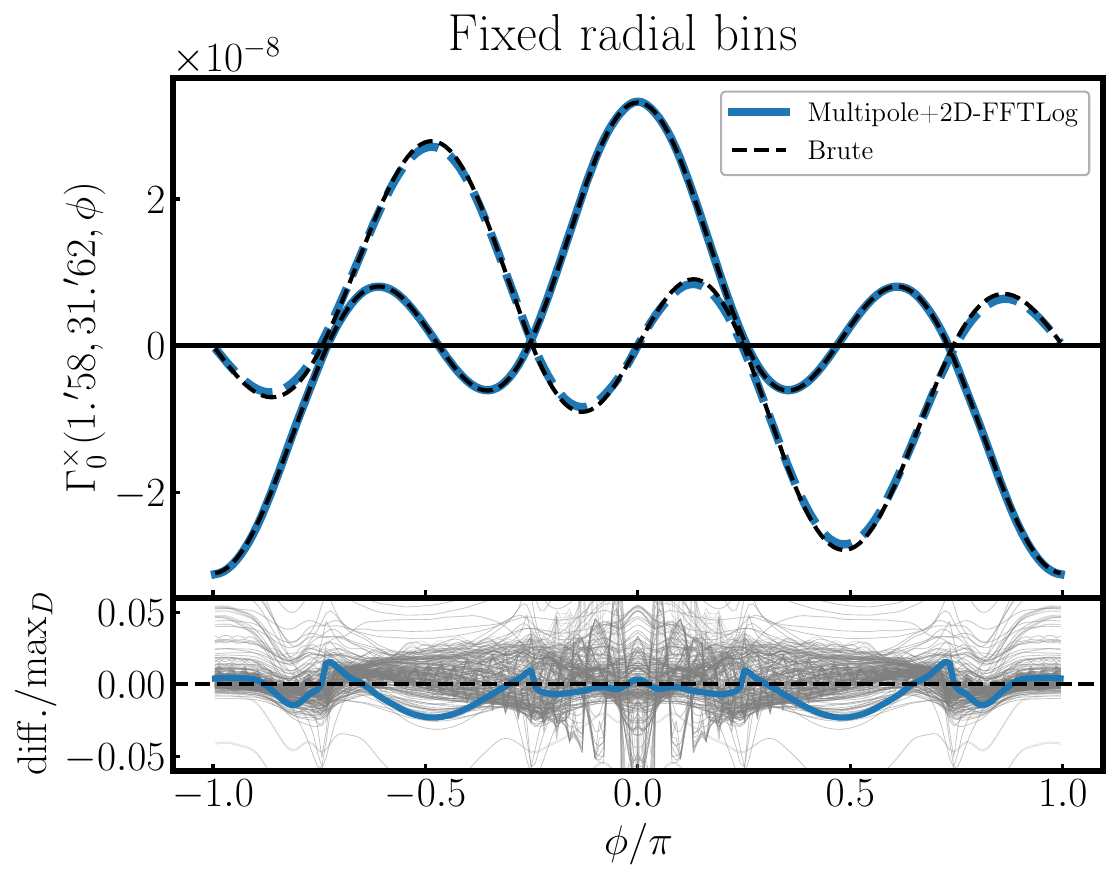}
    \caption{Validation of the multipole method developed in this paper. The blue lines are the 3PCF obtained with the multipole method, and the black lines are the 3PCF obtained with brute-force integration. The solid and dashed line shows the real and imaginary parts of the shear 3PCF. {\it Left-hand side}: The upper panel shows the 3PCF for equilateral triangles with varying scale $\ell$, and the lower panel shows the ratio of the multipole method to brute-force evaluation.
    {\it Right-hand side}: The upper panel shows the 3PCF for fixed-side triangles with varying opening angle $\phi$, and the lower panel shows the residual compared to the brute-force integration. Here, the residual is defined as the difference between the 3PCF obtained with the multipole method and the brute-force integration divided by maximum amplitude of the 3PCF. 
    The gray lines in the lower panel are the residuals for different triangle configurations.
    }
    \label{fig:validation}
\end{figure*}

In this section, we present the validation of the 3PCF obtained with the multipole method developed in this paper. As a reference for validation, we compute the 3PCF by a brute-force integration developed in \cite{Schneider.Lombardi.2003, Heydenreich.Schneider.2022} with further simplifications described in Appendix~\ref{sec:brute-force-natural-component}. We compute the reference values on 20 bins in $\theta_1$ and $\theta_2$ spanning from 1 to $10^2$ arcmins, and 200 bins on $\phi$ spanning from $-\pi$ to $\pi$. We use the fiducial parameters in Table~\ref{tab:setup} for the multipole-based method. After we compute the 3PCF components on the FFT grid $(\theta_1, \theta_2)$, we downsample it to 20 bins in $\theta_1$ and $\theta_2$ spanning from 1 to $10^2$ arcmins to compare with the reference.

Fig.~\ref{fig:validation} shows the validation of the 3PCF obtained with the multipole method. The left panel shows the 3PCF for equilateral triangles with varying scale $\ell$, and the right panel shows the 3PCF for fixed-side triangles with varying opening angle $\alpha$. The upper panels show the 3PCF obtained with the multipole method and the reference 3PCF obtained with brute-force integration. The lower panels compare the two results by showing the ratio and residual on left and right hand sides respectively. Because the 3PCF is an oscillating function of $\phi$ with zero-crossings, we defined the residual on the right panel as the difference between the 3PCFs obtained with the multipole method and the reference divided by the maximum amplitude of the reference for given a set of two side lengths.  We see that the 3PCF obtained with the multipole method is in good agreement with the reference 3PCF, and the residual is $<5\%$ for most of the the triangle configurations. We found that the triangles with $\theta_{1(2)}\ll \theta_{2(1)}$ can have residuals that exceed 5\%. Remembering the rough relation between real and Fourier space, $\theta\sim1/\ell$, the 3PCF with this configuration receives the large contribution from bispectrum multipoles with $\ell_{2(1)}\ll\ell_{1(2)}$, which has contributions from the squeezed limit bispectrum, and hence the finite truncation of multipoles causes relatively large residuals for this configuration.

\subsection{Efficiency of evaluation}\label{sec:efficiency}
\begin{table}[t]
    \centering
    \caption{Computational times of each step of the method developed in this paper. Here the hyper parameters and the triangle bin choice are the same as in Section~\ref{sec:validation}: the hyper parameters on Table~\ref{tab:setup} and $20\times20\times200$ triangle bins on $(\theta_1, \theta_2, \phi)$. The computational time is measured on a MacBook Air 1.1 GHz Quad-Core Intel Core i5 processor. Three different analysis setups using 1, 2, or 4 source redshift bins are considered in different columns.}
    \begin{tabular}{>{\raggedleft\arraybackslash}p{0.3\linewidth} >{\raggedleft\arraybackslash}p{0.2\linewidth} >{\raggedleft\arraybackslash}p{0.2\linewidth} >{\raggedleft\arraybackslash}p{0.2\linewidth}}
        \hline\hline
                               & \textbf{1 $z$-bin} & \textbf{2 $z$-bins} & \textbf{4 $z$-bins} \\
        \hline
        {\tt cosmo      }& 0.002 s & 0.003 s & 0.003 s\\
        {\tt z-kernel   }& 0.000 s & 0.001 s & 0.001 s\\
        {\tt interp bk  }& 5.883 s & 7.238 s & 9.979 s\\
        {\tt interp bkL }& 0.416 s & 2.187 s & 12.383 s\\
        {\tt multipole  }& 0.268 s & 0.864 s & 3.898 s\\
        {\tt HM         }& 0.379 s & 1.786 s & 9.252 s\\
        {\tt GammaM     }& 0.175 s & 0.865 s & 4.693 s\\
        {\tt Gamma      }& 0.007 s & 0.054 s & 0.263 s\\
        {\tt projection }& 0.018 s & 0.089 s & 0.492 s\\
        \hline
        Total            & 7.148 s & 13.086 s & 40.965 s\\
        \hline\hline
    \end{tabular}
    \label{tab:runtime}
\end{table}

In this section, we discuss the efficiency of the evaluation of the 3PCF with the multipole method developed in this paper. 
To assess the computational efficiency, we measure the execution time on a MacBook Air 1.1 GHz Quad-Core Intel Core i5 processor. Because we are interested in the efficiency for Monte-Carlo sampling application, we remove the computational time for the initialization of the code, e.g. the times for the initializing the FFT/interpolation grid, pre-computing the multipole coupling function, and so on.
We divided the remaining computation into 9 steps: {\tt cosmo}, {\tt z-kernel}, {\tt interp bk}, {\tt interp bkL}, {\tt multipole}, {\tt HM}, {\tt GammaM}, {\tt Gamma}, and {\tt projection}. The {\tt cosmo} and {\tt z-kernel} steps compute the redshift-distance relation and lensing kernels in Eq.~(\ref{eq:lensing-efficiency}) from a given set of redshift bins. The {\tt interp bk} and {\tt interp bkL} steps interpolate the bispectrum and the bispectrum multipole on the FFT grid. The {\tt multipole} step finds the bispectrum multipoles from the interpolation. The {\tt HM} step obtains the kernel function $H_M$ by summing the products of the bispectrum multipoles and the multipole coupling functions up to $L_{\rm max}$. The {\tt GammaM} step computes the 3PCF multipoles by performing the double Hankel transformation with 2D-FFTLog for all $|M|<M_{\rm max}$. The {\tt Gamma} step finds the real space 3PCF by summing the multipoles up to $M_{\rm max}$. The {\tt projection} step converts the 3PCF from \xprojection{} to the centroid projection.

We first consider the simplest version of the analysis with just a single source redshift bin. The first column of Table~\ref{tab:runtime} summarize the computational time. In this case, the {\tt interp bk} dominates the total computational time. In this step, we compute the convergence bispectrum performing the line-of-sight integration in Eq.~(\ref{eq:kbispectrum-mbispectrum}). Even though we reduce the redundancy of triangle parametrization using $ruv$ notation, this step still requires the evaluation of the matter bispectrum on $35\times 35\times25\times55\approx10^6$ sampling points for $r, u, v$ and $z$. The next slowest steps are {\tt interp bkL} and {\tt HM}. These steps include only simple operations such as interpolation and linear algebra on an FFT grid, but we repeat the same operation for all the multipoles under consideration $L_{\rm max}$ times and $M_{\rm max}$ times respectively; hence they can be computationally expensive compared to other steps. As expected, the {\tt GammaM} step, where we perform the double Hankel transformation, is not a dominant part of this formalism thanks to the efficiency of 2D-FFTLog. The total computational time for the single source redshift bin is $\sim 7$ seconds, which is of order $10^6$ times faster than the brute-force integration method, which takes more than $\sim 1$ minute for a single triangle configuration and thus around $10^5$ minutes in total.

In weak lensing analysis, it is common to use tomography to probe the growth history of large-scale structure. When we consider a tomographic setup with $n$ source redshift bins, then we will have $n(n+1)(n+2)/6$ auto- and cross-redshift bin combinations, and we need to perform the calculations of the 3PCF for each of these. The second and third columns of Table~\ref{tab:runtime} show the measured computational costs for the cases of $n=2$ and $4$. The computational time of the {\tt interp bk} step is not very affected by the number of source redshift bins, because we reuse the evaluated matter bispectrum for the calculation of the convergence bispectrum, and the additional computational cost is only to perform the line-of-sight integration for additional source redshift bin combinations. The computational costs of the subsequent steps scale linearly with $n(n+1)(n+2)/6$ as expected, and the most time-consuming step is {\tt HM} for the $n=4$ case. The total computational time for the $n=4$ case is $\sim 40$ seconds, which is still acceptable for the application of parameter inference based on Monte-Carlo sampling.

\subsection{Application to third-order aperture mass statistics}
\begin{figure}
    \centering
    \includegraphics[width=\linewidth]{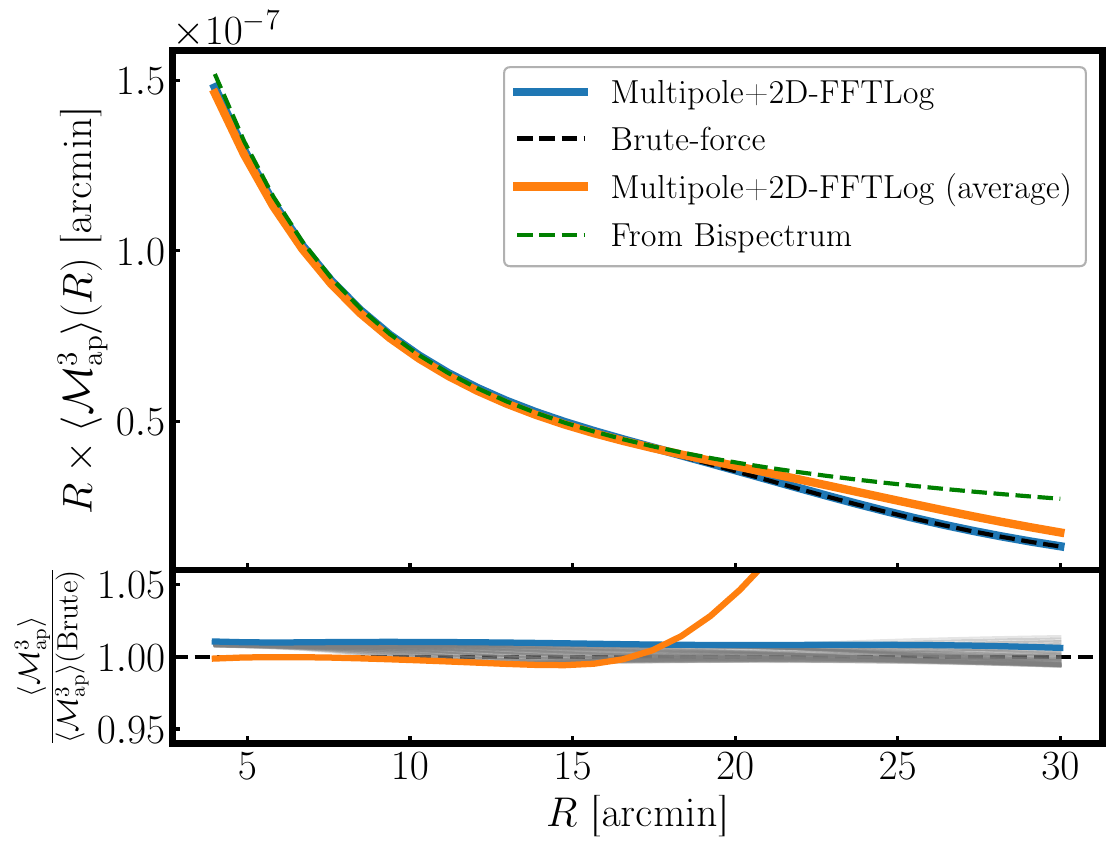}
    \caption{
    The upper panel shows the third-order aperture mass statistics evaluated using the multipole-based method (blue), brute-force integration (black dashed), the multipole-based method with the bin-averaging effect (orange), and the direct prediction from the bispectrum (green dashed). Note that the 3PCF used for the first two aperture mass statistics are evaluated on $20^2$ bins on $\theta_1$ and $\theta_2$ from $[1, 10^2]$ arcmin without including the bin-averaging effect. The lower panel shows the ratio of the aperture mass measures of multipole-based method (blue) compared to brute-force method. The gray lines in the lower panel are the ratios for non-equal aperture radii, where we set $R_1=R, R_2=k_2R, R_3=k_3R$ with $k_2$ and $k_3$ each varying from 1 to 3.}
    \label{fig:map3-validation}
\end{figure}

In this section, we apply the multipole-based method developed in this paper to third-order aperture mass statistics. The aperture mass is the measure of the mass density smoothed within a given aperture radius $R$ with a filter function $U_R(\theta)$ and  defined as \citep{Kaiser.Kaiser.1995,Schneider.Schneider.1996}
\begin{align}
    \mathcal{M}_{\rm ap}(\bm{\theta};R) = \int \dd^2\bm{\theta}' U_R(|\bm{\theta}'|) \kappa(\bm{\theta}+\bm{\theta}').
\end{align}
In the flat-sky approximation, the convergence field and the shear field are related by Eq.~(\ref{eq:shear-convergence-relation}), which leads to an equivalent formulation in terms of the shear field:
\begin{align}
    \mathcal{M}(\bm{\theta};R) = \int \dd^2\bm{\theta}' Q_R(|\bm{\theta}'|) \gamma(\bm{\theta}+\bm{\theta}'),
\end{align}
where the filter $Q_R(\theta)$ is derivable from the choice of $U_R(\theta)$ \citep{Schneider.Schneider.1996}. We use the choice of filter introduced by \citet{Crittenden.Theuns.2002},
\begin{align}
    Q_R(\theta) = \frac{\theta^2}{4\pi R^4}\exp\left[-\frac{\theta^2}{2R^2}\right].
\end{align}
The third-order aperture mass statistics is the mean third moment, or the skewness, of the aperture mass $\mathcal{M}_{\rm ap}(R)$. 

For real convergence fields, the aperture mass is a real quantity.  However, \citet{Schneider.Schneider.1996} introduced the idea of letting it be complex, where the imaginary part is a probe of B-modes in the data, typically taken to be an indication of uncorrected systematic errors:
\begin{align}
    \mathcal{M} = \mathcal{M}_{\rm ap} + i \mathcal{M}_\times,
\end{align}
where the term $\mathcal{M}_\times$ corresponds to the contribution from B-modes.

The third moments of $\mathcal{M}$ are related to the 3PCF of the shear fields as
\begin{align}
    &\langle\mathcal{M}^3\rangle(R_1, R_2, R_3) 
    = \int_0^\infty\dd\theta_1\int_0^\infty\dd\theta_2\int_0^{2\pi}\dd\phi \nonumber\\
    &\hspace{3em} \times \Gamma_0^{\rm cen}(\theta_1, \theta_2, \phi) T_0(R_1,R_2,R_3;\theta_1, \theta_2, \phi),\label{eq:mmm}
\end{align}
\begin{align}
    &\langle\mathcal{M}^*\mathcal{M}^2\rangle(R_1, R_2, R_3) 
    = \int_0^\infty\dd\theta_1\int_0^\infty\dd\theta_2\int_0^{2\pi}\dd\phi \nonumber\\
    &\hspace{3em} \times \Gamma_1^{\rm cen}(\theta_1, \theta_2, \phi) T_1(R_1,R_2,R_3;\theta_1, \theta_2, \phi),\label{eq:mcmm}
\end{align}
\begin{align}
    &\langle\mathcal{M}\mathcal{M}^*\mathcal{M}\rangle(R_1, R_2, R_3) 
    = \int_0^\infty\dd\theta_1\int_0^\infty\dd\theta_2\int_0^{2\pi}\dd\phi \nonumber\\
    &\hspace{3em} \times \Gamma_2^{\rm cen}(\theta_1, \theta_2, \phi) T_2(R_1,R_2,R_3;\theta_1, \theta_2, \phi),\label{eq:mmcm}
\end{align}
\begin{align}
    &\langle\mathcal{M}^2\mathcal{M}^*\rangle(R_1, R_2, R_3) 
    = \int_0^\infty\dd\theta_1\int_0^\infty\dd\theta_2\int_0^{2\pi}\dd\phi \nonumber\\
    &\hspace{3em} \times \Gamma_3^{\rm cen}(\theta_1, \theta_2, \phi) T_3(R_1,R_2,R_3;\theta_1, \theta_2, \phi),\label{eq:mmmc}
\end{align}
where we follow the definition of the 3PCF filter function $T_\mu$ in \citet{Schneider.Lombardi.2003} with generalization to non-equal aperture radii $R_1\neq R_2\neq R_3$. Note that we must use the centroid projection for the shear 3PCF instead of the \xprojection{}. We can obtain the skewness of the real aperture mass $\mathcal{M}_{\rm ap}(R)$ as a linear combination of the above quantities:
\begin{align}
    &\langle\mathcal{M}_{\rm ap}^3\rangle(R_1, R_2, R_3) \nonumber\\
    &\hspace{1em}= \frac{1}{4}\mathcal{R}[\langle\mathcal{M}^3\rangle+\langle\mathcal{M}^*\mathcal{M}^2\rangle\nonumber\\
    &\hspace{2em}+\langle\mathcal{M}\mathcal{M}^*\mathcal{M}\rangle 
    + \langle\mathcal{M}^2\mathcal{M}^*\rangle](R_1, R_2, R_3).
\end{align}

In practice, when we calculate the aperture mass statistics from the shear 3PCF, we perform two operations: the multiplication of the phase factor to change the shear projection from \xprojection{} to centroid projection, and the Riemann summation to evaluate the integral over $\theta_1, \theta_2, \phi$ in Eqs.~(\ref{eq:mmm})--(\ref{eq:mmcm}). These operations are done on the grid of $\theta_1, \theta_2, \phi$ on which the shear 3PCF is computed, and does not necessarily give an unbiased result of the exact integral because the bin-averaging effect is not taken into account for the phase factor for centroid projection as discussed in Section~\ref{sec:bin-averaging} and because the Riemann sum is an approximation of the exact integral. As long as we start from the same observable, in this case the shear 3PCF with \xprojection{}, for measurement and theory, these binning effects can be included automatically on both measurement and theory, and the resultant aperture mass statistics can be fairly compared. 

The aperture mass statistics can be also computed directly from the convergence bispectrum:
\begin{align}
    \langle\mathcal{M}_{\rm ap}^3\rangle(R_1, R_2, R_3) = \int_0^\infty\dd\ell_1\ell_1\int_0^\infty\dd\ell_2\ell_2\int_0^{2\pi}\dd\alpha \nonumber\\
    \times
    \hat{u}(R_1\ell_1)
    \hat{u}(R_2\ell_2)
    \hat{u}(R_3\ell_3)
    b_\kappa(\theta_1, \theta_2, \alpha),\label{eq:map3-bispectrum}
\end{align}
where $\ell_3=\sqrt{\ell_1^2+\ell_2^2-2\ell_1\ell_2\cos(\alpha)}$ and $\hat{u}(x)=x^2/2 \exp(-x^2/2)$. This relation enables us to directly calculate aperture mass statistics from the convergence bispectrum, although it does not take into account the binning effect on the shear 3PCF. For this reason, the aperture mass statistics calculated directly from the convergence bispectrum using Eq.~(\ref{eq:map3-bispectrum}) is not suitable for the comparison with measurement if the bin size is large.

Fig.~\ref{fig:map3-validation} compares the theoretical predictions of aperture mass statistics by the multipole-based method, the brute-force integration described in Appendix~\ref{sec:brute-force-natural-component}, and the direct prediction from the bispectrum using Eq.~(\ref{eq:map3-bispectrum}). 
For the multipole and brute-force methods, we first computed the shear 3PCF on $20^2$ bins on $\theta_1$ and $\theta_2$ spanning from 1 to $10^2$ arcmin, and 200 bins on $\phi$, then converted it to the aperture mass statistics. We find that the multipole-based method can accurately predict the third-order aperture mass statistics at the level of 2\% accuracy. This accuracy is much better than that of the 3PCF in Section~\ref{sec:validation}. This is because the residuals in the 3PCF scatter around zero with 5\% amplitude, which somewhat cancel out when transforming to the aperture mass statistics through integrals. 

The result is also compared with the direct prediction of the aperture mass statistics from the convergence bispectrum. The difference is significant at large $R>15$ arcmin, because the 
contribution of the 3PCF to the aperture mass statistics in Eqs.~(\ref{eq:mmm})--(\ref{eq:mmmc}) is truncated at a finite angular scale $\theta_1, \theta_2<\theta_{\rm max}$. Note that the aperture mass statistics at the aperture radius $R$ has the largest contributions from 3PCF at $\theta\sim4R$, and hence the difference starts to be significant at the scale of $R\sim\theta_{\rm max}/4=25$ \citep[see Fig.~2 of][]{Jarvis.Jain.2003}.

To investigate the bin-averaging effect, we also compute the aperture mass statistics by the multipole method with the bin-averaging effect in Section~\ref{sec:bin-averaging}. At smaller scales, the bin-averaging effect is negligible, while the difference can be larger than 5\% at large $R$. This is again due to the finite maximum angular scale of the 3PCF; by accounting for the bin-averaging effect, we include the contribution of the 3PCF at slightly larger angular scales by $\Delta\!\ln\theta$, and as consequence the aperture mass statistics get larger to be closer to the direct prediction of bispectrum. We can reduce the effect of bin-averaging by using a larger $\theta_{\rm max}$ for the 3PCF, e.g. $\theta_{\rm max}\sim4R_{\rm max}$, but that would increase the computational cost for both the theoretical prediction and the measurements. Therefore, for the comparison to the measurement with a finite maximum angular scale, it is recommended to use the aperture mass statistics calculated from the 3PCF.

\section{Conclusion}\label{sec:conclusion}

In this work, we presented a new method to compute the shear 3PCF from a given model of the convergence bispectrum, based on the multipole expansion of triangles in real and Fourier space. The multipole-based method developed in this paper enables fast and robust computation of theoretical prediction of third-order shear statistics, and makes parameter inference feasible with Mote-Carlo sampling in Stage-III surveys. Because the third-order shear statistics contains complementary information to the 2PCF, the multipole-based method will be a powerful tool for performing joint analyses of second- and third-order shear statistics to put tighter constraints on cosmological parameters and to improve the control of systematic errors. Additionally, utilizing 2D-FFTLog for the double Hankel transformation, our implementation of this method predicts the shear 3PCF on an FFT grid of two triangle side lengths $\theta_1$ and $\theta_2$, which provides useful visualizations of the dependence of the shear 3PCF on the triangle configuration.

We tested the accuracy of the multipole-based method by comparing it to a brute-force calculation, and we found that our method can predict the shear 3PCF at 5\% accuracy for most of the triangle configurations. The residual can be larger than 5\% for the triangles in the squeezed limit. To improve the accuracy further, we would need to capture the missing contributions from the bispectrum in squeezed limit triangles, where the bispectrum diverges, making the computational time significantly longer. \citet{Slepian.Eisenstein.2015} showed that the large amplitude of squeezed limit triangle originates from terms involving $1/k_3^2$ dependence based on the cosmological perturbation theory (where $k_3$ is the side length of triangle opposite to the angle $\phi$ on which the multipole expansion is applied), and that this dependence can be captured in a tailored way of multipole expansion for these specific terms. A similar approach could be applied to the weak lensing bispectrum, because the bispectrum scaling is determined by the same perturbation theory at sufficiently large scale. We leave this for future work.

We also tested the efficiency of the multipole-based method by measuring the computational time of each step of the method. We found that the method can evaluate the 3PCF in about 7 seconds for single source redshift bin and about 40 seconds even for the tomography setup with four source redshift bins, which is 6 orders of magnitude faster than the brute-force method. The method is efficient enough to be applied to the parameter inference based on Monte Carlo sampling.

Finally, we applied the multipole-based method to the third-order aperture mass statistics. We found that the method can predict the aperture mass statistics at 2\% accuracy, and the bin-averaging effect is important unless the maximum angular scale of 3PCF is sufficiently larger than $4R_{\rm max}$.

In this work, we focused on the auto-correlation of three spin-2 shear fields, but the formalism presented in this work can be easily generalized to cross-correlation as well. Especially, the extension to position-position-shear cross correlations is interesting for studying intrinsic alignments, as there is no signal expected from gravitational lensing, so the only source would be from intrinsic alignments. In this paper, we focused on the parity symmetric bispectrum, assuming bispectrum due to weak lensing. By changing the multipole basis from Legendre polynomial to Fourier basis, we can also include the contribution from a parity asymmetric bispectrum, which is especially important for the TATT model of intrinsic alignments. We leave these for future work.

\begin{acknowledgments}
We thank Bhuvnesh Jain, Nickolas Kokron, Daniel Eisenstein, Peter Schneider for useful discussion.
SS is supported by JSPS Overseas Research Fellowships.
\end{acknowledgments}

\bibliography{refs}

\begin{thebibliography}{81}%
\makeatletter
\providecommand \@ifxundefined [1]{%
 \@ifx{#1\undefined}
}%
\providecommand \@ifnum [1]{%
 \ifnum #1\expandafter \@firstoftwo
 \else \expandafter \@secondoftwo
 \fi
}%
\providecommand \@ifx [1]{%
 \ifx #1\expandafter \@firstoftwo
 \else \expandafter \@secondoftwo
 \fi
}%
\providecommand \natexlab [1]{#1}%
\providecommand \enquote  [1]{``#1''}%
\providecommand \bibnamefont  [1]{#1}%
\providecommand \bibfnamefont [1]{#1}%
\providecommand \citenamefont [1]{#1}%
\providecommand \href@noop [0]{\@secondoftwo}%
\providecommand \href [0]{\begingroup \@sanitize@url \@href}%
\providecommand \@href[1]{\@@startlink{#1}\@@href}%
\providecommand \@@href[1]{\endgroup#1\@@endlink}%
\providecommand \@sanitize@url [0]{\catcode `\\12\catcode `\$12\catcode
  `\&12\catcode `\#12\catcode `\^12\catcode `\_12\catcode `\%12\relax}%
\providecommand \@@startlink[1]{}%
\providecommand \@@endlink[0]{}%
\providecommand \url  [0]{\begingroup\@sanitize@url \@url }%
\providecommand \@url [1]{\endgroup\@href {#1}{\urlprefix }}%
\providecommand \urlprefix  [0]{URL }%
\providecommand \Eprint [0]{\href }%
\providecommand \doibase [0]{https://doi.org/}%
\providecommand \selectlanguage [0]{\@gobble}%
\providecommand \bibinfo  [0]{\@secondoftwo}%
\providecommand \bibfield  [0]{\@secondoftwo}%
\providecommand \translation [1]{[#1]}%
\providecommand \BibitemOpen [0]{}%
\providecommand \bibitemStop [0]{}%
\providecommand \bibitemNoStop [0]{.\EOS\space}%
\providecommand \EOS [0]{\spacefactor3000\relax}%
\providecommand \BibitemShut  [1]{\csname bibitem#1\endcsname}%
\let\auto@bib@innerbib\@empty
\bibitem [{\citenamefont {Heymans}\ \emph {et~al.}(2013)\citenamefont
  {Heymans}, \citenamefont {Grocutt}, \citenamefont {Heavens}, \citenamefont
  {Kilbinger}, \citenamefont {Kitching} \emph
  {et~al.}}]{Heymans.Velander.2013}%
  \BibitemOpen
  \bibfield  {author} {\bibinfo {author} {\bibfnamefont {C.}~\bibnamefont
  {Heymans}}, \bibinfo {author} {\bibfnamefont {E.}~\bibnamefont {Grocutt}},
  \bibinfo {author} {\bibfnamefont {A.}~\bibnamefont {Heavens}}, \bibinfo
  {author} {\bibfnamefont {M.}~\bibnamefont {Kilbinger}}, \bibinfo {author}
  {\bibfnamefont {T.~D.}\ \bibnamefont {Kitching}}, \emph {et~al.},\ }\bibfield
   {title} {\bibinfo {title} {{CFHTLenS tomographic weak lensing cosmological
  parameter constraints: Mitigating the impact of intrinsic galaxy
  alignments}},\ }\href {https://doi.org/10.1093/mnras/stt601} {\bibfield
  {journal} {\bibinfo  {journal} {Monthly Notices of the Royal Astronomical
  Society}\ }\textbf {\bibinfo {volume} {432}},\ \bibinfo {pages} {2433}
  (\bibinfo {year} {2013})},\ \Eprint {https://arxiv.org/abs/1303.1808}
  {1303.1808} \BibitemShut {NoStop}%
\bibitem [{\citenamefont {Hildebrandt}\ \emph {et~al.}(2016)\citenamefont
  {Hildebrandt}, \citenamefont {Viola}, \citenamefont {Heymans}, \citenamefont
  {Joudaki}, \citenamefont {Kuijken} \emph
  {et~al.}}]{Hildebrandt.Waerbeke.2016}%
  \BibitemOpen
  \bibfield  {author} {\bibinfo {author} {\bibfnamefont {H.}~\bibnamefont
  {Hildebrandt}}, \bibinfo {author} {\bibfnamefont {M.}~\bibnamefont {Viola}},
  \bibinfo {author} {\bibfnamefont {C.}~\bibnamefont {Heymans}}, \bibinfo
  {author} {\bibfnamefont {S.}~\bibnamefont {Joudaki}}, \bibinfo {author}
  {\bibfnamefont {K.}~\bibnamefont {Kuijken}}, \emph {et~al.},\ }\bibfield
  {title} {\bibinfo {title} {{KiDS-450: cosmological parameter constraints from
  tomographic weak gravitational lensing}},\ }\href
  {https://doi.org/10.1093/mnras/stw2805} {\bibfield  {journal} {\bibinfo
  {journal} {Monthly Notices of the Royal Astronomical Society}\ }\textbf
  {\bibinfo {volume} {465}},\ \bibinfo {pages} {1454} (\bibinfo {year}
  {2016})},\ \Eprint {https://arxiv.org/abs/1606.05338} {1606.05338}
  \BibitemShut {NoStop}%
\bibitem [{\citenamefont {Troxel}\ \emph {et~al.}(2018)\citenamefont {Troxel},
  \citenamefont {MacCrann}, \citenamefont {Zuntz}, \citenamefont {Eifler},
  \citenamefont {Krause} \emph {et~al.}}]{Troxel.Zhang.2018}%
  \BibitemOpen
  \bibfield  {author} {\bibinfo {author} {\bibfnamefont {M.~A.}\ \bibnamefont
  {Troxel}}, \bibinfo {author} {\bibfnamefont {N.}~\bibnamefont {MacCrann}},
  \bibinfo {author} {\bibfnamefont {J.}~\bibnamefont {Zuntz}}, \bibinfo
  {author} {\bibfnamefont {T.~F.}\ \bibnamefont {Eifler}}, \bibinfo {author}
  {\bibfnamefont {E.}~\bibnamefont {Krause}}, \emph {et~al.},\ }\bibfield
  {title} {\bibinfo {title} {{Dark Energy Survey Year 1 Results: Cosmological
  Constraints from Cosmic Shear}},\ }\href
  {https://doi.org/10.1103/physrevd.98.043528} {\bibfield  {journal} {\bibinfo
  {journal} {Physical Review D}\ }\textbf {\bibinfo {volume} {98}},\ \bibinfo
  {pages} {043528} (\bibinfo {year} {2018})},\ \bibinfo {note} {arXiv:
  1708.01538},\ \Eprint {https://arxiv.org/abs/1708.01538} {1708.01538}
  \BibitemShut {NoStop}%
\bibitem [{\citenamefont {Hikage}\ \emph {et~al.}(2019)\citenamefont {Hikage},
  \citenamefont {Oguri}, \citenamefont {Hamana}, \citenamefont {More},
  \citenamefont {Mandelbaum} \emph {et~al.}}]{Hikage.Yamada.2019}%
  \BibitemOpen
  \bibfield  {author} {\bibinfo {author} {\bibfnamefont {C.}~\bibnamefont
  {Hikage}}, \bibinfo {author} {\bibfnamefont {M.}~\bibnamefont {Oguri}},
  \bibinfo {author} {\bibfnamefont {T.}~\bibnamefont {Hamana}}, \bibinfo
  {author} {\bibfnamefont {S.}~\bibnamefont {More}}, \bibinfo {author}
  {\bibfnamefont {R.}~\bibnamefont {Mandelbaum}}, \emph {et~al.},\ }\bibfield
  {title} {\bibinfo {title} {{Cosmology from cosmic shear power spectra with
  Subaru Hyper Suprime-Cam first-year data}},\ }\bibfield  {journal} {\bibinfo
  {journal} {Publications of the Astronomical Society of Japan}\ }\textbf
  {\bibinfo {volume} {71}},\ \href {https://doi.org/10.1093/pasj/psz010}
  {10.1093/pasj/psz010} (\bibinfo {year} {2019}),\ \bibinfo {note} {arXiv:
  1809.09148},\ \Eprint {https://arxiv.org/abs/1809.09148} {1809.09148}
  \BibitemShut {NoStop}%
\bibitem [{\citenamefont {Hamana}\ \emph {et~al.}(2019)\citenamefont {Hamana},
  \citenamefont {Shirasaki}, \citenamefont {Miyazaki}, \citenamefont {Hikage},
  \citenamefont {Oguri} \emph {et~al.}}]{Hamana.Tanaka.2019}%
  \BibitemOpen
  \bibfield  {author} {\bibinfo {author} {\bibfnamefont {T.}~\bibnamefont
  {Hamana}}, \bibinfo {author} {\bibfnamefont {M.}~\bibnamefont {Shirasaki}},
  \bibinfo {author} {\bibfnamefont {S.}~\bibnamefont {Miyazaki}}, \bibinfo
  {author} {\bibfnamefont {C.}~\bibnamefont {Hikage}}, \bibinfo {author}
  {\bibfnamefont {M.}~\bibnamefont {Oguri}}, \emph {et~al.},\ }\bibfield
  {title} {\bibinfo {title} {{Cosmological constraints from cosmic shear
  two-point correlation functions with HSC survey first-year data}},\
  }\bibfield  {journal} {\bibinfo  {journal} {Publications of the Astronomical
  Society of Japan}\ }\textbf {\bibinfo {volume} {72}},\ \href
  {https://doi.org/10.1093/pasj/psz138} {10.1093/pasj/psz138} (\bibinfo {year}
  {2019}),\ \Eprint {https://arxiv.org/abs/1906.06041} {1906.06041}
  \BibitemShut {NoStop}%
\bibitem [{\citenamefont {Asgari}\ \emph {et~al.}(2020)\citenamefont {Asgari},
  \citenamefont {Lin}, \citenamefont {Joachimi}, \citenamefont {Giblin},
  \citenamefont {Heymans} \emph {et~al.}}]{Asgari.Valentijn.2020}%
  \BibitemOpen
  \bibfield  {author} {\bibinfo {author} {\bibfnamefont {M.}~\bibnamefont
  {Asgari}}, \bibinfo {author} {\bibfnamefont {C.-A.}\ \bibnamefont {Lin}},
  \bibinfo {author} {\bibfnamefont {B.}~\bibnamefont {Joachimi}}, \bibinfo
  {author} {\bibfnamefont {B.}~\bibnamefont {Giblin}}, \bibinfo {author}
  {\bibfnamefont {C.}~\bibnamefont {Heymans}}, \emph {et~al.},\ }\bibfield
  {title} {\bibinfo {title} {{KiDS-1000 Cosmology: Cosmic shear constraints and
  comparison between two point statistics}},\ }\bibfield  {journal} {\bibinfo
  {journal} {arXiv}\ }\href {https://doi.org/10.48550/arxiv.2007.15633}
  {10.48550/arxiv.2007.15633} (\bibinfo {year} {2020}),\ \Eprint
  {https://arxiv.org/abs/2007.15633} {2007.15633} \BibitemShut {NoStop}%
\bibitem [{\citenamefont {Amon}\ \emph {et~al.}(2021)\citenamefont {Amon},
  \citenamefont {Gruen}, \citenamefont {Troxel}, \citenamefont {MacCrann},
  \citenamefont {Dodelson} \emph {et~al.}}]{Amon.Weller.2021}%
  \BibitemOpen
  \bibfield  {author} {\bibinfo {author} {\bibfnamefont {A.}~\bibnamefont
  {Amon}}, \bibinfo {author} {\bibfnamefont {D.}~\bibnamefont {Gruen}},
  \bibinfo {author} {\bibfnamefont {M.~A.}\ \bibnamefont {Troxel}}, \bibinfo
  {author} {\bibfnamefont {N.}~\bibnamefont {MacCrann}}, \bibinfo {author}
  {\bibfnamefont {S.}~\bibnamefont {Dodelson}}, \emph {et~al.},\ }\bibfield
  {title} {\bibinfo {title} {{Dark Energy Survey Year 3 Results: Cosmology from
  Cosmic Shear and Robustness to Data Calibration}},\ }\bibfield  {journal}
  {\bibinfo  {journal} {arXiv}\ }\href
  {https://doi.org/10.48550/arxiv.2105.13543} {10.48550/arxiv.2105.13543}
  (\bibinfo {year} {2021}),\ \Eprint {https://arxiv.org/abs/2105.13543}
  {2105.13543} \BibitemShut {NoStop}%
\bibitem [{\citenamefont {Secco}\ \emph {et~al.}(2021)\citenamefont {Secco},
  \citenamefont {Samuroff}, \citenamefont {Krause}, \citenamefont {Jain},
  \citenamefont {Blazek} \emph {et~al.}}]{Secco.To.2021}%
  \BibitemOpen
  \bibfield  {author} {\bibinfo {author} {\bibfnamefont {L.~F.}\ \bibnamefont
  {Secco}}, \bibinfo {author} {\bibfnamefont {S.}~\bibnamefont {Samuroff}},
  \bibinfo {author} {\bibfnamefont {E.}~\bibnamefont {Krause}}, \bibinfo
  {author} {\bibfnamefont {B.}~\bibnamefont {Jain}}, \bibinfo {author}
  {\bibfnamefont {J.}~\bibnamefont {Blazek}}, \emph {et~al.},\ }\bibfield
  {title} {\bibinfo {title} {{Dark Energy Survey Year 3 Results: Cosmology from
  Cosmic Shear and Robustness to Modeling Uncertainty}},\ }\bibfield  {journal}
  {\bibinfo  {journal} {arXiv}\ }\href
  {https://doi.org/10.48550/arxiv.2105.13544} {10.48550/arxiv.2105.13544}
  (\bibinfo {year} {2021}),\ \Eprint {https://arxiv.org/abs/2105.13544}
  {2105.13544} \BibitemShut {NoStop}%
\bibitem [{\citenamefont {Dalal}\ \emph {et~al.}(2023)\citenamefont {Dalal},
  \citenamefont {Li}, \citenamefont {Nicola}, \citenamefont {Zuntz},
  \citenamefont {Strauss} \emph {et~al.}}]{Dalal.Wang.2023}%
  \BibitemOpen
  \bibfield  {author} {\bibinfo {author} {\bibfnamefont {R.}~\bibnamefont
  {Dalal}}, \bibinfo {author} {\bibfnamefont {X.}~\bibnamefont {Li}}, \bibinfo
  {author} {\bibfnamefont {A.}~\bibnamefont {Nicola}}, \bibinfo {author}
  {\bibfnamefont {J.}~\bibnamefont {Zuntz}}, \bibinfo {author} {\bibfnamefont
  {M.~A.}\ \bibnamefont {Strauss}}, \emph {et~al.},\ }\bibfield  {title}
  {\bibinfo {title} {{Hyper Suprime-Cam Year 3 Results: Cosmology from Cosmic
  Shear Power Spectra}},\ }\bibfield  {journal} {\bibinfo  {journal} {arXiv}\
  }\href {https://doi.org/10.48550/arxiv.2304.00701}
  {10.48550/arxiv.2304.00701} (\bibinfo {year} {2023}),\ \Eprint
  {https://arxiv.org/abs/2304.00701} {2304.00701} \BibitemShut {NoStop}%
\bibitem [{\citenamefont {Li}\ \emph {et~al.}(2023)\citenamefont {Li},
  \citenamefont {Zhang}, \citenamefont {Sugiyama}, \citenamefont {Dalal},
  \citenamefont {Terasawa} \emph {et~al.}}]{Li.Wang.2023}%
  \BibitemOpen
  \bibfield  {author} {\bibinfo {author} {\bibfnamefont {X.}~\bibnamefont
  {Li}}, \bibinfo {author} {\bibfnamefont {T.}~\bibnamefont {Zhang}}, \bibinfo
  {author} {\bibfnamefont {S.}~\bibnamefont {Sugiyama}}, \bibinfo {author}
  {\bibfnamefont {R.}~\bibnamefont {Dalal}}, \bibinfo {author} {\bibfnamefont
  {R.}~\bibnamefont {Terasawa}}, \emph {et~al.},\ }\bibfield  {title} {\bibinfo
  {title} {{Hyper Suprime-Cam Year 3 Results: Cosmology from Cosmic Shear
  Two-point Correlation Functions}},\ }\bibfield  {journal} {\bibinfo
  {journal} {arXiv}\ }\href {https://doi.org/10.48550/arxiv.2304.00702}
  {10.48550/arxiv.2304.00702} (\bibinfo {year} {2023}),\ \Eprint
  {https://arxiv.org/abs/2304.00702} {2304.00702} \BibitemShut {NoStop}%
\bibitem [{\citenamefont {Collaboration}\ \emph {et~al.}(2020)\citenamefont
  {Collaboration}, \citenamefont {Aghanim}, \citenamefont {Akrami},
  \citenamefont {Ashdown}, \citenamefont {Aumont} \emph
  {et~al.}}]{Collaboration.Zonca.2020}%
  \BibitemOpen
  \bibfield  {author} {\bibinfo {author} {\bibfnamefont {P.}~\bibnamefont
  {Collaboration}}, \bibinfo {author} {\bibfnamefont {N.}~\bibnamefont
  {Aghanim}}, \bibinfo {author} {\bibfnamefont {Y.}~\bibnamefont {Akrami}},
  \bibinfo {author} {\bibfnamefont {M.}~\bibnamefont {Ashdown}}, \bibinfo
  {author} {\bibfnamefont {J.}~\bibnamefont {Aumont}}, \emph {et~al.},\
  }\bibfield  {title} {\bibinfo {title} {{Planck 2018 results. VI. Cosmological
  parameters}},\ }\href {https://doi.org/10.1051/0004-6361/201833910}
  {\bibfield  {journal} {\bibinfo  {journal} {Astronomy \& Astrophysics}\
  }\textbf {\bibinfo {volume} {641}},\ \bibinfo {pages} {A6} (\bibinfo {year}
  {2020})},\ \bibinfo {note} {arXiv: 1807.06209},\ \Eprint
  {https://arxiv.org/abs/1807.06209} {1807.06209} \BibitemShut {NoStop}%
\bibitem [{\citenamefont {Chang}\ \emph {et~al.}(2018)\citenamefont {Chang},
  \citenamefont {Pujol}, \citenamefont {Mawdsley}, \citenamefont {Bacon},
  \citenamefont {Elvin-Poole} \emph {et~al.}}]{Chang.Collaboration.2018}%
  \BibitemOpen
  \bibfield  {author} {\bibinfo {author} {\bibfnamefont {C.}~\bibnamefont
  {Chang}}, \bibinfo {author} {\bibfnamefont {A.}~\bibnamefont {Pujol}},
  \bibinfo {author} {\bibfnamefont {B.}~\bibnamefont {Mawdsley}}, \bibinfo
  {author} {\bibfnamefont {D.}~\bibnamefont {Bacon}}, \bibinfo {author}
  {\bibfnamefont {J.}~\bibnamefont {Elvin-Poole}}, \emph {et~al.},\ }\bibfield
  {title} {\bibinfo {title} {{Dark Energy Survey Year 1 results: curved-sky
  weak lensing mass map}},\ }\href {https://doi.org/10.1093/mnras/stx3363}
  {\bibfield  {journal} {\bibinfo  {journal} {Monthly Notices of the Royal
  Astronomical Society}\ }\textbf {\bibinfo {volume} {475}},\ \bibinfo {pages}
  {3165} (\bibinfo {year} {2018})},\ \Eprint {https://arxiv.org/abs/1708.01535}
  {1708.01535} \BibitemShut {NoStop}%
\bibitem [{\citenamefont {Gatti}\ \emph {et~al.}(2020)\citenamefont {Gatti},
  \citenamefont {Chang}, \citenamefont {Friedrich}, \citenamefont {Jain},
  \citenamefont {Bacon} \emph {et~al.}}]{Gatti.collaboration.2020}%
  \BibitemOpen
  \bibfield  {author} {\bibinfo {author} {\bibfnamefont {M.}~\bibnamefont
  {Gatti}}, \bibinfo {author} {\bibfnamefont {C.}~\bibnamefont {Chang}},
  \bibinfo {author} {\bibfnamefont {O.}~\bibnamefont {Friedrich}}, \bibinfo
  {author} {\bibfnamefont {B.}~\bibnamefont {Jain}}, \bibinfo {author}
  {\bibfnamefont {D.}~\bibnamefont {Bacon}}, \emph {et~al.},\ }\bibfield
  {title} {\bibinfo {title} {{Dark Energy Survey Year 3 results: cosmology with
  moments of weak lensing mass maps – validation on simulations}},\ }\href
  {https://doi.org/10.1093/mnras/staa2680} {\bibfield  {journal} {\bibinfo
  {journal} {Monthly Notices of the Royal Astronomical Society}\ }\textbf
  {\bibinfo {volume} {498}},\ \bibinfo {pages} {4060} (\bibinfo {year}
  {2020})},\ \Eprint {https://arxiv.org/abs/1911.05568} {1911.05568}
  \BibitemShut {NoStop}%
\bibitem [{\citenamefont {Gatti}\ \emph {et~al.}(2022)\citenamefont {Gatti},
  \citenamefont {Jain}, \citenamefont {Chang}, \citenamefont {Raveri},
  \citenamefont {Z{\"u}rcher} \emph {et~al.}}]{Gatti.Collaboration.2022}%
  \BibitemOpen
  \bibfield  {author} {\bibinfo {author} {\bibfnamefont {M.}~\bibnamefont
  {Gatti}}, \bibinfo {author} {\bibfnamefont {B.}~\bibnamefont {Jain}},
  \bibinfo {author} {\bibfnamefont {C.}~\bibnamefont {Chang}}, \bibinfo
  {author} {\bibfnamefont {M.}~\bibnamefont {Raveri}}, \bibinfo {author}
  {\bibfnamefont {D.}~\bibnamefont {Z{\"u}rcher}}, \emph {et~al.},\ }\bibfield
  {title} {\bibinfo {title} {{Dark Energy Survey Year 3 results: Cosmology with
  moments of weak lensing mass maps}},\ }\href
  {https://doi.org/10.1103/physrevd.106.083509} {\bibfield  {journal} {\bibinfo
   {journal} {Physical Review D}\ }\textbf {\bibinfo {volume} {106}},\ \bibinfo
  {pages} {083509} (\bibinfo {year} {2022})}\BibitemShut {NoStop}%
\bibitem [{\citenamefont {Peel}\ \emph {et~al.}(2018)\citenamefont {Peel},
  \citenamefont {Pettorino}, \citenamefont {Giocoli}, \citenamefont {Starck},\
  and\ \citenamefont {Baldi}}]{Peel.Baldi.2018}%
  \BibitemOpen
  \bibfield  {author} {\bibinfo {author} {\bibfnamefont {A.}~\bibnamefont
  {Peel}}, \bibinfo {author} {\bibfnamefont {V.}~\bibnamefont {Pettorino}},
  \bibinfo {author} {\bibfnamefont {C.}~\bibnamefont {Giocoli}}, \bibinfo
  {author} {\bibfnamefont {J.-L.}\ \bibnamefont {Starck}},\ and\ \bibinfo
  {author} {\bibfnamefont {M.~o.}\ \bibnamefont {Baldi}},\ }\bibfield  {title}
  {\bibinfo {title} {{Breaking degeneracies in modified gravity with higher
  (than 2nd) order weak-lensing statistics}},\ }\bibfield  {journal} {\bibinfo
  {journal} {arXiv}\ }\href {https://doi.org/10.48550/arxiv.1805.05146}
  {10.48550/arxiv.1805.05146} (\bibinfo {year} {2018}),\ \Eprint
  {https://arxiv.org/abs/1805.05146} {1805.05146} \BibitemShut {NoStop}%
\bibitem [{\citenamefont {Petri}\ \emph {et~al.}(2015)\citenamefont {Petri},
  \citenamefont {Liu}, \citenamefont {Haiman}, \citenamefont {May},
  \citenamefont {Hui} \emph {et~al.}}]{Petri.Kratochvil.2015}%
  \BibitemOpen
  \bibfield  {author} {\bibinfo {author} {\bibfnamefont {A.}~\bibnamefont
  {Petri}}, \bibinfo {author} {\bibfnamefont {J.}~\bibnamefont {Liu}}, \bibinfo
  {author} {\bibfnamefont {Z.}~\bibnamefont {Haiman}}, \bibinfo {author}
  {\bibfnamefont {M.}~\bibnamefont {May}}, \bibinfo {author} {\bibfnamefont
  {L.}~\bibnamefont {Hui}}, \emph {et~al.},\ }\bibfield  {title} {\bibinfo
  {title} {{Emulating the CFHTLenS Weak Lensing data: Cosmological Constraints
  from moments and Minkowski functionals}},\ }\href
  {https://doi.org/10.48550/arxiv.1503.06214} {\bibfield  {journal} {\bibinfo
  {journal} {arXiv}\ }\textbf {\bibinfo {volume} {91}},\ \bibinfo {pages}
  {103511} (\bibinfo {year} {2015})},\ \Eprint
  {https://arxiv.org/abs/1503.06214} {1503.06214} \BibitemShut {NoStop}%
\bibitem [{\citenamefont {Porth}\ and\ \citenamefont
  {Smith}(2021)}]{Porth.Smith.2021}%
  \BibitemOpen
  \bibfield  {author} {\bibinfo {author} {\bibfnamefont {L.}~\bibnamefont
  {Porth}}\ and\ \bibinfo {author} {\bibfnamefont {R.~E.}\ \bibnamefont
  {Smith}},\ }\bibfield  {title} {\bibinfo {title} {{Fast estimation of
  aperture mass statistics II: Detectability of higher order statistics in
  current and future surveys}},\ }\bibfield  {journal} {\bibinfo  {journal}
  {arXiv}\ }\href {https://doi.org/10.48550/arxiv.2106.04594}
  {10.48550/arxiv.2106.04594} (\bibinfo {year} {2021}),\ \Eprint
  {https://arxiv.org/abs/2106.04594} {2106.04594} \BibitemShut {NoStop}%
\bibitem [{\citenamefont {Waerbeke}\ \emph {et~al.}(2013)\citenamefont
  {Waerbeke}, \citenamefont {Benjamin}, \citenamefont {Erben}, \citenamefont
  {Heymans}, \citenamefont {Hildebrandt} \emph
  {et~al.}}]{Waerbeke.Velander.2013}%
  \BibitemOpen
  \bibfield  {author} {\bibinfo {author} {\bibfnamefont {L.~V.}\ \bibnamefont
  {Waerbeke}}, \bibinfo {author} {\bibfnamefont {J.}~\bibnamefont {Benjamin}},
  \bibinfo {author} {\bibfnamefont {T.}~\bibnamefont {Erben}}, \bibinfo
  {author} {\bibfnamefont {C.}~\bibnamefont {Heymans}}, \bibinfo {author}
  {\bibfnamefont {H.}~\bibnamefont {Hildebrandt}}, \emph {et~al.},\ }\bibfield
  {title} {\bibinfo {title} {{CFHTLenS: mapping the large-scale structure with
  gravitational lensing}},\ }\href {https://doi.org/10.1093/mnras/stt971}
  {\bibfield  {journal} {\bibinfo  {journal} {Monthly Notices of the Royal
  Astronomical Society}\ }\textbf {\bibinfo {volume} {433}},\ \bibinfo {pages}
  {3373} (\bibinfo {year} {2013})},\ \Eprint {https://arxiv.org/abs/1303.1806}
  {1303.1806} \BibitemShut {NoStop}%
\bibitem [{\citenamefont {Vicinanza}\ \emph {et~al.}(2016)\citenamefont
  {Vicinanza}, \citenamefont {Cardone}, \citenamefont {Maoli}, \citenamefont
  {Scaramella},\ and\ \citenamefont {Er}}]{Vicinanza.Er.2016}%
  \BibitemOpen
  \bibfield  {author} {\bibinfo {author} {\bibfnamefont {M.}~\bibnamefont
  {Vicinanza}}, \bibinfo {author} {\bibfnamefont {V.~F.}\ \bibnamefont
  {Cardone}}, \bibinfo {author} {\bibfnamefont {R.}~\bibnamefont {Maoli}},
  \bibinfo {author} {\bibfnamefont {R.}~\bibnamefont {Scaramella}},\ and\
  \bibinfo {author} {\bibfnamefont {X.~o.}\ \bibnamefont {Er}},\ }\bibfield
  {title} {\bibinfo {title} {{Higher order moments of lensing convergence - I.
  Estimate from simulations}},\ }\bibfield  {journal} {\bibinfo  {journal}
  {arXiv}\ }\href {https://doi.org/10.48550/arxiv.1606.03892}
  {10.48550/arxiv.1606.03892} (\bibinfo {year} {2016}),\ \Eprint
  {https://arxiv.org/abs/1606.03892} {1606.03892} \BibitemShut {NoStop}%
\bibitem [{\citenamefont {Vicinanza}\ \emph {et~al.}(2018)\citenamefont
  {Vicinanza}, \citenamefont {Cardone}, \citenamefont {Maoli}, \citenamefont
  {Scaramella},\ and\ \citenamefont {Er}}]{Vicinanza.Er.2018}%
  \BibitemOpen
  \bibfield  {author} {\bibinfo {author} {\bibfnamefont {M.}~\bibnamefont
  {Vicinanza}}, \bibinfo {author} {\bibfnamefont {V.~F.}\ \bibnamefont
  {Cardone}}, \bibinfo {author} {\bibfnamefont {R.}~\bibnamefont {Maoli}},
  \bibinfo {author} {\bibfnamefont {R.}~\bibnamefont {Scaramella}},\ and\
  \bibinfo {author} {\bibfnamefont {X.~o.}\ \bibnamefont {Er}},\ }\bibfield
  {title} {\bibinfo {title} {{Increasing the lensing figure of merit through
  higher order convergence moments}},\ }\href
  {https://doi.org/10.1103/physrevd.97.023519} {\bibfield  {journal} {\bibinfo
  {journal} {Physical Review D}\ }\textbf {\bibinfo {volume} {97}},\ \bibinfo
  {pages} {023519} (\bibinfo {year} {2018})},\ \Eprint
  {https://arxiv.org/abs/1802.02963} {1802.02963} \BibitemShut {NoStop}%
\bibitem [{\citenamefont {Ajani}\ \emph {et~al.}(2020)\citenamefont {Ajani},
  \citenamefont {Peel}, \citenamefont {Pettorino}, \citenamefont {Starck},
  \citenamefont {Li} \emph {et~al.}}]{Ajani.Liu.2020}%
  \BibitemOpen
  \bibfield  {author} {\bibinfo {author} {\bibfnamefont {V.}~\bibnamefont
  {Ajani}}, \bibinfo {author} {\bibfnamefont {A.}~\bibnamefont {Peel}},
  \bibinfo {author} {\bibfnamefont {V.}~\bibnamefont {Pettorino}}, \bibinfo
  {author} {\bibfnamefont {J.-L.}\ \bibnamefont {Starck}}, \bibinfo {author}
  {\bibfnamefont {Z.}~\bibnamefont {Li}}, \emph {et~al.},\ }\bibfield  {title}
  {\bibinfo {title} {{Constraining neutrino masses with weak-lensing multiscale
  peak counts}},\ }\href {https://doi.org/10.1103/physrevd.102.103531}
  {\bibfield  {journal} {\bibinfo  {journal} {Physical Review D}\ }\textbf
  {\bibinfo {volume} {102}},\ \bibinfo {pages} {103531} (\bibinfo {year}
  {2020})},\ \Eprint {https://arxiv.org/abs/2001.10993} {2001.10993}
  \BibitemShut {NoStop}%
\bibitem [{\citenamefont {Dietrich}\ and\ \citenamefont
  {Hartlap}(2010)}]{Dietrich.Hartlap.2010}%
  \BibitemOpen
  \bibfield  {author} {\bibinfo {author} {\bibfnamefont {J.~P.}\ \bibnamefont
  {Dietrich}}\ and\ \bibinfo {author} {\bibfnamefont {J.}~\bibnamefont
  {Hartlap}},\ }\bibfield  {title} {\bibinfo {title} {{Cosmology with the
  shear‐peak statistics}},\ }\href
  {https://doi.org/10.1111/j.1365-2966.2009.15948.x} {\bibfield  {journal}
  {\bibinfo  {journal} {Monthly Notices of the Royal Astronomical Society}\
  }\textbf {\bibinfo {volume} {402}},\ \bibinfo {pages} {1049} (\bibinfo {year}
  {2010})},\ \Eprint {https://arxiv.org/abs/0906.3512} {0906.3512} \BibitemShut
  {NoStop}%
\bibitem [{\citenamefont {Harnois-D{\'e}raps}\ \emph
  {et~al.}(2021)\citenamefont {Harnois-D{\'e}raps}, \citenamefont {Martinet},\
  and\ \citenamefont {Reischke}}]{Harnois-Deraps.Reischke.2021}%
  \BibitemOpen
  \bibfield  {author} {\bibinfo {author} {\bibfnamefont {J.}~\bibnamefont
  {Harnois-D{\'e}raps}}, \bibinfo {author} {\bibfnamefont {N.}~\bibnamefont
  {Martinet}},\ and\ \bibinfo {author} {\bibfnamefont {R.}~\bibnamefont
  {Reischke}},\ }\bibfield  {title} {\bibinfo {title} {{Cosmic shear beyond
  2-point statistics: Accounting for galaxy intrinsic alignment with projected
  tidal fields}},\ }\href {https://doi.org/10.1093/mnras/stab3222} {\bibfield
  {journal} {\bibinfo  {journal} {Monthly Notices of the Royal Astronomical
  Society}\ }\textbf {\bibinfo {volume} {509}},\ \bibinfo {pages} {3868}
  (\bibinfo {year} {2021})},\ \Eprint {https://arxiv.org/abs/2107.08041}
  {2107.08041} \BibitemShut {NoStop}%
\bibitem [{\citenamefont {Kacprzak}\ \emph {et~al.}(2016)\citenamefont
  {Kacprzak}, \citenamefont {Kirk}, \citenamefont {Friedrich}, \citenamefont
  {Amara}, \citenamefont {Refregier} \emph
  {et~al.}}]{Kacprzak.Collaboration.2016}%
  \BibitemOpen
  \bibfield  {author} {\bibinfo {author} {\bibfnamefont {T.}~\bibnamefont
  {Kacprzak}}, \bibinfo {author} {\bibfnamefont {D.}~\bibnamefont {Kirk}},
  \bibinfo {author} {\bibfnamefont {O.}~\bibnamefont {Friedrich}}, \bibinfo
  {author} {\bibfnamefont {A.}~\bibnamefont {Amara}}, \bibinfo {author}
  {\bibfnamefont {A.}~\bibnamefont {Refregier}}, \emph {et~al.},\ }\bibfield
  {title} {\bibinfo {title} {{Cosmology constraints from shear peak statistics
  in Dark Energy Survey Science Verification data}},\ }\href
  {https://doi.org/10.1093/mnras/stw2070} {\bibfield  {journal} {\bibinfo
  {journal} {Monthly Notices of the Royal Astronomical Society}\ }\textbf
  {\bibinfo {volume} {463}},\ \bibinfo {pages} {3653} (\bibinfo {year}
  {2016})},\ \Eprint {https://arxiv.org/abs/1603.05040} {1603.05040}
  \BibitemShut {NoStop}%
\bibitem [{\citenamefont {Kratochvil}\ \emph {et~al.}(2010)\citenamefont
  {Kratochvil}, \citenamefont {Haiman},\ and\ \citenamefont
  {May}}]{Kratochvil.May.2010}%
  \BibitemOpen
  \bibfield  {author} {\bibinfo {author} {\bibfnamefont {J.~M.}\ \bibnamefont
  {Kratochvil}}, \bibinfo {author} {\bibfnamefont {Z.}~\bibnamefont {Haiman}},\
  and\ \bibinfo {author} {\bibfnamefont {M.}~\bibnamefont {May}},\ }\bibfield
  {title} {\bibinfo {title} {{Probing cosmology with weak lensing peak
  counts}},\ }\href {https://doi.org/10.1103/physrevd.81.043519} {\bibfield
  {journal} {\bibinfo  {journal} {Physical Review D}\ }\textbf {\bibinfo
  {volume} {81}},\ \bibinfo {pages} {043519} (\bibinfo {year} {2010})},\
  \Eprint {https://arxiv.org/abs/0907.0486} {0907.0486} \BibitemShut {NoStop}%
\bibitem [{\citenamefont {Liu}\ \emph {et~al.}(2015)\citenamefont {Liu},
  \citenamefont {Petri}, \citenamefont {Haiman}, \citenamefont {Hui},
  \citenamefont {Kratochvil} \emph {et~al.}}]{Liu.May.2015}%
  \BibitemOpen
  \bibfield  {author} {\bibinfo {author} {\bibfnamefont {J.}~\bibnamefont
  {Liu}}, \bibinfo {author} {\bibfnamefont {A.}~\bibnamefont {Petri}}, \bibinfo
  {author} {\bibfnamefont {Z.}~\bibnamefont {Haiman}}, \bibinfo {author}
  {\bibfnamefont {L.}~\bibnamefont {Hui}}, \bibinfo {author} {\bibfnamefont
  {J.~M.}\ \bibnamefont {Kratochvil}}, \emph {et~al.},\ }\bibfield  {title}
  {\bibinfo {title} {{Cosmology constraints from the weak lensing peak counts
  and the power spectrum in CFHTLenS data}},\ }\href
  {https://doi.org/10.1103/physrevd.91.063507} {\bibfield  {journal} {\bibinfo
  {journal} {Physical Review D}\ }\textbf {\bibinfo {volume} {91}},\ \bibinfo
  {pages} {063507} (\bibinfo {year} {2015})},\ \Eprint
  {https://arxiv.org/abs/1412.0757} {1412.0757} \BibitemShut {NoStop}%
\bibitem [{\citenamefont {Martinet}\ \emph {et~al.}(2017)\citenamefont
  {Martinet}, \citenamefont {Schneider}, \citenamefont {Hildebrandt},
  \citenamefont {Shan}, \citenamefont {Asgari} \emph
  {et~al.}}]{Martinet.Nakajima.2017}%
  \BibitemOpen
  \bibfield  {author} {\bibinfo {author} {\bibfnamefont {N.}~\bibnamefont
  {Martinet}}, \bibinfo {author} {\bibfnamefont {P.}~\bibnamefont {Schneider}},
  \bibinfo {author} {\bibfnamefont {H.}~\bibnamefont {Hildebrandt}}, \bibinfo
  {author} {\bibfnamefont {H.}~\bibnamefont {Shan}}, \bibinfo {author}
  {\bibfnamefont {M.}~\bibnamefont {Asgari}}, \emph {et~al.},\ }\bibfield
  {title} {\bibinfo {title} {{KiDS-450: cosmological constraints from
  weak-lensing peak statistics – II: Inference from shear peaks using N-body
  simulations}},\ }\href {https://doi.org/10.1093/mnras/stx2793} {\bibfield
  {journal} {\bibinfo  {journal} {Monthly Notices of the Royal Astronomical
  Society}\ }\textbf {\bibinfo {volume} {474}},\ \bibinfo {pages} {712}
  (\bibinfo {year} {2017})},\ \Eprint {https://arxiv.org/abs/1709.07678}
  {1709.07678} \BibitemShut {NoStop}%
\bibitem [{\citenamefont {Shan}\ \emph {et~al.}(2017)\citenamefont {Shan},
  \citenamefont {Liu}, \citenamefont {Hildebrandt}, \citenamefont {Pan},
  \citenamefont {Martinet} \emph {et~al.}}]{Shan.Wang.2017}%
  \BibitemOpen
  \bibfield  {author} {\bibinfo {author} {\bibfnamefont {H.}~\bibnamefont
  {Shan}}, \bibinfo {author} {\bibfnamefont {X.}~\bibnamefont {Liu}}, \bibinfo
  {author} {\bibfnamefont {H.}~\bibnamefont {Hildebrandt}}, \bibinfo {author}
  {\bibfnamefont {C.}~\bibnamefont {Pan}}, \bibinfo {author} {\bibfnamefont
  {N.}~\bibnamefont {Martinet}}, \emph {et~al.},\ }\bibfield  {title} {\bibinfo
  {title} {{KiDS-450: cosmological constraints from weak lensing peak
  statistics – I. Inference from analytical prediction of high
  signal-to-noise ratio convergence peaks}},\ }\href
  {https://doi.org/10.1093/mnras/stx2837} {\bibfield  {journal} {\bibinfo
  {journal} {Monthly Notices of the Royal Astronomical Society}\ }\textbf
  {\bibinfo {volume} {474}},\ \bibinfo {pages} {1116} (\bibinfo {year}
  {2017})},\ \Eprint {https://arxiv.org/abs/1709.07651} {1709.07651}
  \BibitemShut {NoStop}%
\bibitem [{\citenamefont {Z{\"u}rcher}\ \emph {et~al.}(2023)\citenamefont
  {Z{\"u}rcher}, \citenamefont {Fluri}, \citenamefont {Ajani}, \citenamefont
  {Fischbacher}, \citenamefont {Refregier} \emph
  {et~al.}}]{Zurcher.Kacprzak.2023}%
  \BibitemOpen
  \bibfield  {author} {\bibinfo {author} {\bibfnamefont {D.}~\bibnamefont
  {Z{\"u}rcher}}, \bibinfo {author} {\bibfnamefont {J.}~\bibnamefont {Fluri}},
  \bibinfo {author} {\bibfnamefont {V.}~\bibnamefont {Ajani}}, \bibinfo
  {author} {\bibfnamefont {S.}~\bibnamefont {Fischbacher}}, \bibinfo {author}
  {\bibfnamefont {A.}~\bibnamefont {Refregier}}, \emph {et~al.},\ }\bibfield
  {title} {\bibinfo {title} {{Towards a full wCDM map-based analysis for weak
  lensing surveys}},\ }\href {https://doi.org/10.1093/mnras/stad2212}
  {\bibfield  {journal} {\bibinfo  {journal} {Monthly Notices of the Royal
  Astronomical Society}\ }\textbf {\bibinfo {volume} {525}},\ \bibinfo {pages}
  {761} (\bibinfo {year} {2023})},\ \Eprint {https://arxiv.org/abs/2206.01450}
  {2206.01450} \BibitemShut {NoStop}%
\bibitem [{\citenamefont {Z{\"u}rcher}\ \emph {et~al.}(2020)\citenamefont
  {Z{\"u}rcher}, \citenamefont {Fluri}, \citenamefont {Sgier}, \citenamefont
  {Kacprzak},\ and\ \citenamefont {Refregier}}]{Zurcher.Refregier.2020}%
  \BibitemOpen
  \bibfield  {author} {\bibinfo {author} {\bibfnamefont {D.}~\bibnamefont
  {Z{\"u}rcher}}, \bibinfo {author} {\bibfnamefont {J.}~\bibnamefont {Fluri}},
  \bibinfo {author} {\bibfnamefont {R.}~\bibnamefont {Sgier}}, \bibinfo
  {author} {\bibfnamefont {T.}~\bibnamefont {Kacprzak}},\ and\ \bibinfo
  {author} {\bibfnamefont {A.~o.}\ \bibnamefont {Refregier}},\ }\bibfield
  {title} {\bibinfo {title} {{Cosmological Forecast for non-Gaussian Statistics
  in large-scale weak Lensing Surveys}},\ }\bibfield  {journal} {\bibinfo
  {journal} {arXiv}\ }\href {https://doi.org/10.48550/arxiv.2006.12506}
  {10.48550/arxiv.2006.12506} (\bibinfo {year} {2020}),\ \Eprint
  {https://arxiv.org/abs/2006.12506} {2006.12506} \BibitemShut {NoStop}%
\bibitem [{\citenamefont {Barthelemy}\ \emph {et~al.}(2020)\citenamefont
  {Barthelemy}, \citenamefont {Codis}, \citenamefont {Uhlemann}, \citenamefont
  {Bernardeau},\ and\ \citenamefont {Gavazzi}}]{Barthelemy.Gavazzi.2020}%
  \BibitemOpen
  \bibfield  {author} {\bibinfo {author} {\bibfnamefont {A.}~\bibnamefont
  {Barthelemy}}, \bibinfo {author} {\bibfnamefont {S.}~\bibnamefont {Codis}},
  \bibinfo {author} {\bibfnamefont {C.}~\bibnamefont {Uhlemann}}, \bibinfo
  {author} {\bibfnamefont {F.}~\bibnamefont {Bernardeau}},\ and\ \bibinfo
  {author} {\bibfnamefont {R.~o.}\ \bibnamefont {Gavazzi}},\ }\bibfield
  {title} {\bibinfo {title} {{A nulling strategy for modelling lensing
  convergence in cones with large deviation theory}},\ }\href
  {https://doi.org/10.1093/mnras/staa053} {\bibfield  {journal} {\bibinfo
  {journal} {Monthly Notices of the Royal Astronomical Society}\ }\textbf
  {\bibinfo {volume} {492}},\ \bibinfo {pages} {3420} (\bibinfo {year}
  {2020})},\ \Eprint {https://arxiv.org/abs/1909.02615} {1909.02615}
  \BibitemShut {NoStop}%
\bibitem [{\citenamefont {Boyle}\ \emph {et~al.}(2021)\citenamefont {Boyle},
  \citenamefont {Uhlemann}, \citenamefont {Friedrich}, \citenamefont
  {Barthelemy}, \citenamefont {Codis} \emph {et~al.}}]{Boyle.Baldi.2021}%
  \BibitemOpen
  \bibfield  {author} {\bibinfo {author} {\bibfnamefont {A.}~\bibnamefont
  {Boyle}}, \bibinfo {author} {\bibfnamefont {C.}~\bibnamefont {Uhlemann}},
  \bibinfo {author} {\bibfnamefont {O.}~\bibnamefont {Friedrich}}, \bibinfo
  {author} {\bibfnamefont {A.}~\bibnamefont {Barthelemy}}, \bibinfo {author}
  {\bibfnamefont {S.}~\bibnamefont {Codis}}, \emph {et~al.},\ }\bibfield
  {title} {\bibinfo {title} {{Nuw CDM cosmology from the weak-lensing
  convergence PDF}},\ }\href {https://doi.org/10.1093/mnras/stab1381}
  {\bibfield  {journal} {\bibinfo  {journal} {Monthly Notices of the Royal
  Astronomical Society}\ }\textbf {\bibinfo {volume} {505}},\ \bibinfo {pages}
  {2886} (\bibinfo {year} {2021})},\ \Eprint {https://arxiv.org/abs/2012.07771}
  {2012.07771} \BibitemShut {NoStop}%
\bibitem [{\citenamefont {Thiele}\ \emph {et~al.}(2020)\citenamefont {Thiele},
  \citenamefont {Hill},\ and\ \citenamefont {Smith}}]{Thiele.Smith.2020}%
  \BibitemOpen
  \bibfield  {author} {\bibinfo {author} {\bibfnamefont {L.}~\bibnamefont
  {Thiele}}, \bibinfo {author} {\bibfnamefont {J.~C.}\ \bibnamefont {Hill}},\
  and\ \bibinfo {author} {\bibfnamefont {K.~M.}\ \bibnamefont {Smith}},\
  }\bibfield  {title} {\bibinfo {title} {{Accurate analytic model for the weak
  lensing convergence one-point probability distribution function and its
  autocovariance}},\ }\href {https://doi.org/10.1103/physrevd.102.123545}
  {\bibfield  {journal} {\bibinfo  {journal} {Physical Review D}\ }\textbf
  {\bibinfo {volume} {102}},\ \bibinfo {pages} {123545} (\bibinfo {year}
  {2020})},\ \Eprint {https://arxiv.org/abs/2009.06547} {2009.06547}
  \BibitemShut {NoStop}%
\bibitem [{\citenamefont {Grewal}\ \emph {et~al.}(2022)\citenamefont {Grewal},
  \citenamefont {Zuntz}, \citenamefont {Tr{\"o}ster},\ and\ \citenamefont
  {Amon}}]{Grewal.Amon.2022}%
  \BibitemOpen
  \bibfield  {author} {\bibinfo {author} {\bibfnamefont {N.}~\bibnamefont
  {Grewal}}, \bibinfo {author} {\bibfnamefont {J.}~\bibnamefont {Zuntz}},
  \bibinfo {author} {\bibfnamefont {T.}~\bibnamefont {Tr{\"o}ster}},\ and\
  \bibinfo {author} {\bibfnamefont {A.}~\bibnamefont {Amon}},\ }\bibfield
  {title} {\bibinfo {title} {{Minkowski Functionals in Joint Galaxy Clustering
  \& Weak Lensing Analyses}},\ }\bibfield  {journal} {\bibinfo  {journal}
  {arXiv}\ }\href {https://doi.org/10.48550/arxiv.2206.03877}
  {10.48550/arxiv.2206.03877} (\bibinfo {year} {2022}),\ \Eprint
  {https://arxiv.org/abs/2206.03877} {2206.03877} \BibitemShut {NoStop}%
\bibitem [{\citenamefont {Kratochvil}\ \emph {et~al.}(2012)\citenamefont
  {Kratochvil}, \citenamefont {Lim}, \citenamefont {Wang}, \citenamefont
  {Haiman}, \citenamefont {May} \emph {et~al.}}]{Kratochvil.Huffenberger.2012}%
  \BibitemOpen
  \bibfield  {author} {\bibinfo {author} {\bibfnamefont {J.~M.}\ \bibnamefont
  {Kratochvil}}, \bibinfo {author} {\bibfnamefont {E.~A.}\ \bibnamefont {Lim}},
  \bibinfo {author} {\bibfnamefont {S.}~\bibnamefont {Wang}}, \bibinfo {author}
  {\bibfnamefont {Z.}~\bibnamefont {Haiman}}, \bibinfo {author} {\bibfnamefont
  {M.}~\bibnamefont {May}}, \emph {et~al.},\ }\bibfield  {title} {\bibinfo
  {title} {{Probing cosmology with weak lensing Minkowski functionals}},\
  }\href {https://doi.org/10.1103/physrevd.85.103513} {\bibfield  {journal}
  {\bibinfo  {journal} {Physical Review D}\ }\textbf {\bibinfo {volume} {85}},\
  \bibinfo {pages} {103513} (\bibinfo {year} {2012})},\ \Eprint
  {https://arxiv.org/abs/1109.6334} {1109.6334} \BibitemShut {NoStop}%
\bibitem [{\citenamefont {Parroni}\ \emph {et~al.}(2019)\citenamefont
  {Parroni}, \citenamefont {Cardone}, \citenamefont {Maoli},\ and\
  \citenamefont {Scaramella}}]{Parroni.Scaramella.2019}%
  \BibitemOpen
  \bibfield  {author} {\bibinfo {author} {\bibfnamefont {C.}~\bibnamefont
  {Parroni}}, \bibinfo {author} {\bibfnamefont {V.~F.}\ \bibnamefont
  {Cardone}}, \bibinfo {author} {\bibfnamefont {R.}~\bibnamefont {Maoli}},\
  and\ \bibinfo {author} {\bibfnamefont {R.}~\bibnamefont {Scaramella}},\
  }\bibfield  {title} {\bibinfo {title} {{Going deep with Minkowski functionals
  of convergence maps}},\ }\bibfield  {journal} {\bibinfo  {journal} {arXiv}\
  }\href {https://doi.org/10.48550/arxiv.1911.06243}
  {10.48550/arxiv.1911.06243} (\bibinfo {year} {2019}),\ \Eprint
  {https://arxiv.org/abs/1911.06243} {1911.06243} \BibitemShut {NoStop}%
\bibitem [{\citenamefont {Vicinanza}\ \emph {et~al.}(2019)\citenamefont
  {Vicinanza}, \citenamefont {Cardone}, \citenamefont {Maoli}, \citenamefont
  {Scaramella}, \citenamefont {Er} \emph {et~al.}}]{Vicinanza.Tereno.2019}%
  \BibitemOpen
  \bibfield  {author} {\bibinfo {author} {\bibfnamefont {M.}~\bibnamefont
  {Vicinanza}}, \bibinfo {author} {\bibfnamefont {V.~F.}\ \bibnamefont
  {Cardone}}, \bibinfo {author} {\bibfnamefont {R.}~\bibnamefont {Maoli}},
  \bibinfo {author} {\bibfnamefont {R.}~\bibnamefont {Scaramella}}, \bibinfo
  {author} {\bibfnamefont {X.}~\bibnamefont {Er}}, \emph {et~al.},\ }\bibfield
  {title} {\bibinfo {title} {{Minkowski functionals of convergence maps and the
  lensing figure of merit}},\ }\href
  {https://doi.org/10.1103/physrevd.99.043534} {\bibfield  {journal} {\bibinfo
  {journal} {Physical Review D}\ }\textbf {\bibinfo {volume} {99}},\ \bibinfo
  {pages} {043534} (\bibinfo {year} {2019})},\ \Eprint
  {https://arxiv.org/abs/1905.00410} {1905.00410} \BibitemShut {NoStop}%
\bibitem [{\citenamefont {Feldbrugge}\ \emph {et~al.}(2019)\citenamefont
  {Feldbrugge}, \citenamefont {Engelen}, \citenamefont {Weygaert},
  \citenamefont {Pranav},\ and\ \citenamefont
  {Vegter}}]{Feldbrugge.Vegter.2019}%
  \BibitemOpen
  \bibfield  {author} {\bibinfo {author} {\bibfnamefont {J.}~\bibnamefont
  {Feldbrugge}}, \bibinfo {author} {\bibfnamefont {M.~v.}\ \bibnamefont
  {Engelen}}, \bibinfo {author} {\bibfnamefont {R.~v.~d.}\ \bibnamefont
  {Weygaert}}, \bibinfo {author} {\bibfnamefont {P.}~\bibnamefont {Pranav}},\
  and\ \bibinfo {author} {\bibfnamefont {G.~o.}\ \bibnamefont {Vegter}},\
  }\bibfield  {title} {\bibinfo {title} {{Stochastic Homology of Gaussian vs.
  non-Gaussian Random Fields: Graphs towards Betti Numbers and Persistence
  Diagrams}},\ }\bibfield  {journal} {\bibinfo  {journal} {arXiv}\ }\href
  {https://doi.org/10.48550/arxiv.1908.01619} {10.48550/arxiv.1908.01619}
  (\bibinfo {year} {2019}),\ \Eprint {https://arxiv.org/abs/1908.01619}
  {1908.01619} \BibitemShut {NoStop}%
\bibitem [{\citenamefont {Parroni}\ \emph {et~al.}(2020)\citenamefont
  {Parroni}, \citenamefont {Tollet}, \citenamefont {Cardone}, \citenamefont
  {Maoli},\ and\ \citenamefont {Scaramella}}]{Parroni.Scaramella.2020}%
  \BibitemOpen
  \bibfield  {author} {\bibinfo {author} {\bibfnamefont {C.}~\bibnamefont
  {Parroni}}, \bibinfo {author} {\bibfnamefont {E.}~\bibnamefont {Tollet}},
  \bibinfo {author} {\bibfnamefont {V.~F.}\ \bibnamefont {Cardone}}, \bibinfo
  {author} {\bibfnamefont {R.}~\bibnamefont {Maoli}},\ and\ \bibinfo {author}
  {\bibfnamefont {R.~o.}\ \bibnamefont {Scaramella}},\ }\bibfield  {title}
  {\bibinfo {title} {{Higher order statistics of shear field: a machine
  learning approach}},\ }\bibfield  {journal} {\bibinfo  {journal} {arXiv}\
  }\href {https://doi.org/10.48550/arxiv.2011.10438}
  {10.48550/arxiv.2011.10438} (\bibinfo {year} {2020}),\ \Eprint
  {https://arxiv.org/abs/2011.10438} {2011.10438} \BibitemShut {NoStop}%
\bibitem [{\citenamefont {Heydenreich}\ \emph
  {et~al.}(2022{\natexlab{a}})\citenamefont {Heydenreich}, \citenamefont
  {Br{\"u}ck}, \citenamefont {Burger}, \citenamefont {Harnois-D{\'e}raps},
  \citenamefont {Unruh} \emph {et~al.}}]{Heydenreich.Martinet.2022}%
  \BibitemOpen
  \bibfield  {author} {\bibinfo {author} {\bibfnamefont {S.}~\bibnamefont
  {Heydenreich}}, \bibinfo {author} {\bibfnamefont {B.}~\bibnamefont
  {Br{\"u}ck}}, \bibinfo {author} {\bibfnamefont {P.}~\bibnamefont {Burger}},
  \bibinfo {author} {\bibfnamefont {J.}~\bibnamefont {Harnois-D{\'e}raps}},
  \bibinfo {author} {\bibfnamefont {S.}~\bibnamefont {Unruh}}, \emph {et~al.},\
  }\bibfield  {title} {\bibinfo {title} {{Persistent homology in cosmic shear
  II: A tomographic analysis of DES-Y1}},\ }\bibfield  {journal} {\bibinfo
  {journal} {arXiv}\ }\href {https://doi.org/10.48550/arxiv.2204.11831}
  {10.48550/arxiv.2204.11831} (\bibinfo {year} {2022}{\natexlab{a}}),\ \Eprint
  {https://arxiv.org/abs/2204.11831} {2204.11831} \BibitemShut {NoStop}%
\bibitem [{\citenamefont {Heydenreich}\ \emph
  {et~al.}(2022{\natexlab{b}})\citenamefont {Heydenreich}, \citenamefont
  {Linke}, \citenamefont {Burger},\ and\ \citenamefont
  {Schneider}}]{Heydenreich.Schneider.2022}%
  \BibitemOpen
  \bibfield  {author} {\bibinfo {author} {\bibfnamefont {S.}~\bibnamefont
  {Heydenreich}}, \bibinfo {author} {\bibfnamefont {L.}~\bibnamefont {Linke}},
  \bibinfo {author} {\bibfnamefont {P.}~\bibnamefont {Burger}},\ and\ \bibinfo
  {author} {\bibfnamefont {P.}~\bibnamefont {Schneider}},\ }\bibfield  {title}
  {\bibinfo {title} {{A roadmap to cosmological parameter analysis with
  third-order shear statistics I: Modelling and validation}},\ }\bibfield
  {journal} {\bibinfo  {journal} {arXiv}\ }\href
  {https://doi.org/10.48550/arxiv.2208.11686} {10.48550/arxiv.2208.11686}
  (\bibinfo {year} {2022}{\natexlab{b}}),\ \Eprint
  {https://arxiv.org/abs/2208.11686} {2208.11686} \BibitemShut {NoStop}%
\bibitem [{\citenamefont {Cheng}\ \emph {et~al.}(2020)\citenamefont {Cheng},
  \citenamefont {Ting}, \citenamefont {M{\'e}nard},\ and\ \citenamefont
  {Bruna}}]{Cheng.Bruna.2020}%
  \BibitemOpen
  \bibfield  {author} {\bibinfo {author} {\bibfnamefont {S.}~\bibnamefont
  {Cheng}}, \bibinfo {author} {\bibfnamefont {Y.-S.}\ \bibnamefont {Ting}},
  \bibinfo {author} {\bibfnamefont {B.}~\bibnamefont {M{\'e}nard}},\ and\
  \bibinfo {author} {\bibfnamefont {J.}~\bibnamefont {Bruna}},\ }\bibfield
  {title} {\bibinfo {title} {{A new approach to observational cosmology using
  the scattering transform}},\ }\href {https://doi.org/10.1093/mnras/staa3165}
  {\bibfield  {journal} {\bibinfo  {journal} {Monthly Notices of the Royal
  Astronomical Society}\ }\textbf {\bibinfo {volume} {499}},\ \bibinfo {pages}
  {5902} (\bibinfo {year} {2020})},\ \Eprint {https://arxiv.org/abs/2006.08561}
  {2006.08561} \BibitemShut {NoStop}%
\bibitem [{\citenamefont {Valogiannis}\ and\ \citenamefont
  {Dvorkin}(2022{\natexlab{a}})}]{Valogiannis.Dvorkin.2022}%
  \BibitemOpen
  \bibfield  {author} {\bibinfo {author} {\bibfnamefont {G.}~\bibnamefont
  {Valogiannis}}\ and\ \bibinfo {author} {\bibfnamefont {C.}~\bibnamefont
  {Dvorkin}},\ }\bibfield  {title} {\bibinfo {title} {{Going beyond the galaxy
  power spectrum: An analysis of BOSS data with wavelet scattering
  transforms}},\ }\href {https://doi.org/10.1103/physrevd.106.103509}
  {\bibfield  {journal} {\bibinfo  {journal} {Physical Review D}\ }\textbf
  {\bibinfo {volume} {106}},\ \bibinfo {pages} {103509} (\bibinfo {year}
  {2022}{\natexlab{a}})},\ \Eprint {https://arxiv.org/abs/2204.13717}
  {2204.13717} \BibitemShut {NoStop}%
\bibitem [{\citenamefont {Valogiannis}\ and\ \citenamefont
  {Dvorkin}(2022{\natexlab{b}})}]{Valogiannis.Dvorkin.2022m5}%
  \BibitemOpen
  \bibfield  {author} {\bibinfo {author} {\bibfnamefont {G.}~\bibnamefont
  {Valogiannis}}\ and\ \bibinfo {author} {\bibfnamefont {C.}~\bibnamefont
  {Dvorkin}},\ }\bibfield  {title} {\bibinfo {title} {{Towards an optimal
  estimation of cosmological parameters with the wavelet scattering
  transform}},\ }\href {https://doi.org/10.1103/physrevd.105.103534} {\bibfield
   {journal} {\bibinfo  {journal} {Physical Review D}\ }\textbf {\bibinfo
  {volume} {105}},\ \bibinfo {pages} {103534} (\bibinfo {year}
  {2022}{\natexlab{b}})},\ \Eprint {https://arxiv.org/abs/2108.07821}
  {2108.07821} \BibitemShut {NoStop}%
\bibitem [{\citenamefont {Allys}\ \emph {et~al.}(2020)\citenamefont {Allys},
  \citenamefont {Marchand}, \citenamefont {Cardoso}, \citenamefont
  {Villaescusa-Navarro}, \citenamefont {Ho} \emph
  {et~al.}}]{Allys.Mallat.2020}%
  \BibitemOpen
  \bibfield  {author} {\bibinfo {author} {\bibfnamefont {E.}~\bibnamefont
  {Allys}}, \bibinfo {author} {\bibfnamefont {T.}~\bibnamefont {Marchand}},
  \bibinfo {author} {\bibfnamefont {J.~F.}\ \bibnamefont {Cardoso}}, \bibinfo
  {author} {\bibfnamefont {F.}~\bibnamefont {Villaescusa-Navarro}}, \bibinfo
  {author} {\bibfnamefont {S.}~\bibnamefont {Ho}}, \emph {et~al.},\ }\bibfield
  {title} {\bibinfo {title} {{New interpretable statistics for large-scale
  structure analysis and generation}},\ }\href
  {https://doi.org/10.1103/physrevd.102.103506} {\bibfield  {journal} {\bibinfo
   {journal} {Physical Review D}\ }\textbf {\bibinfo {volume} {102}},\ \bibinfo
  {pages} {103506} (\bibinfo {year} {2020})},\ \Eprint
  {https://arxiv.org/abs/2006.06298} {2006.06298} \BibitemShut {NoStop}%
\bibitem [{\citenamefont {Anbajagane}\ \emph {et~al.}(2023)\citenamefont
  {Anbajagane}, \citenamefont {Chang}, \citenamefont {Banerjee}, \citenamefont
  {Abel}, \citenamefont {Gatti} \emph {et~al.}}]{Anbajagane.Wiseman.2023}%
  \BibitemOpen
  \bibfield  {author} {\bibinfo {author} {\bibfnamefont {D.}~\bibnamefont
  {Anbajagane}}, \bibinfo {author} {\bibfnamefont {C.}~\bibnamefont {Chang}},
  \bibinfo {author} {\bibfnamefont {A.}~\bibnamefont {Banerjee}}, \bibinfo
  {author} {\bibfnamefont {T.}~\bibnamefont {Abel}}, \bibinfo {author}
  {\bibfnamefont {M.}~\bibnamefont {Gatti}}, \emph {et~al.},\ }\bibfield
  {title} {\bibinfo {title} {{Beyond the 3rd moment: A practical study of using
  lensing convergence CDFs for cosmology with DES Y3}},\ }\bibfield  {journal}
  {\bibinfo  {journal} {arXiv}\ }\href
  {https://doi.org/10.48550/arxiv.2308.03863} {10.48550/arxiv.2308.03863}
  (\bibinfo {year} {2023}),\ \Eprint {https://arxiv.org/abs/2308.03863}
  {2308.03863} \BibitemShut {NoStop}%
\bibitem [{\citenamefont {Banerjee}\ and\ \citenamefont
  {Abel}(2022)}]{Banerjee.Abel.2022}%
  \BibitemOpen
  \bibfield  {author} {\bibinfo {author} {\bibfnamefont {A.}~\bibnamefont
  {Banerjee}}\ and\ \bibinfo {author} {\bibfnamefont {T.}~\bibnamefont
  {Abel}},\ }\bibfield  {title} {\bibinfo {title} {{Tracer-field
  cross-correlations with k-nearest neighbour distributions}},\ }\href
  {https://doi.org/10.1093/mnras/stac3813} {\bibfield  {journal} {\bibinfo
  {journal} {Monthly Notices of the Royal Astronomical Society}\ }\textbf
  {\bibinfo {volume} {519}},\ \bibinfo {pages} {4856} (\bibinfo {year}
  {2022})},\ \Eprint {https://arxiv.org/abs/2210.05140} {2210.05140}
  \BibitemShut {NoStop}%
\bibitem [{\citenamefont {Boruah}\ \emph {et~al.}(2022)\citenamefont {Boruah},
  \citenamefont {Lavaux},\ and\ \citenamefont {Hudson}}]{Boruah.Hudson.2022}%
  \BibitemOpen
  \bibfield  {author} {\bibinfo {author} {\bibfnamefont {S.~S.}\ \bibnamefont
  {Boruah}}, \bibinfo {author} {\bibfnamefont {G.}~\bibnamefont {Lavaux}},\
  and\ \bibinfo {author} {\bibfnamefont {M.~J.}\ \bibnamefont {Hudson}},\
  }\bibfield  {title} {\bibinfo {title} {{Bayesian reconstruction of dark
  matter distribution from peculiar velocities: accounting for inhomogeneous
  Malmquist bias}},\ }\href {https://doi.org/10.1093/mnras/stac2985} {\bibfield
   {journal} {\bibinfo  {journal} {Monthly Notices of the Royal Astronomical
  Society}\ }\textbf {\bibinfo {volume} {517}},\ \bibinfo {pages} {4529}
  (\bibinfo {year} {2022})},\ \Eprint {https://arxiv.org/abs/2111.15535}
  {2111.15535} \BibitemShut {NoStop}%
\bibitem [{\citenamefont {Porqueres}\ \emph {et~al.}(2021)\citenamefont
  {Porqueres}, \citenamefont {Heavens}, \citenamefont {Mortlock},\ and\
  \citenamefont {Lavaux}}]{Porqueres.Lavaux.2021}%
  \BibitemOpen
  \bibfield  {author} {\bibinfo {author} {\bibfnamefont {N.}~\bibnamefont
  {Porqueres}}, \bibinfo {author} {\bibfnamefont {A.}~\bibnamefont {Heavens}},
  \bibinfo {author} {\bibfnamefont {D.}~\bibnamefont {Mortlock}},\ and\
  \bibinfo {author} {\bibfnamefont {G.}~\bibnamefont {Lavaux}},\ }\bibfield
  {title} {\bibinfo {title} {{Lifting weak lensing degeneracies with a
  field-based likelihood}},\ }\href {https://doi.org/10.1093/mnras/stab3234}
  {\bibfield  {journal} {\bibinfo  {journal} {Monthly Notices of the Royal
  Astronomical Society}\ }\textbf {\bibinfo {volume} {509}},\ \bibinfo {pages}
  {3194} (\bibinfo {year} {2021})},\ \Eprint {https://arxiv.org/abs/2108.04825}
  {2108.04825} \BibitemShut {NoStop}%
\bibitem [{\citenamefont {Fluri}\ \emph {et~al.}(2019)\citenamefont {Fluri},
  \citenamefont {Kacprzak}, \citenamefont {Lucchi}, \citenamefont {Refregier},
  \citenamefont {Amara} \emph {et~al.}}]{Fluri.Schneider.2019}%
  \BibitemOpen
  \bibfield  {author} {\bibinfo {author} {\bibfnamefont {J.}~\bibnamefont
  {Fluri}}, \bibinfo {author} {\bibfnamefont {T.}~\bibnamefont {Kacprzak}},
  \bibinfo {author} {\bibfnamefont {A.}~\bibnamefont {Lucchi}}, \bibinfo
  {author} {\bibfnamefont {A.}~\bibnamefont {Refregier}}, \bibinfo {author}
  {\bibfnamefont {A.}~\bibnamefont {Amara}}, \emph {et~al.},\ }\bibfield
  {title} {\bibinfo {title} {{Cosmological constraints with deep learning from
  KiDS-450 weak lensing maps}},\ }\href
  {https://doi.org/10.1103/physrevd.100.063514} {\bibfield  {journal} {\bibinfo
   {journal} {Physical Review D}\ }\textbf {\bibinfo {volume} {100}},\ \bibinfo
  {pages} {063514} (\bibinfo {year} {2019})},\ \Eprint
  {https://arxiv.org/abs/1906.03156} {1906.03156} \BibitemShut {NoStop}%
\bibitem [{\citenamefont {Fluri}\ \emph {et~al.}(2018)\citenamefont {Fluri},
  \citenamefont {Kacprzak}, \citenamefont {Refregier}, \citenamefont {Amara},
  \citenamefont {Lucchi} \emph {et~al.}}]{Fluri.Hofmann.2018}%
  \BibitemOpen
  \bibfield  {author} {\bibinfo {author} {\bibfnamefont {J.}~\bibnamefont
  {Fluri}}, \bibinfo {author} {\bibfnamefont {T.}~\bibnamefont {Kacprzak}},
  \bibinfo {author} {\bibfnamefont {A.}~\bibnamefont {Refregier}}, \bibinfo
  {author} {\bibfnamefont {A.}~\bibnamefont {Amara}}, \bibinfo {author}
  {\bibfnamefont {A.}~\bibnamefont {Lucchi}}, \emph {et~al.},\ }\bibfield
  {title} {\bibinfo {title} {{Cosmological constraints from noisy convergence
  maps through deep learning}},\ }\href
  {https://doi.org/10.1103/physrevd.98.123518} {\bibfield  {journal} {\bibinfo
  {journal} {Physical Review D}\ }\textbf {\bibinfo {volume} {98}},\ \bibinfo
  {pages} {123518} (\bibinfo {year} {2018})},\ \Eprint
  {https://arxiv.org/abs/1807.08732} {1807.08732} \BibitemShut {NoStop}%
\bibitem [{\citenamefont {Jeffrey}\ \emph {et~al.}(2020)\citenamefont
  {Jeffrey}, \citenamefont {Alsing},\ and\ \citenamefont
  {Lanusse}}]{Jeffrey.Lanusse.2020}%
  \BibitemOpen
  \bibfield  {author} {\bibinfo {author} {\bibfnamefont {N.}~\bibnamefont
  {Jeffrey}}, \bibinfo {author} {\bibfnamefont {J.}~\bibnamefont {Alsing}},\
  and\ \bibinfo {author} {\bibfnamefont {F.}~\bibnamefont {Lanusse}},\
  }\bibfield  {title} {\bibinfo {title} {{Likelihood-free inference with neural
  compression of DES SV weak lensing map statistics}},\ }\href
  {https://doi.org/10.1093/mnras/staa3594} {\bibfield  {journal} {\bibinfo
  {journal} {Monthly Notices of the Royal Astronomical Society}\ }\textbf
  {\bibinfo {volume} {501}},\ \bibinfo {pages} {954} (\bibinfo {year}
  {2020})},\ \Eprint {https://arxiv.org/abs/2009.08459} {2009.08459}
  \BibitemShut {NoStop}%
\bibitem [{\citenamefont {Lu}\ \emph {et~al.}(2023)\citenamefont {Lu},
  \citenamefont {Haiman},\ and\ \citenamefont {Li}}]{Lu.Li.2023}%
  \BibitemOpen
  \bibfield  {author} {\bibinfo {author} {\bibfnamefont {T.}~\bibnamefont
  {Lu}}, \bibinfo {author} {\bibfnamefont {Z.}~\bibnamefont {Haiman}},\ and\
  \bibinfo {author} {\bibfnamefont {X.}~\bibnamefont {Li}},\ }\bibfield
  {title} {\bibinfo {title} {{Cosmological constraints from HSC survey
  first-year data using deep learning}},\ }\href
  {https://doi.org/10.1093/mnras/stad686} {\bibfield  {journal} {\bibinfo
  {journal} {Monthly Notices of the Royal Astronomical Society}\ }\textbf
  {\bibinfo {volume} {521}},\ \bibinfo {pages} {2050} (\bibinfo {year}
  {2023})},\ \Eprint {https://arxiv.org/abs/2301.01354} {2301.01354}
  \BibitemShut {NoStop}%
\bibitem [{\citenamefont {Ribli}\ \emph {et~al.}(2019)\citenamefont {Ribli},
  \citenamefont {Pataki},\ and\ \citenamefont {Csabai}}]{Ribli.Csabai.2019}%
  \BibitemOpen
  \bibfield  {author} {\bibinfo {author} {\bibfnamefont {D.}~\bibnamefont
  {Ribli}}, \bibinfo {author} {\bibfnamefont {B.~{\'A}.}\ \bibnamefont
  {Pataki}},\ and\ \bibinfo {author} {\bibfnamefont {I.}~\bibnamefont
  {Csabai}},\ }\bibfield  {title} {\bibinfo {title} {{An improved cosmological
  parameter inference scheme motivated by deep learning}},\ }\href
  {https://doi.org/10.1038/s41550-018-0596-8} {\bibfield  {journal} {\bibinfo
  {journal} {Nature Astronomy}\ }\textbf {\bibinfo {volume} {3}},\ \bibinfo
  {pages} {93} (\bibinfo {year} {2019})}\BibitemShut {NoStop}%
\bibitem [{\citenamefont {Schneider}\ and\ \citenamefont
  {Lombardi}(2002)}]{Schneider.Lombardi.2002}%
  \BibitemOpen
  \bibfield  {author} {\bibinfo {author} {\bibfnamefont {P.}~\bibnamefont
  {Schneider}}\ and\ \bibinfo {author} {\bibfnamefont {M.}~\bibnamefont
  {Lombardi}},\ }\bibfield  {title} {\bibinfo {title} {{The three-point
  correlation function of cosmic shear: I. The natural components}},\
  }\bibfield  {journal} {\bibinfo  {journal} {arXiv}\ }\href
  {https://doi.org/10.48550/arxiv.astro-ph/0207454}
  {10.48550/arxiv.astro-ph/0207454} (\bibinfo {year} {2002}),\ \Eprint
  {https://arxiv.org/abs/astro-ph/0207454} {astro-ph/0207454} \BibitemShut
  {NoStop}%
\bibitem [{\citenamefont {Schneider}\ \emph {et~al.}(2003)\citenamefont
  {Schneider}, \citenamefont {Kilbinger},\ and\ \citenamefont
  {Lombardi}}]{Schneider.Lombardi.2003}%
  \BibitemOpen
  \bibfield  {author} {\bibinfo {author} {\bibfnamefont {P.}~\bibnamefont
  {Schneider}}, \bibinfo {author} {\bibfnamefont {M.}~\bibnamefont
  {Kilbinger}},\ and\ \bibinfo {author} {\bibfnamefont {M.}~\bibnamefont
  {Lombardi}},\ }\bibfield  {title} {\bibinfo {title} {{The three-point
  correlation function of cosmic shear. II: Relation to the bispectrum of the
  projected mass density and generalized third-order aperture measures}},\
  }\bibfield  {journal} {\bibinfo  {journal} {arXiv}\ }\href
  {https://doi.org/10.48550/arxiv.astro-ph/0308328}
  {10.48550/arxiv.astro-ph/0308328} (\bibinfo {year} {2003}),\ \Eprint
  {https://arxiv.org/abs/astro-ph/0308328} {astro-ph/0308328} \BibitemShut
  {NoStop}%
\bibitem [{\citenamefont {Zaldarriaga}\ and\ \citenamefont
  {Scoccimarro}(2003)}]{Zaldarriaga.Scoccimarro.2003}%
  \BibitemOpen
  \bibfield  {author} {\bibinfo {author} {\bibfnamefont {M.}~\bibnamefont
  {Zaldarriaga}}\ and\ \bibinfo {author} {\bibfnamefont {R.}~\bibnamefont
  {Scoccimarro}},\ }\bibfield  {title} {\bibinfo {title} {{Higher Order Moments
  of the Cosmic Shear and Other Spin-2 Fields}},\ }\href
  {https://doi.org/10.1086/345789} {\bibfield  {journal} {\bibinfo  {journal}
  {The Astrophysical Journal}\ }\textbf {\bibinfo {volume} {584}},\ \bibinfo
  {pages} {559} (\bibinfo {year} {2003})},\ \Eprint
  {https://arxiv.org/abs/astro-ph/0208075} {astro-ph/0208075} \BibitemShut
  {NoStop}%
\bibitem [{\citenamefont {Takada}\ and\ \citenamefont
  {Jain}(2004)}]{Takada.Jain.2004}%
  \BibitemOpen
  \bibfield  {author} {\bibinfo {author} {\bibfnamefont {M.}~\bibnamefont
  {Takada}}\ and\ \bibinfo {author} {\bibfnamefont {B.}~\bibnamefont {Jain}},\
  }\bibfield  {title} {\bibinfo {title} {{Cosmological parameters from lensing
  power spectrum and bispectrum tomography}},\ }\href
  {https://doi.org/10.1111/j.1365-2966.2004.07410.x} {\bibfield  {journal}
  {\bibinfo  {journal} {Monthly Notices of the Royal Astronomical Society}\
  }\textbf {\bibinfo {volume} {348}},\ \bibinfo {pages} {897} (\bibinfo {year}
  {2004})},\ \bibinfo {note} {arXiv: astro-ph/0310125},\ \Eprint
  {https://arxiv.org/abs/astro-ph/0310125} {astro-ph/0310125} \BibitemShut
  {NoStop}%
\bibitem [{\citenamefont {Secco}\ \emph {et~al.}(2022)\citenamefont {Secco},
  \citenamefont {Jarvis}, \citenamefont {Jain}, \citenamefont {Chang},
  \citenamefont {Gatti} \emph {et~al.}}]{Secco.Weller.2022}%
  \BibitemOpen
  \bibfield  {author} {\bibinfo {author} {\bibfnamefont {L.~F.}\ \bibnamefont
  {Secco}}, \bibinfo {author} {\bibfnamefont {M.}~\bibnamefont {Jarvis}},
  \bibinfo {author} {\bibfnamefont {B.}~\bibnamefont {Jain}}, \bibinfo {author}
  {\bibfnamefont {C.}~\bibnamefont {Chang}}, \bibinfo {author} {\bibfnamefont
  {M.}~\bibnamefont {Gatti}}, \emph {et~al.},\ }\bibfield  {title} {\bibinfo
  {title} {{Dark Energy Survey Year 3 Results: Three-Point Shear Correlations
  and Mass Aperture Moments}},\ }\bibfield  {journal} {\bibinfo  {journal}
  {arXiv}\ }\href {https://doi.org/10.48550/arxiv.2201.05227}
  {10.48550/arxiv.2201.05227} (\bibinfo {year} {2022}),\ \Eprint
  {https://arxiv.org/abs/2201.05227} {2201.05227} \BibitemShut {NoStop}%
\bibitem [{\citenamefont {Porth}\ \emph {et~al.}(2023)\citenamefont {Porth},
  \citenamefont {Heydenreich}, \citenamefont {Burger}, \citenamefont {Linke},\
  and\ \citenamefont {Schneider}}]{Porth.Schneider.2023}%
  \BibitemOpen
  \bibfield  {author} {\bibinfo {author} {\bibfnamefont {L.}~\bibnamefont
  {Porth}}, \bibinfo {author} {\bibfnamefont {S.}~\bibnamefont {Heydenreich}},
  \bibinfo {author} {\bibfnamefont {P.}~\bibnamefont {Burger}}, \bibinfo
  {author} {\bibfnamefont {L.}~\bibnamefont {Linke}},\ and\ \bibinfo {author}
  {\bibfnamefont {P.~o.}\ \bibnamefont {Schneider}},\ }\bibfield  {title}
  {\bibinfo {title} {{A roadmap to cosmological parameter analysis with
  third-order shear statistics III: Efficient estimation of third-order shear
  correlation functions and an application to the KiDS-1000 data}},\
  }\href@noop {} {\bibfield  {journal} {\bibinfo  {journal} {arXiv}\ }
  (\bibinfo {year} {2023})},\ \Eprint {https://arxiv.org/abs/2309.08601}
  {2309.08601} \BibitemShut {NoStop}%
\bibitem [{\citenamefont {Burger}\ \emph {et~al.}(2023)\citenamefont {Burger},
  \citenamefont {Porth}, \citenamefont {Heydenreich}, \citenamefont {Linke},
  \citenamefont {Wielders} \emph {et~al.}}]{Burger.Martinet.2023}%
  \BibitemOpen
  \bibfield  {author} {\bibinfo {author} {\bibfnamefont {P.~A.}\ \bibnamefont
  {Burger}}, \bibinfo {author} {\bibfnamefont {L.}~\bibnamefont {Porth}},
  \bibinfo {author} {\bibfnamefont {S.}~\bibnamefont {Heydenreich}}, \bibinfo
  {author} {\bibfnamefont {L.}~\bibnamefont {Linke}}, \bibinfo {author}
  {\bibfnamefont {N.}~\bibnamefont {Wielders}}, \emph {et~al.},\ }\bibfield
  {title} {\bibinfo {title} {{KiDS-1000 cosmology: Combined second- and
  third-order shear statistics}},\ }\bibfield  {journal} {\bibinfo  {journal}
  {arXiv}\ }\href {https://doi.org/10.48550/arxiv.2309.08602}
  {10.48550/arxiv.2309.08602} (\bibinfo {year} {2023}),\ \Eprint
  {https://arxiv.org/abs/2309.08602} {2309.08602} \BibitemShut {NoStop}%
\bibitem [{Note1()}]{Note1}%
  \BibitemOpen
  \bibinfo {note} {Formally, the implementation given in \protect \citet
  {Porth.Schneider.2023} is $O(N^2)$, however, in practice using hierarchical
  grids or a tree structure, the algorithm can be converted to $O(N \log
  N)$}\BibitemShut {NoStop}%
\bibitem [{\citenamefont {Chen}\ and\ \citenamefont
  {Szapudi}(2005)}]{Chen.Szapudi.2005}%
  \BibitemOpen
  \bibfield  {author} {\bibinfo {author} {\bibfnamefont {G.}~\bibnamefont
  {Chen}}\ and\ \bibinfo {author} {\bibfnamefont {I.}~\bibnamefont {Szapudi}},\
  }\bibfield  {title} {\bibinfo {title} {{Measuring the Three Point Correlation
  Function of the Cosmic Microwave Background}},\ }\bibfield  {journal}
  {\bibinfo  {journal} {arXiv}\ }\href
  {https://doi.org/10.48550/arxiv.astro-ph/0508316}
  {10.48550/arxiv.astro-ph/0508316} (\bibinfo {year} {2005}),\ \Eprint
  {https://arxiv.org/abs/astro-ph/0508316} {astro-ph/0508316} \BibitemShut
  {NoStop}%
\bibitem [{\citenamefont {Slepian}\ and\ \citenamefont
  {Eisenstein}()}]{Slepian.Eisenstein.2015}%
  \BibitemOpen
  \bibfield  {author} {\bibinfo {author} {\bibfnamefont {Z.}~\bibnamefont
  {Slepian}}\ and\ \bibinfo {author} {\bibfnamefont {D.~J.}\ \bibnamefont
  {Eisenstein}},\ }\bibfield  {title} {\bibinfo {title} {{{Computing the
  Three-Point Correlation Function of Galaxies in O(N 2) Time}}, year =
  {2015}},\ }\bibfield  {journal} {\bibinfo  {journal} {arXiv}\ }\href
  {https://doi.org/10.48550/arxiv.1506.02040} {10.48550/arxiv.1506.02040},\
  \Eprint {https://arxiv.org/abs/1506.02040} {1506.02040} \BibitemShut
  {NoStop}%
\bibitem [{\citenamefont {Philcox}\ \emph {et~al.}(2021)\citenamefont
  {Philcox}, \citenamefont {Slepian}, \citenamefont {Hou}, \citenamefont
  {Warner}, \citenamefont {Cahn} \emph {et~al.}}]{Philcox.Eisenstein.2021}%
  \BibitemOpen
  \bibfield  {author} {\bibinfo {author} {\bibfnamefont {O.~H.~E.}\
  \bibnamefont {Philcox}}, \bibinfo {author} {\bibfnamefont {Z.}~\bibnamefont
  {Slepian}}, \bibinfo {author} {\bibfnamefont {J.}~\bibnamefont {Hou}},
  \bibinfo {author} {\bibfnamefont {C.}~\bibnamefont {Warner}}, \bibinfo
  {author} {\bibfnamefont {R.~N.}\ \bibnamefont {Cahn}}, \emph {et~al.},\
  }\bibfield  {title} {\bibinfo {title} {{encore : an O ( N g2) estimator for
  galaxy N -point correlation functions}},\ }\href
  {https://doi.org/10.1093/mnras/stab3025} {\bibfield  {journal} {\bibinfo
  {journal} {Monthly Notices of the Royal Astronomical Society}\ }\textbf
  {\bibinfo {volume} {509}},\ \bibinfo {pages} {2457} (\bibinfo {year}
  {2021})},\ \Eprint {https://arxiv.org/abs/2105.08722} {2105.08722}
  \BibitemShut {NoStop}%
\bibitem [{\citenamefont {Umeh}(2020)}]{Umeh.Umeh.2020}%
  \BibitemOpen
  \bibfield  {author} {\bibinfo {author} {\bibfnamefont {O.}~\bibnamefont
  {Umeh}},\ }\bibfield  {title} {\bibinfo {title} {{Optimal computation of
  anisotropic galaxy three point correlation function multipoles using 2DFFTLOG
  formalism}},\ }\bibfield  {journal} {\bibinfo  {journal} {arXiv}\ }\href
  {https://doi.org/10.48550/arxiv.2011.05889} {10.48550/arxiv.2011.05889}
  (\bibinfo {year} {2020}),\ \Eprint {https://arxiv.org/abs/2011.05889}
  {2011.05889} \BibitemShut {NoStop}%
\bibitem [{\citenamefont {Guidi}\ \emph {et~al.}(2022)\citenamefont {Guidi},
  \citenamefont {Veropalumbo}, \citenamefont {Branchini}, \citenamefont
  {Eggemeier},\ and\ \citenamefont {Carbone}}]{Guidi.Carbone.2022}%
  \BibitemOpen
  \bibfield  {author} {\bibinfo {author} {\bibfnamefont {M.}~\bibnamefont
  {Guidi}}, \bibinfo {author} {\bibfnamefont {A.}~\bibnamefont {Veropalumbo}},
  \bibinfo {author} {\bibfnamefont {E.}~\bibnamefont {Branchini}}, \bibinfo
  {author} {\bibfnamefont {A.}~\bibnamefont {Eggemeier}},\ and\ \bibinfo
  {author} {\bibfnamefont {C.~o.}\ \bibnamefont {Carbone}},\ }\bibfield
  {title} {\bibinfo {title} {{Modelling the next-to-leading order matter
  three-point correlation function using FFTLog}},\ }\bibfield  {journal}
  {\bibinfo  {journal} {arXiv}\ }\href
  {https://doi.org/10.48550/arxiv.2212.07382} {10.48550/arxiv.2212.07382}
  (\bibinfo {year} {2022}),\ \Eprint {https://arxiv.org/abs/2212.07382}
  {2212.07382} \BibitemShut {NoStop}%
\bibitem [{\citenamefont {Aviles}\ and\ \citenamefont
  {Niz}(2023)}]{Aviles.Niz.2023}%
  \BibitemOpen
  \bibfield  {author} {\bibinfo {author} {\bibfnamefont {A.}~\bibnamefont
  {Aviles}}\ and\ \bibinfo {author} {\bibfnamefont {G.}~\bibnamefont {Niz}},\
  }\bibfield  {title} {\bibinfo {title} {{On the galaxy 3-point correlation
  function in Modified Gravity}},\ }\bibfield  {journal} {\bibinfo  {journal}
  {arXiv}\ }\href {https://doi.org/10.48550/arxiv.2301.07240}
  {10.48550/arxiv.2301.07240} (\bibinfo {year} {2023}),\ \Eprint
  {https://arxiv.org/abs/2301.07240} {2301.07240} \BibitemShut {NoStop}%
\bibitem [{\citenamefont {Takahashi}\ \emph {et~al.}(2019)\citenamefont
  {Takahashi}, \citenamefont {Nishimichi}, \citenamefont {Namikawa},
  \citenamefont {Taruya}, \citenamefont {Kayo} \emph
  {et~al.}}]{Takahashi.Shirasaki.2019}%
  \BibitemOpen
  \bibfield  {author} {\bibinfo {author} {\bibfnamefont {R.}~\bibnamefont
  {Takahashi}}, \bibinfo {author} {\bibfnamefont {T.}~\bibnamefont
  {Nishimichi}}, \bibinfo {author} {\bibfnamefont {T.}~\bibnamefont
  {Namikawa}}, \bibinfo {author} {\bibfnamefont {A.}~\bibnamefont {Taruya}},
  \bibinfo {author} {\bibfnamefont {I.}~\bibnamefont {Kayo}}, \emph {et~al.},\
  }\bibfield  {title} {\bibinfo {title} {{Fitting the nonlinear matter
  bispectrum by the Halofit approach}},\ }\bibfield  {journal} {\bibinfo
  {journal} {arXiv}\ }\href {https://doi.org/10.48550/arxiv.1911.07886}
  {10.48550/arxiv.1911.07886} (\bibinfo {year} {2019}),\ \Eprint
  {https://arxiv.org/abs/1911.07886} {1911.07886} \BibitemShut {NoStop}%
\bibitem [{Note2()}]{Note2}%
  \BibitemOpen
  \bibinfo {note} {\protect \url
  {https://github.com/rmjarvis/TreeCorr}}\BibitemShut {NoStop}%
\bibitem [{\citenamefont {Jarvis}\ \emph {et~al.}(2003)\citenamefont {Jarvis},
  \citenamefont {Bernstein},\ and\ \citenamefont {Jain}}]{Jarvis.Jain.2003}%
  \BibitemOpen
  \bibfield  {author} {\bibinfo {author} {\bibfnamefont {M.}~\bibnamefont
  {Jarvis}}, \bibinfo {author} {\bibfnamefont {G.}~\bibnamefont {Bernstein}},\
  and\ \bibinfo {author} {\bibfnamefont {B.}~\bibnamefont {Jain}},\ }\bibfield
  {title} {\bibinfo {title} {{The Skewness of the Aperture Mass Statistic}},\
  }\bibfield  {journal} {\bibinfo  {journal} {arXiv}\ }\href
  {https://doi.org/10.48550/arxiv.astro-ph/0307393}
  {10.48550/arxiv.astro-ph/0307393} (\bibinfo {year} {2003}),\ \Eprint
  {https://arxiv.org/abs/astro-ph/0307393} {astro-ph/0307393} \BibitemShut
  {NoStop}%
\bibitem [{\citenamefont {Schmitz}\ \emph {et~al.}(2018)\citenamefont
  {Schmitz}, \citenamefont {Hirata}, \citenamefont {Blazek},\ and\
  \citenamefont {Krause}}]{Schmitz.Krause.2018}%
  \BibitemOpen
  \bibfield  {author} {\bibinfo {author} {\bibfnamefont {D.~M.}\ \bibnamefont
  {Schmitz}}, \bibinfo {author} {\bibfnamefont {C.~M.}\ \bibnamefont {Hirata}},
  \bibinfo {author} {\bibfnamefont {J.}~\bibnamefont {Blazek}},\ and\ \bibinfo
  {author} {\bibfnamefont {E.}~\bibnamefont {Krause}},\ }\bibfield  {title}
  {\bibinfo {title} {{Time evolution of intrinsic alignments of galaxies}},\
  }\bibfield  {journal} {\bibinfo  {journal} {arXiv}\ }\href
  {https://doi.org/10.48550/arxiv.1805.02649} {10.48550/arxiv.1805.02649}
  (\bibinfo {year} {2018}),\ \Eprint {https://arxiv.org/abs/1805.02649}
  {1805.02649} \BibitemShut {NoStop}%
\bibitem [{\citenamefont {Fang}\ \emph {et~al.}(2020)\citenamefont {Fang},
  \citenamefont {Eifler},\ and\ \citenamefont {Krause}}]{Fang.Krause.2020}%
  \BibitemOpen
  \bibfield  {author} {\bibinfo {author} {\bibfnamefont {X.}~\bibnamefont
  {Fang}}, \bibinfo {author} {\bibfnamefont {T.}~\bibnamefont {Eifler}},\ and\
  \bibinfo {author} {\bibfnamefont {E.}~\bibnamefont {Krause}},\ }\bibfield
  {title} {\bibinfo {title} {{2D-FFTLog: efficient computation of real-space
  covariance matrices for galaxy clustering and weak lensing}},\ }\href
  {https://doi.org/10.1093/mnras/staa1726} {\bibfield  {journal} {\bibinfo
  {journal} {Monthly Notices of the Royal Astronomical Society}\ }\textbf
  {\bibinfo {volume} {497}},\ \bibinfo {pages} {2699} (\bibinfo {year}
  {2020})},\ \Eprint {https://arxiv.org/abs/2004.04833} {2004.04833}
  \BibitemShut {NoStop}%
\bibitem [{\citenamefont {McEwen}\ \emph {et~al.}(2016)\citenamefont {McEwen},
  \citenamefont {Fang}, \citenamefont {Hirata},\ and\ \citenamefont
  {Blazek}}]{McEwen.Blazek.2016}%
  \BibitemOpen
  \bibfield  {author} {\bibinfo {author} {\bibfnamefont {J.~E.}\ \bibnamefont
  {McEwen}}, \bibinfo {author} {\bibfnamefont {X.}~\bibnamefont {Fang}},
  \bibinfo {author} {\bibfnamefont {C.~M.}\ \bibnamefont {Hirata}},\ and\
  \bibinfo {author} {\bibfnamefont {J.~A.}\ \bibnamefont {Blazek}},\ }\bibfield
   {title} {\bibinfo {title} {{FAST-PT: a novel algorithm to calculate
  convolution integrals in cosmological perturbation theory}},\ }\href
  {https://doi.org/10.1088/1475-7516/2016/09/015} {\bibfield  {journal}
  {\bibinfo  {journal} {Journal of Cosmology and Astroparticle Physics}\
  }\textbf {\bibinfo {volume} {2016}}\bibfield  {number} {\bibinfo  {number} {
  (09)},\ \bibinfo {pages} {015}},\ }\bibinfo {note} {arXiv: 1603.04826},\
  \Eprint {https://arxiv.org/abs/1603.04826} {1603.04826} \BibitemShut
  {NoStop}%
\bibitem [{Note3()}]{Note3}%
  \BibitemOpen
  \bibinfo {note} {\protect \url
  {https://github.com/xfangcosmo/2DFFTLog}}\BibitemShut {NoStop}%
\bibitem [{Note4()}]{Note4}%
  \BibitemOpen
  \bibinfo {note} {\protect \url
  {https://github.com/git-sunao/fastnc}}\BibitemShut {NoStop}%
\bibitem [{Note5()}]{Note5}%
  \BibitemOpen
  \bibinfo {note} {\protect \url
  {https://pypi.org/project/fastnc/}}\BibitemShut {NoStop}%
\bibitem [{Note6()}]{Note6}%
  \BibitemOpen
  \bibinfo {note} {\protect \url
  {https://anaconda.org/ssunao/fastnc}}\BibitemShut {NoStop}%
\bibitem [{\citenamefont {Kaiser}(1995)}]{Kaiser.Kaiser.1995}%
  \BibitemOpen
  \bibfield  {author} {\bibinfo {author} {\bibfnamefont {N.}~\bibnamefont
  {Kaiser}},\ }\bibfield  {title} {\bibinfo {title} {{Nonlinear cluster lens
  reconstruction}},\ }\href {https://doi.org/10.1086/187730} {\bibfield
  {journal} {\bibinfo  {journal} {The Astrophysical Journal}\ }\textbf
  {\bibinfo {volume} {439}},\ \bibinfo {pages} {L1} (\bibinfo {year} {1995})},\
  \Eprint {https://arxiv.org/abs/astro-ph/9408092} {astro-ph/9408092}
  \BibitemShut {NoStop}%
\bibitem [{\citenamefont {Schneider}(1996)}]{Schneider.Schneider.1996}%
  \BibitemOpen
  \bibfield  {author} {\bibinfo {author} {\bibfnamefont {P.}~\bibnamefont
  {Schneider}},\ }\bibfield  {title} {\bibinfo {title} {{Detection of (dark)
  matter concentrations via weak gravitational lensing}},\ }\href
  {https://doi.org/10.1093/mnras/283.3.837} {\bibfield  {journal} {\bibinfo
  {journal} {Monthly Notices of the Royal Astronomical Society}\ }\textbf
  {\bibinfo {volume} {283}},\ \bibinfo {pages} {837} (\bibinfo {year}
  {1996})},\ \Eprint {https://arxiv.org/abs/astro-ph/9601039}
  {astro-ph/9601039} \BibitemShut {NoStop}%
\bibitem [{\citenamefont {Crittenden}\ \emph {et~al.}(2002)\citenamefont
  {Crittenden}, \citenamefont {Natarajan}, \citenamefont {Pen},\ and\
  \citenamefont {Theuns}}]{Crittenden.Theuns.2002}%
  \BibitemOpen
  \bibfield  {author} {\bibinfo {author} {\bibfnamefont {R.~G.}\ \bibnamefont
  {Crittenden}}, \bibinfo {author} {\bibfnamefont {P.}~\bibnamefont
  {Natarajan}}, \bibinfo {author} {\bibfnamefont {U.-L.}\ \bibnamefont {Pen}},\
  and\ \bibinfo {author} {\bibfnamefont {T.}~\bibnamefont {Theuns}},\
  }\bibfield  {title} {\bibinfo {title} {{Discriminating Weak Lensing from
  Intrinsic Spin Correlations Using the Curl-Gradient Decomposition}},\ }\href
  {https://doi.org/10.1086/338838} {\bibfield  {journal} {\bibinfo  {journal}
  {The Astrophysical Journal}\ }\textbf {\bibinfo {volume} {568}},\ \bibinfo
  {pages} {20} (\bibinfo {year} {2002})},\ \Eprint
  {https://arxiv.org/abs/astro-ph/0012336} {astro-ph/0012336} \BibitemShut
  {NoStop}%
\end{thebibliography}%

\appendix
\section{Accurate and efficient evaluation of Legendre multipoles}\label{sec:numerical-multipole-expansion}
\begin{figure}[t]
    \centering
    \includegraphics[width=\linewidth]{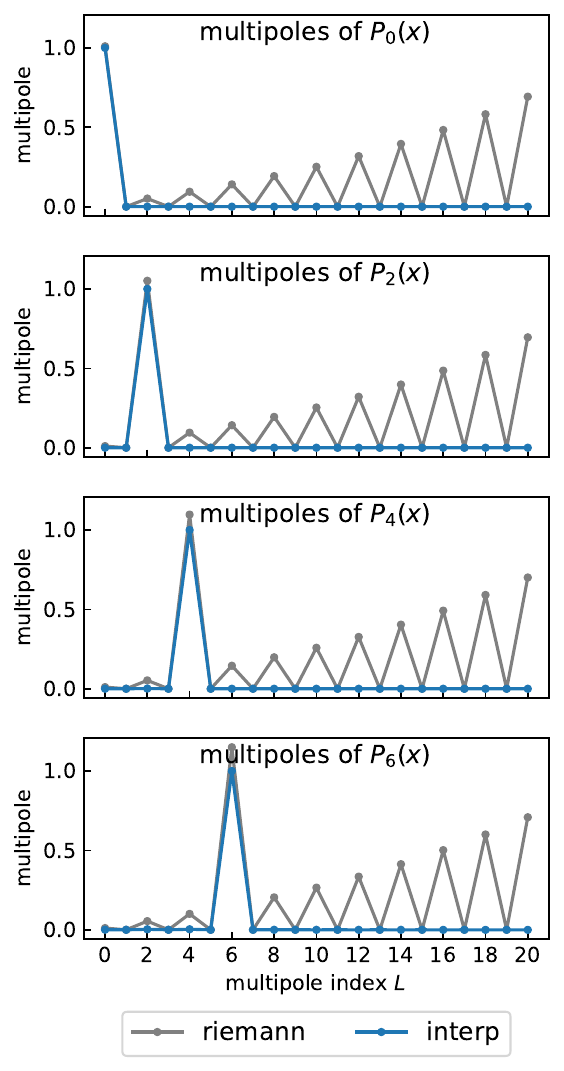}
    \caption{Demonstration of multipole estimation. The numerical estimates of multipoles are plotted as a function of multipole index $L$. In this example, we consider four Legendre polynomials as smooth functions, $P_L(\mu)$ where $L=0,2,4$ and $6$ from top to bottom panels. We sampled 100 points with $\mu$ equally spaced over $[-1,1]$. The gray points show the multipoles estimated by Riemann sum in Eq.~(\ref{eq:multipole-riemann}), showing artificial multipoles. The blue points show the multipoles estimated using the interpolation coefficients in Eq.(\ref{eq:multipole-interp}).}
    \label{fig:multipole-estimate}
\end{figure}

The multipoles of function $f(\mu)$ by Legendre function are defined as
\begin{align}
    f_L = \frac{2L+1}{2}\int_{-1}^{1}\dd \mu P_L(\mu)f(\mu).
\end{align}
One of the simplest numerical estimators of the multipoles is to approximate the above integral by a Riemann sum as 
\begin{align}
    \hat{f}_L^{N} = \frac{2L+1}{2}\sum_{i=0}^{N-1}\Delta x P_L(\mu_i)f(\mu_i),
    \label{eq:multipole-riemann}
\end{align}
where $N$ is the number of bins over $\mu$, $\Delta \mu=2/N$, and $\mu_i=i\Delta\mu$. This estimator is simple but gives an inaccurate estimate for high $L$ with a fixed $N$. To obtain an accurate multipole estimate with this Riemann sum, we need dense binning so that $\Delta\mu\sim1/L$, because the $L$th Legendre polynomial has an oscillatory feature with wavelength $\sim1/L$, which is inefficient. 

Another way to see the inaccuracy of the Riemann sum estimate is that it gives a non-zero multipole estimates when $f(\mu)$ is itself a Legendre function, $f(\mu) = P_{L^\prime}(\mu)$.  In this case, the multipoles should be zero for all $L \neq L^\prime$.  Fig.~\ref{fig:multipole-estimate} shows the result of the Riemann sum multipoles for several cases where $f(\mu)$ is a Legendre polynomial, which give very inaccurate results for $L \neq L^\prime$, especially as $L$ gets large. This is a big problem for our bispectrum case, because the bispectrum is a smooth function of the inner angle $\alpha$ and we expect that the higher $L$ multipoles to be small so that truncation works at some finite $L_{\rm max}$.

When the function $f(\mu)$ is a smooth function of $\mu$ so that its features can be captured by a modest number of multipoles, we can construct a much more efficient estimator of $f_L$ using coefficients of linear interpolation. Note that the cosmological bispectrum is usually a smooth function of $\mu=\cos\alpha$. We start with the $N$ data points $f(\mu_i)$ evaluated at each point $\mu_i$. We then approximate the function $f(\mu)$ by linear interpolation of the data as $a_i \mu+b_i$ within each bin, $\mu\in[\mu_i, \mu_{i+1}]$, where coefficients $a_i$ and $b_i$ are calculated from the data points. Using this linear interpolated function, we can analytically calculate the multipole coefficients of function $f(\mu)$ as
\begin{align}
    \hat{f}_L^{N} 
    &= \frac{2L+1}{2}\sum_{i=0}^{N-1}\int_{\mu_i}^{\mu_{i+1}}\dd\mu P_L(\mu)(a_i\mu+b_i) \nonumber\\
    &= \frac{1}{2}\sum_{i=0}^{N-1}(a_i p_L^{1i} + b_ip_L^{0i}).
    \label{eq:multipole-interp}
\end{align}
Here the integrated bases $p_L^{0i}$ and $p_L^{1i}$ are analytically given by
\begin{align}
    p_L^{0i} &= (2L+1)\int_{\mu_i}^{\mu_{i+1}}\dd\mu P_L(\mu) \nonumber\\
    &= \left[P_L(\mu)-P_{L-1}(\mu)\right]|_{\mu_i}^{\mu_{i+1}} \hspace{2em} (L\geq1) \\
    p_L^{1i} &= (2L+1)\int_{\mu_i}^{\mu_{i+1}}\dd\mu P_L(\mu)\mu \nonumber\\
    &=\frac{P_L(\mu)-P_{L-2}(\mu)}{2L-1} + \mu[P_{L+1}(\mu)-P_{L-1}(\mu)] \nonumber\\
    &\hspace{2em}+\left.\frac{P_L(\mu)-P_{L+2}(\mu)}{2L+3}\right|_{\mu_i}^{\mu_{i+1}} \hspace{2em} (L\geq2), 
\end{align}
and 
\begin{align}
    p_0^{0i} &= \mu|_{\mu_i}^{\mu_{i+1}} \\
    p_0^{1i} &= \mu^2/2|_{\mu_i}^{\mu_{i+1}}\\
    p_1^{1i} &= \mu^3|_{\mu_i}^{\mu_{i+1}},
\end{align}
for lower moments. In a similar way, we could extend this methodology to a higher-order interpolation scheme, e.g. cubic interpolation, if necessary. In this paper, we use linear interpolation for simplicity. 

In Fig.~\ref{fig:multipole-estimate}, we compare the performances of Eq.~(\ref{eq:multipole-riemann}) and Eq.~(\ref{eq:multipole-interp}). In this demonstration, we choose the Legendre polynomials with $L=0,2,4$ and $6$ as smooth test functions. By definition, the $L$-th multipole of Legendre polynomial $P_{L'}(\mu)$ is equal to one when $L=L'$ and zero otherwise. We sampled 100 points on $\mu$ equally spaced over $[-1,1]$. As already mentioned above, the multipole estimate from the Riemann sum approximation in Eq.~(\ref{eq:multipole-riemann}) gives inaccurate values, especially at high $L$. The method using interpolation coefficients in Eq.~(\ref{eq:multipole-interp}) gives accurate results for all $L$ values.

\section{Brute-force calculation of the natural components}
\label{sec:brute-force-natural-component}
In this appendix, we review the natural components derived in \citet{Schneider.Lombardi.2003} and \citet{Heydenreich.Schneider.2022}, and present a way to reduce the  computational time. We start with Eq.~(19) of \cite{Heydenreich.Schneider.2022}:
\begin{align}
\begin{split}
    \Gamma_0^\times(\theta_1, \theta_2, \phi) 
    &= \frac{-1}{2(2\pi)^3}\int_0^{2\pi}\dd(\Delta\beta)\int_0^{\pi/2}\dd\psi\int_0^\infty\dd\ln\ell\\
    &\times \sin(2\psi) e^{2\bar\beta} e^{-6i\alpha}\\
    &\times \ell^4 b_\kappa(\ell\sin\psi, \ell\cos\psi, \Delta\beta)J_6(\ell A),
    \label{eq:Gamma0-J6-def}
\end{split}
\end{align}
where $A$ and $\alpha$ are functions of $\ell_1, \ell_2, \theta_1, \theta_2, \Delta\beta$ and $\phi$
\begin{align}
    &\ell A = [(\ell_1\theta_1)^2+(\ell_2\theta_2)^2 \nonumber\\
    &\hspace{4em}+2\ell_1\ell_2\theta_1\theta_2\cos(\Delta\beta+\phi)]^{1/2}\\
    &A\cos\alpha = (\ell_1\theta_1+\ell_2\theta_2)\cos\left(\frac{\Delta\beta+\phi}{2}\right)\\
    &A\sin\alpha = (\ell_1\theta_1-\ell_2\theta_2)\sin\left(\frac{\Delta\beta+\phi}{2}\right).
\end{align}
Note that the phase factors, e.g. $e^{i(\phi_1-\phi_2)}$ in their paper, are removed because of the difference in shear projections. Also note that three different terms in their paper are all identical, because the bispectrum is invariant under any exchange of triangle sides, and we took the first term expression here. Introducing a polar coordinate for real space, $\theta_1=\theta\cos\tau, \theta_2=\theta\sin\tau$, we obtain 
\begin{align}
\begin{split}
    A &= \theta r(\psi, \Delta\beta; \tau, \phi) \\
    &\equiv \theta[(\cos\tau\cos\psi)^2+(\sin\tau\sin\psi)^2 \\
    &\hspace{5em}+\sin(2\tau)\sin(2\psi)\cos(\Delta\beta+\phi)/2]^{1/2}\\
\end{split}
\end{align}
and the function $\alpha(\psi, \Delta\beta; \tau, \phi)$ is given by
\begin{align}
    &A\cos\alpha = \theta\ell\cos(\tau-\psi)\cos\left(\frac{\Delta\beta+\phi}{2}\right)\\
    &A\sin\alpha = \theta\ell\cos(\tau+\psi)\sin\left(\frac{\Delta\beta+\phi}{2}\right).
\end{align}
In Eq.~(\ref{eq:Gamma0-J6-def}), the $\ell$ integration has the form of
\begin{align}
\begin{split}
    f(A; \psi, \Delta\beta) &= \int_0^\infty\dd\ln\ell~\ell^4J_6(\ell A)\\
    &\hspace{2em}\times b_\kappa(\ell\cos\psi,\ell\sin\psi, \Delta\beta).
    \label{eq:onej-integral}
\end{split}
\end{align}
For a given set of $(\psi, \Delta\beta)$, the $\ell$ integration can be efficiently performed using FFTLog. Once we evaluate the function on the FFT grid on $A$ for a given set of $(\psi, \Delta\beta)$, we can easily obtain the value of $f(A; \psi, \Delta\beta)$ for any entries of $A$ by interpolating the points on the FFT grid. The function $f$ has the following symmetry because of the mirror symmetry of the bispectrum due to weak lensing:
\begin{align}
    f(A; \pi/2-\psi,\Delta\beta) = f(A;\psi,2\pi-\Delta\beta) = f(A; \psi, \Delta\beta).
\end{align}
This is a similar symmetry to that of $\bar\beta$ (see the main text below Eq.~\ref{eq:angbar-sincos}). This property can be used to reduce the integration domain as $\varphi\in[0,2\pi]\rightarrow[0,\pi]$ and $\psi\in[0,\pi/2]\rightarrow[0,\pi/4]$, resulting in a fourfold increase in efficiency. The final expression is then
\begin{align}
    &\Gamma_0^\times(\theta_1,\theta_2,\phi)\nonumber\\
    &= \frac{-1}{2(2\pi)^3}\int_0^\pi\dd(\Delta\beta)\int_0^{\pi/4}\dd\psi\sin(2\psi)e^{2i\bar\beta}\nonumber\\
    &\hspace{1em} \times 
    [f(\theta r(\psi,\Delta\beta);\psi,\Delta\beta)\times e^{-6i\alpha(\psi,\Delta\beta)}\\
    &\hspace{1.5em}+f(\theta r(\pi/2-\psi,\Delta\beta);\psi,\Delta\beta)e^{-6i\alpha(\pi/2-\psi,\Delta\beta)}\nonumber\\
    &\hspace{1.5em}+f(\theta r(\psi,-\Delta\beta);\psi,\Delta\beta)e^{-6i\alpha(\psi,-\Delta\beta)}\nonumber\\
    &\hspace{1.5em}+f(\theta r(\pi/2-\psi,-\Delta\beta);\psi,\Delta\beta)e^{-6i\alpha(\pi/2-\psi,-\Delta\beta)}].\nonumber
\end{align}
Here, we omit the dependence of the functions $r$ and $\alpha$ on $\tau$ and $\phi$ for the sake of clarity.

\section{Multipole coupling functions with other multipole expansion bases}\label{sec:other-coupling-functions}
In this paper, we used Legendre polynomials as the basis of the bispectrum multipole expansion and derived 
the mode couping function of multipoles in Fourier and configuration spaces.
In this appendix, we consider the multipole expansion of the bispectrum using other bases and derive the 
corresponding multipole-coupling functions. To avoid confusion, we denote the multipole coupling function in Eq.~(\ref{eq:GLM}) as $G_{LM}^{P}(\psi)$. 
We start with the Fourier basis. With the Fourier basis, we define the multipole of the bispectrum as 
\begin{align}
    b_{\kappa}(\ell_1, \ell_2, \alpha) = \sum_{L=-\infty}^{\infty} b_\kappa^{(L)}(\ell_1, \ell_2) e^{iL\alpha}.
\end{align}
Then the multipole coupling function with this basis can be obtained by replacing the Legendre polynomial in Eq.~(\ref{eq:GLM}) with the Fourier basis:
\begin{align}
    &\int\dd(\dbeta)\dd\beta e^{iL\alpha}e^{i(6+m+n)[\beta-(\varphi_1+\varphi_2)/2]}\nonumber\\
    &\hspace{3em}\times
    e^{i[2\bar\beta+(m-n)(\dbeta+\pi-\phi)/2]}\nonumber\\
    =&(-1)^Le^{iM\phi}\delta^{K}_{6+m+n,0}\nonumber\\
    &\hspace{2em}\times2\pi\int_0^{2\pi}\dd(\dbeta) e^{i(2\bar\beta+(M-L)\dbeta)}\nonumber\\
    \equiv& (-1)^Le^{iM\phi}\delta^{K}_{6+m+n,0}G_{LM}^{F}(\psi).
\end{align}
We note that the multipole coupling function $G_{LM}^{F}$ depends on $L$ and $M$ only through their difference $L-M$.

We can also derive the relation between the multipole coupling functions $G_{LM}^{P}$ and $G_{LM}^{F}$ using the relationship between the two bases. 
The Legendre polynomial can be expanded with the Fourier basis using the coefficients $p_{LL'}$ as
\begin{align}
    P_L(\cos\alpha) = \sum_{L'=0}^{L}p_{LL'}e^{iL'\alpha}.
\end{align}
Here, since the Legendre polynomial is an even function of $\alpha$, the coefficients satisfy $p_{LL'}=p_{L(-L')}$ and the coefficients have non-zero values only when $L-L'$ is even.
Substituting the above equation into Eq.~(\ref{eq:GLM}), we obtain
\begin{align}
    G_{LM}^{P}(\psi) = \sum_{L'=-L}^{L}p_{LL'}(-1)^{L-L'}G_{L'M}^{F}(\psi).
\end{align}
In Section~\ref{sec:multipole-coupling}, we saw that there is a similar $\psi$ dependence if $|L-M|$ is constant. This can be understood from the above equation. Because the $L$-th Legendre polynomial has the largest contribution from the Fourier basis with $L'=L$, meaning $p_{LL}$ is the largest coefficient, most of the contribution comes from $G_{LM}^{F}$ which depends on $L$ and $M$ only through their difference. To prove Eq.~(\ref{eq:GLM-property-even-zero}), we start with the definition of $G_{LM}^{F}$. When $M-L$ is even, $M-L'$ is also even because of the $p_{LL'}$ property. For $\psi=0$, $\bar\beta(\psi, \dbeta)=\dbeta$. Then the coupling function is
\begin{align}
    G_{L'M}^{F}(\psi=0) = \int_0^{2\pi}\dd(\dbeta) e^{i(M-L'+1)\dbeta}=0
\end{align}
where the last equality holds because $M-L'+1$ is an odd number. In the same way, we can prove prove Eq.~(\ref{eq:GLM-property-odd-zero}). When $M-L$ is odd, $M-L'$ is also odd. For $\psi=\pi/4$, $\bar\beta(\psi, \dbeta)=0$. Then the coupling function is 
\begin{align}
    G_{L'M}^{F}(\psi=\pi/4) = \int_0^{2\pi}\dd(\dbeta) e^{i(M-L')\dbeta}=0.
\end{align}

In the same way, we can derive the multipole coupling functions with cosine and sine basis. We define these expansion as 
\begin{align}
    b_{\kappa}(\ell_1, \ell_2, \alpha) &= \sum_{L=-\infty}^{\infty} b_\kappa^{(L)}(\ell_1, \ell_2) \cos(L\alpha)\\
    b_{\kappa}(\ell_1, \ell_2, \alpha) &= \sum_{L=-\infty}^{\infty} b_\kappa^{(L)}(\ell_1, \ell_2)i\sin(L\alpha)
\end{align}
Note that we include the imagirnary unit in the sine basis. The multipole coupling functions with these bases are then given by
\begin{align}
    G_{LM}^{\cos}(\psi) &= 4\pi\int_0^\pi \dd(\dbeta) \cos(L\dbeta)\nonumber\\
    &\hspace{2em}\times\cos(M\dbeta+2\bar\beta) \\
    G_{LM}^{\sin}(\psi) &= 4\pi\int_0^\pi \dd(\dbeta) \sin(L\dbeta)\nonumber\\
    &\hspace{2em}\times\sin(M\dbeta+2\bar\beta).
\end{align}
For the application to the three-point correlation function including contributions from the asymmetric bispectrum, e.g. for the TATT model of intrinsic alignment, we have to use the basis including mirror asymmetric terms as well, e.g. the Fourier basis.

\section{Useful identities}\label{sec:identities}
Here, we present some useful identities that we used when deriving the multipole method for the 3PCF in this paper. 

The partial-wave expansion of a plane wave in cylindrical coordinates is 
\begin{align}
    e^{i\bm{\ell}\cdot \bm{x}} = \sum_{m=-\infty}^{\infty}i^m J_{m}(\ell x) e^{im(\beta-\varphi)}.
    \label{eq:plane-wave-exp}
\end{align}

The Bessel function of order $-n$ is related to the Bessel function of order $n$ as 
\begin{align}
    J_{-n}(x) = (-1)^nJ_n(x)
\end{align}

The Legendre polynomial is an even (odd) function when $L$ is even (odd).
\begin{align}
    P_L(-x) = (-1)^LP_L(x)
\end{align}

\end{document}